\documentclass[journal ]{new-aiaa}
\usepackage[utf8]{inputenc}
\usepackage{textcomp}

\usepackage{graphicx}
\usepackage{amsmath}
\usepackage{nth}
\usepackage[version=4]{mhchem}
\usepackage{siunitx}
\usepackage{longtable,tabularx}
\usepackage[labelformat=empty, position=top]{subcaption}
\usepackage[export]{adjustbox}

\title{Effect of Tripping and Domain-Width on Transonic Buffet on Periodic NASA-CRM Airfoils}

\author{David J. Lusher\footnote{JSPS Research Fellow, Aircraft Lifecycle Innovation Hub, JAXA Chofu Aerospace Center, Tokyo.}, Andrea Sansica\footnote{Researcher, Aircraft Lifecycle Innovation Hub, JAXA Chofu Aerospace Center, Tokyo.}, and Atsushi Hashimoto \footnote{Manager, Aircraft Lifecycle Innovation Hub, JAXA Chofu Aerospace Center, Tokyo.}}
\affil{Japan Aerospace Exploration Agency (JAXA), 7-44-1 Jindaiji Higashi-machi, Chofu-shi, Tokyo 182-8522, Japan}

\begin{document}

\maketitle

\begin{abstract} 
Transonic buffet is an instability characterized by shock-oscillations and separated boundary-layers. High-fidelity simulations have typically been limited to narrow domains to be computationally feasible, overly constraining the flow and introducing modelling errors. Depending on the boundary-layer state upstream of the interaction, different buffet features are observed. High-fidelity simulations were performed on the periodic NASA-CRM wing at moderate Reynolds number to assess sensitivity of the two-dimensional transonic buffet to boundary-layer state and domain width. Simulations were cross-validated against RANS/URANS and global stability analysis and excellent agreement was found near the onset. By varying the boundary-layer tripping amplitude, laminar, transitional, and turbulent buffet interactions were obtained. All cases consisted of a single shock and low-frequency oscillations $(St \approx 0.07)$. The transitional interaction also exhibited reduced shock movement, a $15\%$ increase in $\overline{C_L}$, and energy content at higher frequencies $(St \approx 1.3)$. Span-wise domain studies showed sensitivity at the shock location and near the trailing edge. We conclude that the span-width must be greater than the trailing-edge boundary-layer thickness to obtain span-independent solutions. For largely separated cases, the sensitivity to span-width increased and variations across the span were observed. This was found to be associated to a loss of two-dimensionality of the flow.
\end{abstract}

\section*{Nomenclature}
{\renewcommand\arraystretch{1.0}
\noindent\begin{longtable*}{@{}l @{\quad=\quad} l@{}}
$c$  & Airfoil Chord Length \\
$u, v, w$ & Velocity Components\\
$x_t$, $A$, $\sigma$ & Trip Location, Trip Amplitude, and Gaussian Scaling Factor\\
$x, y, z$ & Cartesian Coordinates\\
$AR$ & Aspect Ratio of the Airfoil, $AR=L_z / c$\\
$C_L, C_D, C_f$ & Lift, Drag, and Skin-friction Coefficients\\
$C_{Dp}, C_{Df}$ & Pressure and Skin-friction Drag components \\
$L_z$ & Airfoil Spanwise Length \\
$M_{\infty}$ & Freestream Mach number\\
$Pr$ & Prandtl number\\
$Re$ & Reynolds number\\
$St$ & Strouhal Number (non-dimensional frequency)\\
$T_{\textrm{ref}}, T_{\textrm{Suth}}$ & Reference and Sutherland's temperature \\
$\alpha$	& Angle of Attack (degrees)\\
$\gamma$ & Ratio of Specific Heat Capacities\\
$k_i, \Phi_i, \omega_i$ & Tripping Wavenumbers, Phases, and Frequencies\\
$\rho, p, T, \rho E, \mu$ & Density, Pressure, Temperature, Total Energy, and Viscosity\\
$\xi, \eta, \zeta$ & Curvilinear Coordinates\\
$\Delta t$ & Computational Time-step
\end{longtable*}}

\section{Introduction}
\label{sec:introduction}
\lettrine{T}ransonic buffet is an aerodynamic instability related to shockwave/boundary-layer interactions (SBLI) \citep{D2001} and is characterized by periodic self-sustained shock oscillations and boundary-layer separation \citep{L1990,L2001}. This instability is relevant at the boundaries of the flight envelope of commercial aircraft, namely for high speeds and high angles of attack. Transonic shock buffet can cause large amplitude oscillations in lift and drag, leading to reduced control and increased fatigue and failure rates \citep{Giannelis_buffet_review}. Buffet is known to consist of both a two-dimensional (2D) chord-wise shock oscillation instability, and three-dimensional (3D) cross-flow outboard propagating cellular separation patterns, known as buffet/stall-cells. While these two instabilities can coexist under specific conditions on 3D idealized wings \citep{plante2020towards, CGMT2009, CGS2019,PBDSR2019}, they have been traditionally categorized based on geometrical features. The 2D shock-oscillation instability is generally linked to 2D airfoils and 3D unswept infinite wings \citep{JMDMS2009,CGMT2009,AK2023}, and the 3D buffet cells one is related to 3D swept infinite wings \cite{IR2015,CGS2019,PDL2020,PDBLS2021,SHKK2022}, finite wings \cite{OIH2018} and full aircraft \cite{MTP2020,T2020,SNKNNA2021,SH2023} configurations. The debate on the applicability of state-of-the-art solvers for industry relevant configurations, sustained by communities such as the AIAA Drag Prediction Workshop \cite{T2019dpw,dpw-summ1}, is driving the interest towards 3D configurations, hence with more focus on the buffet cells instability. However, the fundamental mechanisms of the 2D shock-oscillations are not yet fully understood and require further scrutiny.

While earlier studies proposed feedback loop models based on upstream and downstream travelling waves from the trailing edge and shock foot \citep{L1990} to explain the 2D shock oscillations, more recent studies have linked the origins of transonic buffet to a global instability \citep{CGMT2009,SMS2015}. In this scenario, the onset of the unsteadiness is the result of a Hopf bifurcation \citep{CGMT2009}, and the instability is localized to the region around the shock and partially in the separated shear layer \citep{SMS2015}. Despite some rectifications of the feedback loop model \cite{D2005,JMDMS2009,HFS2013}, the explanation based on global stability remains the most popular but does not clarify the full mechanisms that lead to self-sustained shock oscillations. In this regard, \citet{KAWAI_RESOLVENT2023} recently used resolvent analysis to suggest that shock induced separation height and pressure dynamics around the shock-wave both contribute to the self-sustained oscillations.

To accurately calculate the 2D buffet instability, the prediction of the separation size is of fundamental importance and a variety of approaches have been extensively used for this purpose. The computational cost for scale-resolving methods applied to buffet can be prohibitive, and most numerical studies have therefore been limited to using low-fidelity approaches such as those used in Reynolds-Averaged Navier-Stokes (RANS) solvers, or mixed-fidelity hybrid RANS/Large-Eddy Simulation (LES). Examples of low-fidelity/reduced-cost methods applied to two-dimensional buffet include: Unsteady RANS (URANS) \citep{TC2006}, modal decomposition and system identification \citep{CGMT2009,CGS2019,PRD2019,SLKHR2022} and hybrid methods, such as Detached Detached-Eddy Simulation (DDES) \citep{GBH2014} and Zonal DES (ZDES) \citep{D2005}, among others. Lower-fidelity methods are generally suitable to predict 2D buffet characteristics, and have shown good agreement to experiment \citep{D2005,JMDMS2009}. However, as shown by \citet{TC2006}, predictions obtained by RANS-based solvers can be sensitive to the turbulence model used. The review of \citet{Giannelis_buffet_review} further commented that URANS simulations exhibit high sensitivity to simulation parameters, turbulence model, and both the spatial and temporal discretisation methods used. Finally and just as importantly, RANS models are also known to perform poorly in the presence of large flow separation. These factors all motivate the need for high-fidelity studies of buffet, which explicitly resolve the energy-containing eddies for more accurate predictions of SBLI and unsteady shock-induced separation \citep{FK2018}.

The work of \citet{MBP2018} compared URANS and DDES methods to scale-resolving Implicit LES (ILES), for the V2C supercritical laminar wing at Reynolds number ($Re$, based on free-stream quantities and airfoil chord, $c$) of $Re=3\times10^{6}$. ILES \citep{grinstein2007implicit} (or, alternatively, under-resolved Direct Numerical Simulation) is the approach taken in the current work. ILES does not require additional turbulence modelling, as the governing equations are solved directly, with the numerical dissipation from the shock-capturing scheme and filters acting as a sub-grid scale model \citep{Garnier_ILES_1999, grinstein2007implicit, MBP2018,fu2023review}. For high-fidelity airfoil simulations at both low \citep{AK2023} and high \citep{GD2010} speeds, the size of the separation is shown to be sensitive to the airfoil span-width ($L_z$), that therefore needs to be appropriately selected. At high angle of attack low-speed stall, \citet{AK2023} tested domain sensitivity for separated airfoils at aspect ratios ($AR=L_z / c$) of $AR=0.05, 0.1, 0.2$. It was found that the narrow domains caused excessive vertical movement of the separated shear-layer and the overly-constrained wake vortices failed to become fully three-dimensional, leading to large fluctuations in aerodynamic coefficients and predictions sensitive to the domain width used. Similarly, the stall simulations of \citet{TJW2020} tested span-widths up to $100\:\%$ of chord length ($AR=1$), to assess the effect of domain width on aeroacoustic predictions. It was shown that even relatively wide $AR=0.2$ domains over-predicted noise generation by $10 \: dB$ compared to $AR=1$. Aside from the domain-width tests between $AR=0.0365$ and $AR=0.073$ of \citet{GD2010}, less attention has been given to domain-width sensitivity in the high-speed buffet regime, motivating the need for the present work. Most of the studies in the literature set airfoil span-widths based on general know-how or referring to studies on similar flow configurations, without directly addressing the domain width sensitivity problem. Due to computational cost, these kind of high-fidelity simulations of periodic wings have typically been limited to narrow aspect ratios. Some relevant examples for 2D buffet studies used $AR=0.0365 - 0.073$ \citep{GD2010}, $AR=0.065$ \citep{FK2018,Nguyen2022}, $AR=0.05$ \citep{moise_zauner_sandham_2022,Moise2023_AIAAJ}, and, most recently, $AR=0.25$ \citep{LongWong2024_Laminar_buffet}. \citet{GD2010} doubled the spanwise width of their simulations from 3.65\% to 7.3\% and showed significant reduction of the pressure fluctuations at the trailing edge \citep{Giannelis_buffet_review}. The wider domain `better captures trailing edge pressures by allowing three-dimensional coherent structures to develop' \citep{Giannelis_buffet_review}. This highlights that while scale-resolving simulations explicitly resolve turbulence and should lead to more accurate buffet predictions, this cannot be at the expense of overly-narrow domains which artificially constrain the problem and introduce an alternative form of modelling error by imposed two-dimensionality on turbulent structures. Two wide-span high-fidelity outliers are the ILES of \citet{MBP2018} ($AR=0.5$) and that of \citet{MarkusPRF_2020} ($AR=1$), although the upstream boundary-layer was un-tripped in these cases.

Another important consideration that affects separation size (and consequently the selection of an appropriate airfoil domain width) is the nature of the incoming boundary-layer. In the present work, we categorize the interaction type by the state of the boundary-layer upstream of the shock-wave. Based on this definition, transonic buffet can be separated into laminar buffet \citep{dandois_mary_brion_2018,MarkusPRF_2020,moise_zauner_sandham_2022,LongWong2024_Laminar_buffet}, and turbulent buffet \citep{FK2018,Nguyen2022}. Turbulent buffet is characterized by streamwise shock-wave oscillations in the Strouhal number range of $St=[0.06, 0.1]$ \citep{plante2020towards} ($St$, based on the freestream velocity $U_{\infty}^{*}$ and airfoil chord $c^{*}$). The shock-wave terminates a large region of supersonic flow on the suction side of the airfoil. A complex inviscid/viscous coupling between the shock and the shear-layer lead to a periodically separating and reattaching boundary-layer. In contrast to turbulent buffet, laminar buffet typically has more extensive regions of flow separation, and additional higher frequencies present ($St=1-1.2$), which have been observed both experimentally \citep{Brion2017_Laminar_Experiment} and computationally \citep{dandois_mary_brion_2018} at $Re=3\times 10^6$. These higher frequencies have been linked to a breathing of the separation bubble at the shock foot. Later complementary experiments \citep{Brion2020_Laminar_Experiment} showed the same high-frequency instability, accompanied by a low-frequency shock oscillation with $St = 0.05$. The low-frequency shock oscillations were similar to those observed in the turbulent interaction. In a separate computational study, \citet{Moise2023_AIAAJ} simulated cases at a moderate Reynolds number of $5\times 10^5$ with both free- and forced-transition, observing similar ranges of low-frequency peaks for the two-dimensional instability in both cases. Despite the aforementioned differences between laminar and turbulent buffet, \citet{Moise2023_AIAAJ} argued that the underlying mechanism shares the same origin. However, it should also be noted that in the free-transition simulations of \citet{moise_zauner_sandham_2022,Moise2023_AIAAJ}, the laminar interaction was comprised of multiple shock-wave/expansions across the suction side of the airfoil, which is in contrast to the single shock-wave configurations found in laminar simulations and experiments of \citep{Brion2017_Laminar_Experiment,dandois_mary_brion_2018,Brion2020_Laminar_Experiment}. The authors of \cite{moise_zauner_sandham_2022} showed that the number of shocks decreased with increasing Reynolds number, and suggested the differences in shock topology in the laminar cases may be due to a Reynolds number effect, consistent with studies at higher Reynolds number \citep{Brion2017_Laminar_Experiment,dandois_mary_brion_2018,Brion2020_Laminar_Experiment}. \citet{Garbaruk2021} applied RANS and GSA to assess the impact of extended laminar flow on infinite-span wings at buffet onset. They found that extending the laminar flow region on the suction side led to a $5-6\%$ increase in lift, while the boundary-layer state on the pressure side had a negligible effect. The present study focuses mainly on turbulent transonic buffet, as it is the most commonly found in experiments and industrial aviation applications \citep{MTP2020, SKNNNA2018, SNKNNA2021, SKK2022}. However, the sensitivity and response of buffet to upstream boundary-layer state are not entirely clear. Given the potential fuel efficiency improvements associated with the use of laminar wings to meet climate objectives, laminar and transitional interactions in the transonic regime are of great importance and require further attention.


\noindent Within the framework presented above, the focus in this paper is on determining the sensitivity of 2D transonic buffet to numerical tripping and domain width, for cases characterized by different separation sizes. The objectives of this work are to assess:
\begin{itemize}
\item{The sensitivity of turbulent transonic buffet to tripping strength and, as a result, to different separation sizes determined by laminar, transitional, and turbulent shock-wave/boundary-layer interactions.}
\item{The potential influence of spanwise domain width on the 2D buffet shock-oscillation instability in high-fidelity 3D (ILES) simulations, for buffet conditions with moderate and large separations.}
\item{The extent to which low-fidelity industry-relevant RANS/URANS and modal stability (GSA) methods can correlate to data from high-fidelity simulations at greatly reduced computational cost.}
\end{itemize}

\noindent The current work is organized as follows. Section~\ref{sec:opensbli} and Section~\ref{sec:fastar} describe the solvers used for ILES (OpenSBLI \citep{OpenSBLI_2021_CPC}) and RANS/URANS/GSA (FaSTAR \citep{FaSTAR_Hashimoto_2012}), respectively. Section~\ref{sec:mesh} provides information about the computational meshes used, with simulation parameters and flow conditions given in Section~\ref{sec:flow_conditions}. Section~\ref{sec:finding_buffet} performs initial ILES simulations to identify pre- and post-onset 2D buffet conditions. This is then compared to steady-RANS, GSA and URANS predictions in Section~\ref{sec:RANS_compare}. ILES simulations are then used in Section~\ref{sec:tripping_parametric} and Section~\ref{sec:spanwidth_sensitivity} to assess sensitivity to tripping strength and span-wise domain width, respectively. The sensitivity to domain span-width is first determined at moderate angles of attack (Section~\ref{sec:spanwidth_moderate}), where the flow is mostly attached in a time-averaged sense, and then at a higher angle of attack (Section~\ref{sec:spanwidth_large}), where there is extensive time-averaged flow separation present. Finally, conclusions are given in Section~\ref{sec:conclusions}.

\section{Computational Method}\label{sec:computational_method}

\subsection{OpenSBLI high-fidelity (ILES) solver}\label{sec:opensbli}

All high-fidelity simulations in this work were performed in OpenSBLI \citep{OpenSBLI_2021_CPC}, an open-source high-order compressible multi-block flow solver on structured curvilinear meshes. OpenSBLI was developed at the University of Southampton \citep{LUSHER201817,OpenSBLI_2021_CPC} and the Japan Aerospace Exploration Agency (JAXA) \citep{LZSH2023} to perform high-speed aerospace research. Written in Python, OpenSBLI utilises symbolic algebra to automatically generate a complete finite-difference CFD solver in the Oxford Parallel Structured (OPS) \citep{Reguly_2014_OPSC} Domain-Specific Language (DSL). The base governing equations are the non-dimensional compressible Navier-Stokes equations for an ideal fluid. Applying conservation of mass, momentum, and energy, in the three spatial directions $x_i$ $\left(i=0, 1, 2\right)$, results in a system of five partial differential equations to solve. These equations are defined for a density $\rho$, pressure $p$, temperature $T$, total energy $E$, and velocity components $u_k$ as
\begin{align}\label{ns_eqn}
\frac{\partial \rho}{\partial t} &+ \frac{\partial}{\partial x_k} \left(\rho u_k \right) = 0,\\
\frac{\partial}{\partial t}\left(\rho u_i\right) &+ \frac{\partial}{\partial x_k} \left(\rho u_i u_k + p \delta_{ik} - \tau_{ik}\right) = 0,\\
\frac{\partial}{\partial t}\left(\rho E\right) &+ \frac{\partial}{\partial x_k} \left(\rho u_k \left(E + \frac{p}{\rho}\right) + q_k - u_i \tau_{ik}\right) = 0,
\end{align}
with heat flux $q_k$ and stress tensor $\tau_{ij}$ defined as 
\begin{equation}\label{heat_flux}
q_k = \frac{-\mu}{\left(\gamma - 1\right) M_{\rm{ref}}^{2} Pr Re}\frac{\partial T}{\partial x_k},
\end{equation}
\begin{equation}\label{stress_tensor}
\tau_{ik} = \frac{\mu}{Re} \left(\frac{\partial u_i}{\partial x_k} + \frac{\partial u_k}{\partial x_i} - \frac{2}{3}\frac{\partial u_j}{\partial x_j} \delta_{ik}\right).
\end{equation}
$Pr$, $Re$, and $\gamma=1.4$ are the Prandtl number, Reynolds number and ratio of specific heat capacities for an ideal gas, respectively. Support for curvilinear meshes is provided by using body-fitted meshes with a coordinate transformation. The equations are non-dimensionalized by a reference velocity, density and temperature $\left(U^{*}_{\rm{ref}}, \rho^{*}_{\rm{ref}}, T^{*}_{\rm{ref}}\right)$. For a reference Mach number $M_{\rm{ref}}$, the pressure is defined as 
\begin{equation}\label{pressure_eqn}
p = \left(\gamma - 1\right) \left(\rho E - \frac{1}{2} \rho u_i u_i\right) = \frac{1}{\gamma M^{2}_{\rm{ref}}} \rho T,
\end{equation}

OpenSBLI is explicit in both space and time, with a range of different discretisation options available to users. Spatial discretisation is performed in this work by \nth{4} order central differences recast in a cubic split form \citep{Coppola_CubicSplit_2019} to boost numerical stability. Time-advancement is performed by a \nth{4}-order 5-stage low-storage Runge-Kutta scheme \citep{carpenter_kennedy_1994}. Dispersion Relation Preserving (DRP) filters \citep{BOGEY2004194} are applied to the freestream using a targeted filter approach which turns the filter off in well-resolved regions to further reduce numerical dissipation \citep{LZSH2023}. The DRP filters are also only applied once every 25 iterations. Shock-capturing is performed via Weighted Essentially Non-Oscillatory (WENO) schemes, specifically the \nth{5}-order WENO-Z variant by \cite{Borges2008}. The effectiveness and resolution of the underlying shock-capturing schemes in OpenSBLI was assessed for the compressible Taylor-Green vortex case involving shockwaves and transition to turbulence in \citep{lusher2021assessment} and for compressible wall-bounded turbulence in \citep{hamzehloo2021_IJNMF}. The shock-capturing scheme is applied within a characteristic-based filter framework \citep{Yee1999_filter_schemes,Yee2018}. The dissipative part of the WENO-Z reconstruction is applied at the end of the full time-step to capture shocks based on a modified version of the Ducros sensor \citep{DUCROS1999517,Bhagatwala2009}. Similar central plus filter step shock-capturing schemes have been used for a number of buffet studies, most recently in the turbulent cases of \citet{Moise2023_AIAAJ}. The numerical methods in OpenSBLI were validated for laminar-transitional buffet cases compared to literature on the V2C airfoil geometry in \cite{LZSH2023}.

\subsection{FaSTAR low-fidelity (RANS/URANS/GSA) solver}\label{sec:fastar}
Comparisons to the high-fidelity ILES data are provided by the FaSTAR unstructured mesh CFD solver \citep{FaSTAR_Hashimoto_2012,IIHAT2016} developed at JAXA. A cell-centered finite volume method is used for the spatial discretisation of the compressible 3D RANS equations. The numerical fluxes are computed by the Harten-Lax-van Leer-Einfeldt-Wada (HLLEW) scheme \cite{O1995} and the weighted Green-Gauss method is used for the gradient computation \cite{M2003}. For the mean flow and transport equations, the spatial accuracy is set to the second and first order, respectively. The turbulence model selected for the present simulations is the Spalart-Allmaras turbulence model \cite{SA1992} with rotation correction (SA-R) \cite{DM1995} and quadratic constitutive relation 2000 version (SA-R-QCR2000) \cite{S2000}. No-slip velocity and adiabatic temperature boundary conditions are imposed on the wing walls; far-field boundary conditions are employed at the outer boundaries, and the angle of attack is applied to the incoming flow. 

To compute the steady solutions, the Lower/Upper Symmetric Gauss-Seidel (LU-SGS) \cite{Setail1998} scheme with a Courant–Friedrichs–Lewy (CFL)-fixed local time step is used. The CFL number is 50 and all simulations are run until convergence. Iterative convergence is here defined as a) when the residuals of the state variables in the $L_2$-norm are at least below $10^{-8}$ and b) the variations of the lift-coefficient are below the fourth decimal digit. To promote iterative convergence, the selective frequency damping (SFD) method \cite{ABHHMS2006, RLM2016} is used in the LU-SGS pseudo-time integration, as previously done by \citet{PDBLS2021}. For the unsteady calculations, dual-time stepping \cite{Vetal2000} is used to improve accuracy of the implicit time integration method. The LU-SGS scheme is used for the pseudo-time sub-iterations and the physical time derivative is approximated by the three-point backward difference.

To perform GSA, the linearized URANS equations need to be derived and solved. These equations assume the existence of a steady equilibrium solution, referred to as base flow, to perform the stability calculation. In the URANS-based GSA framework, the base flow coincides with the (steady) RANS solution. The details and validation of the solver linearization and Jacobian-free Newton-Krylov method used for the modal decomposition are thoroughly described in \citet{SH2023}.

\subsection{Geometry and mesh configuration}\label{sec:mesh}

\begin{figure}[h]
\begin{center}
\includegraphics[width=0.497\textwidth]{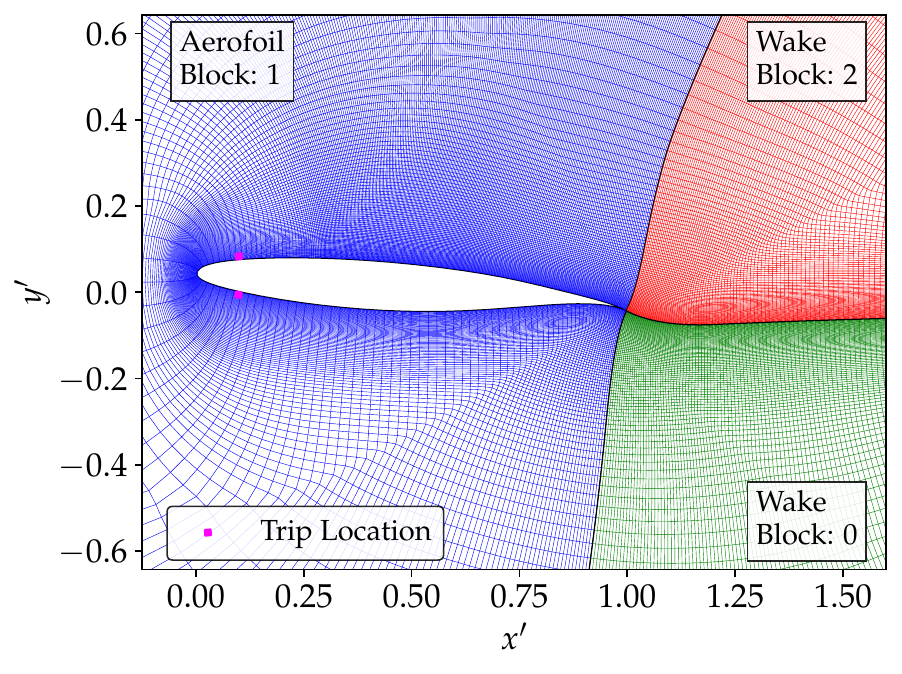}
\includegraphics[width=0.497\textwidth]{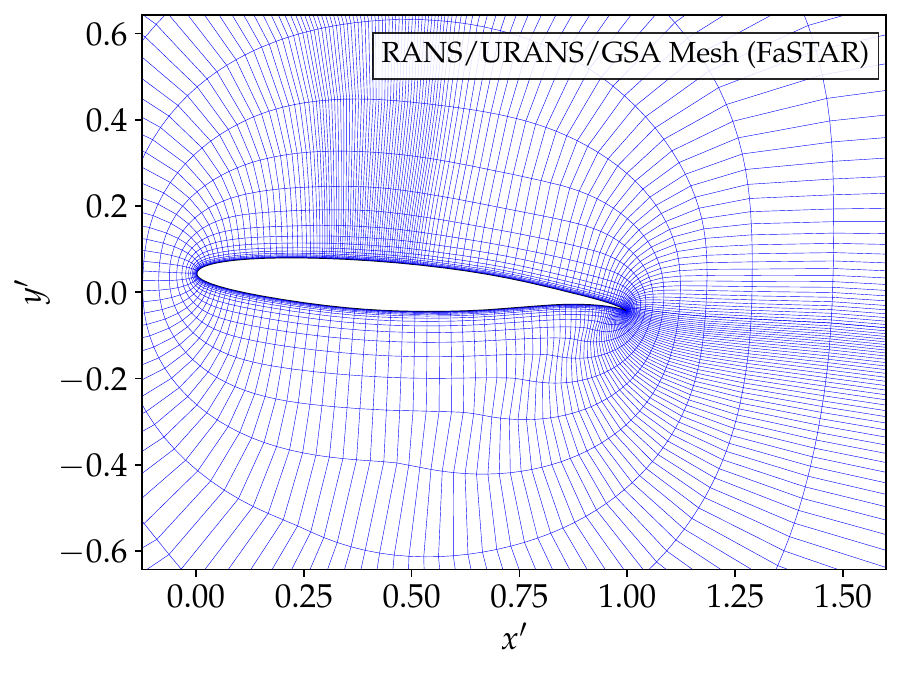}
\caption{Visualisation of the mesh configurations for the NASA-CRM-65 \citep{CRM} airfoil used in this study at $\alpha = 5^{\circ}$. The ILES grid is comprised of a C-mesh and two wake blocks, with boundary-layer trips placed at $0.1c$. The URANS grid is comprised of a single block O-mesh. For illustrative purposes, the ILES and URANS grids are plotted at every \nth{7} and \nth{5} grid line, respectively.}
\label{fig:mesh_blocks}
\end{center}
\end{figure}

The selected geometry is the 65\% semispan station of the NASA Common Research Model (CRM) wing, commonly used for turbulent transonic buffet research. Two-dimensional body-fitted structured meshes are created in Pointwise\texttrademark. The three-dimensional mesh is generated by extruding the two-dimensional grid in the spanwise direction with uniform spacing. Figure \ref{fig:mesh_blocks} shows the meshes used by the (a) ILES simulations in OpenSBLI, and (b) RANS/URANS/GSA simulations in FaSTAR, plotted at every \nth{7} and \nth{5} line respectively for visualization purposes. 

In the case of OpenSBLI, an airfoil C-mesh is connected to two wake blocks with a sharp trailing edge configuration. The in-flow boundary is set at a distance of $25c$ with the outlet $5c$ downstream of the airfoil. The inflow is set to be uniform $u_{\infty}=1$, with the angle of attack prescribed by rotating the airfoil within the mesh. For each case, the near-wake mesh is also slightly modified to take into account the deflection of the wake based on the AoA. In the $\xi$ and $\eta$ directions clockwise around the airfoil and normal to the surface, the airfoil and wake blocks have $\left(2249, 681\right)$ and $\left(701, 681\right)$ points, respectively. Around the airfoil, the pressure and suction sides have 500 and 1749 points in the $\xi$ direction, respectively. The $\xi$ distribution is refined between $0.3 < x < 0.6$ on the suction side to improve the resolution at the main shock-wave and SBLI. The domain width is initially set to a narrow value of $5\%$ chord length with $N_z = 50$ points to compare to steady RANS predictions. Upstream of the main shock-wave at $x=0.4$, the grid has wall units of $\left(\Delta x^{+} , \Delta y^{+}, \Delta z^{+} \right) = \left(6.1, 2.2, 14.8\right)$, and in the attached turbulent region downstream of the shock reaches a maximum at $x=0.7$ of $\left(\Delta x^{+} , \Delta y^{+}, \Delta z^{+} \right) = \left(3.9, 1.1, 7.6\right)$. In addition to wall criteria, it is important to maintain good resolution throughout the entire boundary-layer by applying only weak grid stretching. At $x=0.4$ and $x=0.7$ there are 80 and 195 points in the boundary-layer, respectively. A spanwise grid study at buffet conditions is shown in appendix~\ref{sec:grid_study}, with the medium resolution selected for the cases shown in the rest of this work. Aerodynamic coefficients, pressure distributions and skin-friction are all span-averaged unless otherwise mentioned.

In the case of the cell-centered finite volume FaSTAR solver, the blunt trailing-edge version of the CRM wing is used. The numerical mesh is obtained by first defining the distribution of cells around the airfoil and then by normal extrusion to obtain a single block O-grid. The number of cells in the $\xi$ and $\eta$ directions is $\left(1050, 162\right)$. The distribution around the airfoil consists of $600$ and $400$ cells on suction and pressure sides respectively, and $50$ cells are used to discretize the blunt trailing edge. A region of chord-wise width equal to $0.2 c$ is refined around the shock and counts $200$ cells. To account for different shock locations, this refinement region changes chord-wise position depending on the angle of attack. The domain boundaries extend to about $100$ chords from the airfoil in all directions. Unless specified otherwise, all FaSTAR grids are purely 2D (i.e. only one cell in the spanwise direction).

\subsection{Flow parameters, computational setup and initial conditions}\label{sec:flow_conditions}
All simulations were performed at a moderate Reynolds number of $Re=500,000$ based on airfoil chord length and freestream Mach number of $M_\infty=0.72$. The non-dimensional time-step is set as $\Delta t = 5\times 10^{-5}$ for all OpenSBLI ILES cases. The ILES simulations are advanced from uniform flow conditions for 20 time units until the boundary-layer is fully turbulent and the buffet unsteadiness fully develops.

In order to investigate turbulent transonic buffet, numerical tripping must be applied to the oncoming boundary-layer to promote a fast transition to turbulence upstream of the shockwave. This is achieved by forcing a set of unstable modes as a time-varying blowing/suction strip near the leading edge of the airfoil. This type of forcing is commonly used in CFD research as a method to mimic arrays of tripping dots used in experiments \citep{SKNNNA2018,SKK2022}. The forcing strip is centred around the $0.1c$ location on both the suction and pressure sides of the airfoil. The forcing is applied to the wall-normal velocity component, which is then used to set the momentum and total energy on the wall. Outside of the forcing strip the wall is a standard isothermal no-slip viscous boundary condition. The forcing is taken to be a modified form of the one given in \cite{Moise2023_AIAAJ} as
\begin{equation}\label{eq:tripping_eqn} 
\rho v_w=\sum_{i=1}^3 A \exp \left(-\frac{\left(x-x_t\right)^2}{2 \sigma^2}\right) \sin \left(\frac{k_i z}{0.05c}\right) \sin \left(\omega_i t+\Phi_i\right),
\end{equation}
for simulation time $t$, trip location $x_t$, and Gaussian scaling factor $\sigma=0.00833$. The three modes $\left(0, 1, 2\right)$ have spatial wavenumbers of $k_i = \left(6\pi, 8\pi, 8\pi\right)$, phases $\Phi_i = \left(0, \pi, -\pi/2\right)$, and temporal frequencies of $\omega_i = \left(26, 88, 200\right)$. The tripping strength is initially set to $7.5\%$ of the freestream ($A=0.075$), to initiate the transition to turbulence. The sensitivity of the 2D buffet instability to this tripping strength parameter is investigated in section \ref{sec:tripping_parametric} over a range of $0.5\%$ to $10\%$ of freestream velocity.

\section{Identifying buffet conditions on the NASA-CRM profile \label{sec:finding_buffet}}

\begin{figure}[h]
\begin{center}
  \includegraphics[width=0.497\columnwidth]{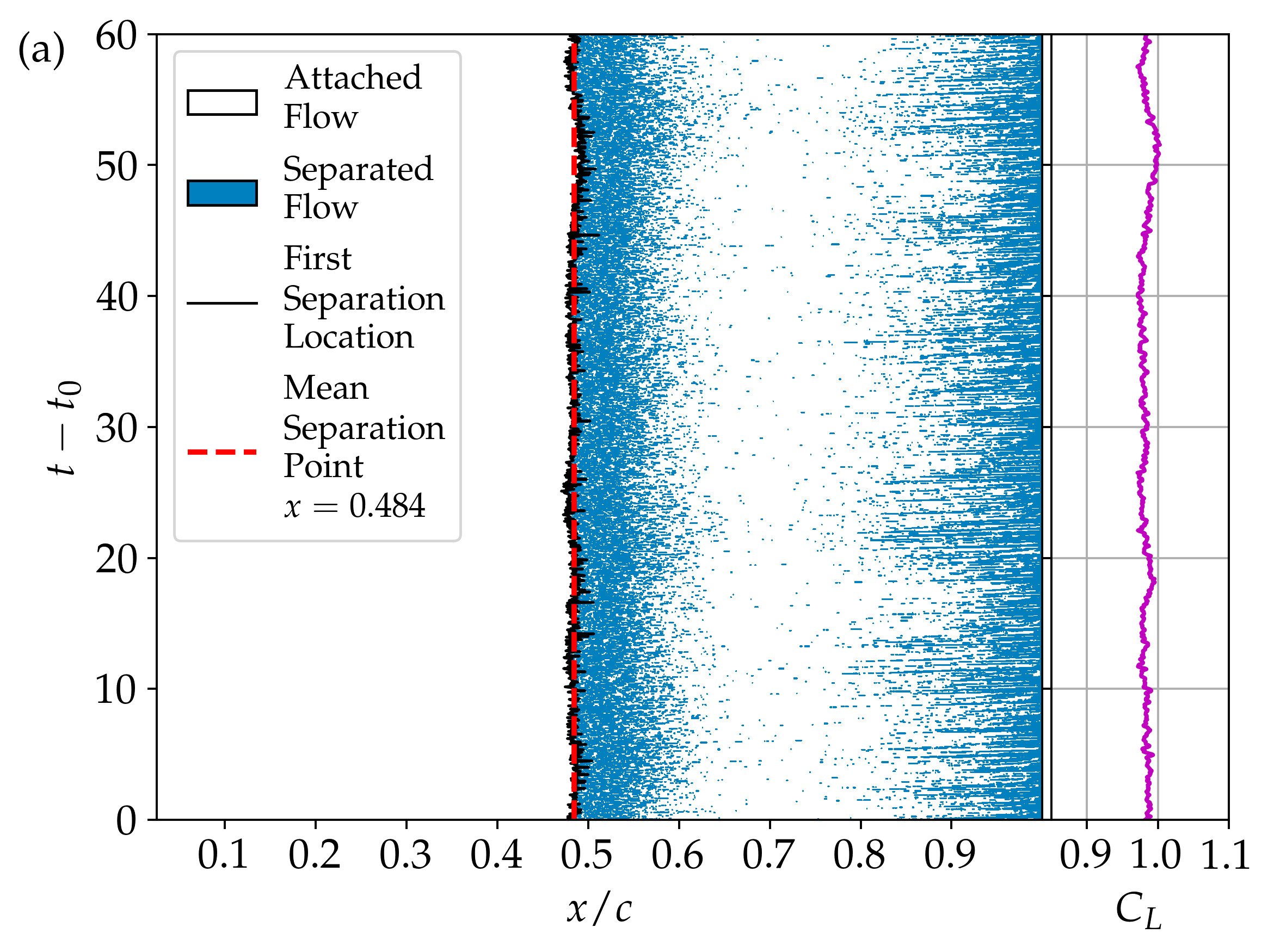}
  \includegraphics[width=0.497\columnwidth]{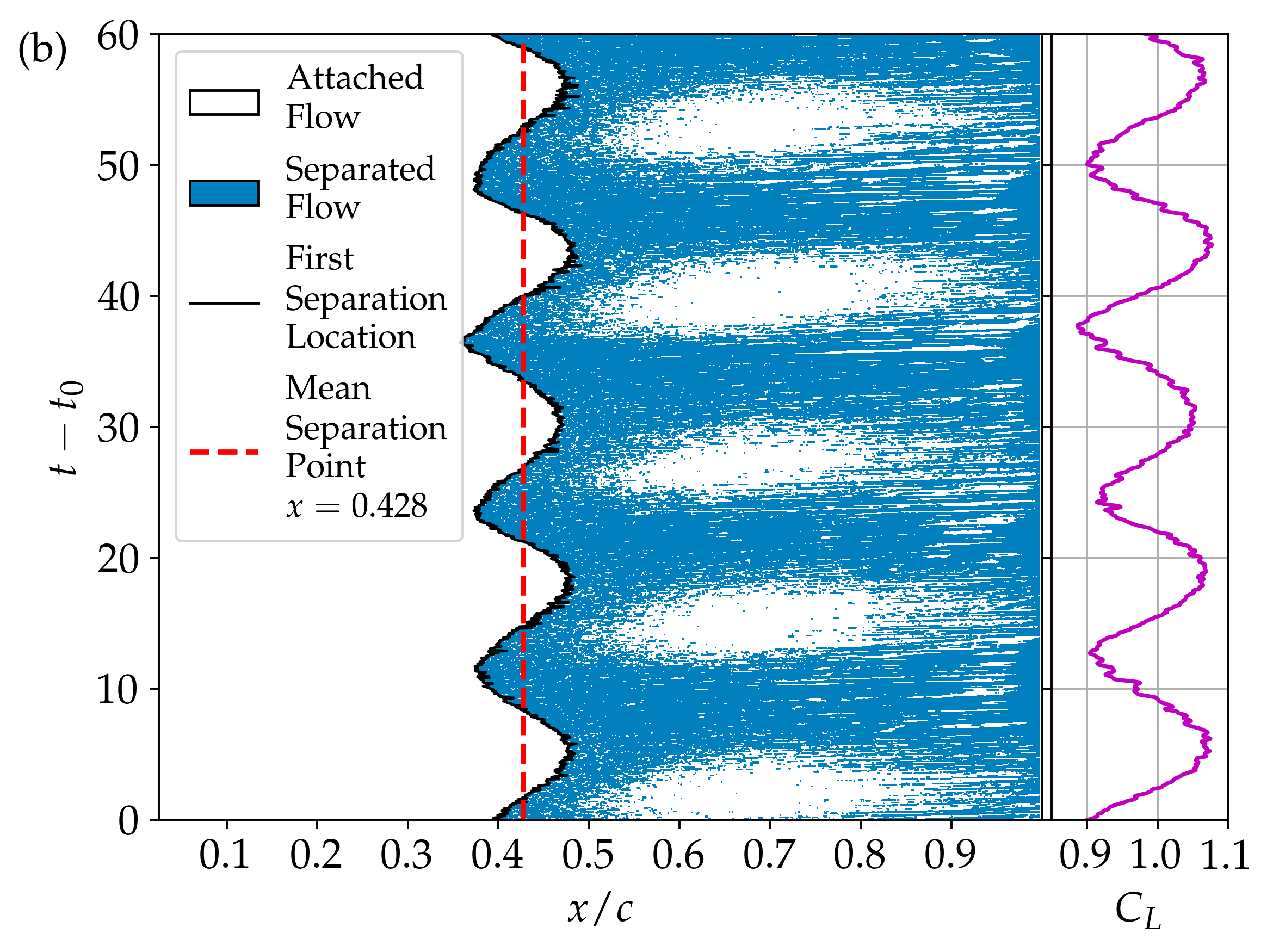}
  \caption{Skin-friction $\left(x-t\right)$ diagrams showing the position of the shockwave and separated/attached flow at (a) $\alpha=4^{\circ}$ (pre-onset) and (b) $\alpha=5^{\circ}$ (buffet) conditions at $AR=0.05$. The side plot shows the lift time-histories.}
\label{fig:OpenSBLI_onset_XT_diagram}
\end{center}
\end{figure}

High-fidelity simulations are first carried out on narrow-span ($AR=0.05$) domains to identify the Angle of Attack (AoA, $\alpha$) required to reach buffet onset for the 2D periodic shock oscillation instability. Figure~\ref{fig:OpenSBLI_onset_XT_diagram} shows $x-t$ diagrams of skin-friction for (a) $\alpha = 4^{\circ}$ (pre-onset) and (b) $\alpha = 5^{\circ}$ (post-onset) conditions. The diagrams show the time history of attached ($C_f \geq 0$, white) and separated ($C_f < 0$, blue) flow regions averaged over the span. The solid black line represents the first streamwise location of flow separation at that time instance. The side plot shows the corresponding lift coefficient history. The $C_L$ comparison shows that for the current conditions on the CRM profile, the 2D buffet onset occurs between $\alpha =  [4^{\circ}, 5^{\circ}]$. Once the AoA is raised sufficiently high, large periodic oscillations are observed, consistent with previous buffet studies \citep{Giannelis_buffet_review,Moise2023_AIAAJ}. At $\alpha = 4^{\circ}$, the shock location corresponds to the vertical dashed red line showing a mean separation location of $x=0.484$. There is a small region of flow separation at the shock position, before the flow reattaches at around $x=0.6$. The flow then remains attached until close to the trailing edge, where it separates due to the curved geometry at the back of the airfoil. As the angle of attack is increased to $\alpha = 5^{\circ}$, despite a similar mean separation location, the shock and separation point begin to oscillate periodically across the surface. Furthermore, the lift coefficient shows oscillations of $\pm 10\%$ of mean $C_L$. At post-onset conditions the region immediately downstream of the shock-wave ($0.5 < x < 0.8$) undergoes periodic separation and reattachment as part of the 2D buffet instability. The lift coefficient plot shows these correspond to the low- and high-lift phases of the buffet cycle, respectively.

\begin{figure}[h]
\begin{center}
  \includegraphics[width=0.497\columnwidth]{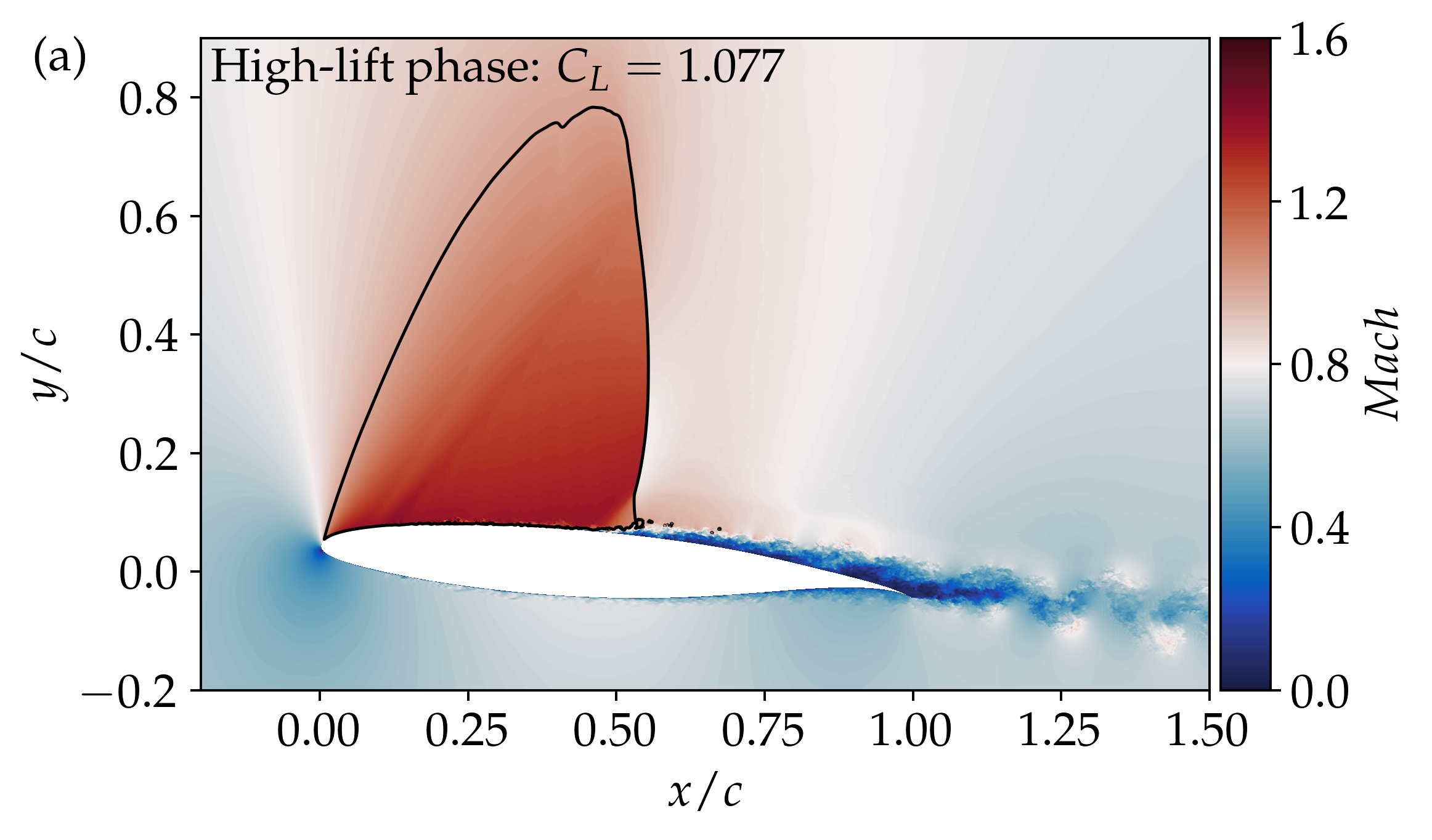}
  \includegraphics[width=0.497\columnwidth]{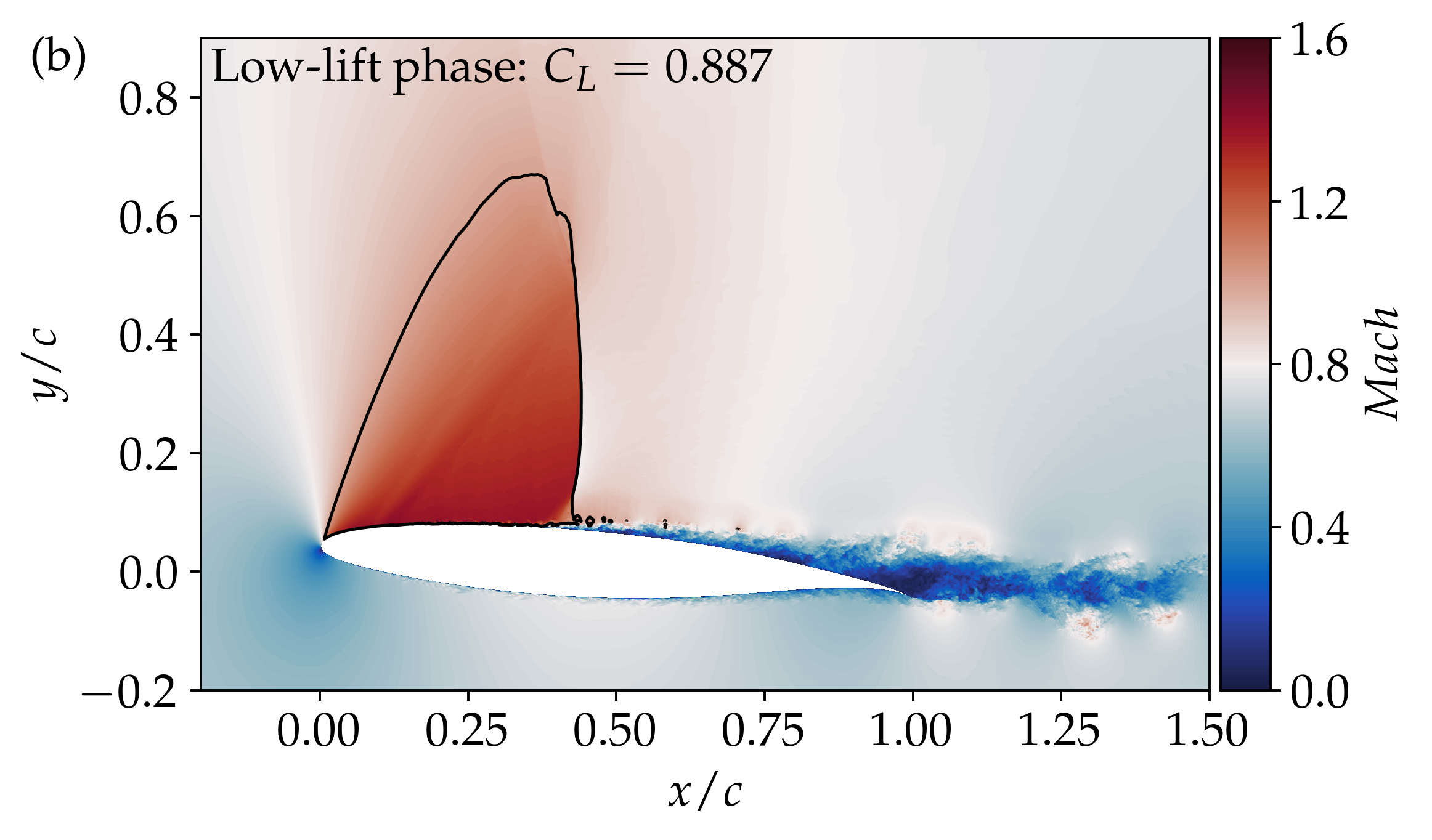}
  \caption{Illustration of the (a) high- and (b) low-lift phases of the buffet cycle for the $\alpha=5^{\circ}$ case on the $AR=0.05$ narrow-span domain. The solid black line outlines the $M=1$ sonic condition.}
\label{fig:OpenSBLI_onset_side_view}
\end{center}
\end{figure}

Figure~\ref{fig:OpenSBLI_onset_side_view} shows instantaneous Mach number contours for the $\alpha = 5^{\circ}$ case at buffet conditions. The solid black line shows the $M=1$ sonic condition that borders the supersonic flow on the suction side of the aerofoil. In transonic buffet, the shock-wave oscillates periodically on the surface of the aerofoil, resulting in large oscillations in lift and drag. The high- and low-lift phases are shown in Figure~\ref{fig:OpenSBLI_onset_side_view} (a) and (b), respectively. During the low-lift phase, the supersonic region decreases in extent as the shock-wave moves forward and the amount of flow separation increases. The flow then reattaches and the lift generation increases during the next phase of the cycle. These periodic oscillations repeat without any external forcing and are associated with the non-linear saturation of a linear global instability. The AR005-AoA5 ($AR=0.05$ and $\alpha=5^\circ$) case is used as a baseline narrow domain result at onset conditions. Before proceeding to investigations of upstream boundary-layer state and buffet with wider span-widths, the next section first cross-validates the data and 2D buffet onset condition to those obtained by RANS-based solutions and predictions from GSA.

\subsection{Cross-validation with low-fidelity methods}\label{sec:RANS_compare}

Having obtained a baseline ILES case from OpenSBLI \citep{LUSHER201817,OpenSBLI_2021_CPC}, this section compares the data to low-fidelity solutions from the FaSTAR \citep{FaSTAR_Hashimoto_2012,IIHAT2016} solver. The aim is both to cross-validate the data before proceeding with the simulations for the rest of this study, and also to assess how well low-fidelity RANS/URANS and stability methods agree with the more expensive ILES approach. We first compare time-averaged pressure distributions from the narrow-span ILES ($AR=0.05$) to steady RANS solutions at pre- and post-onset. The RANS solutions are then used as a base flows to perform GSA in the range $\alpha = \left[4, 5^{\circ}\right]$, to obtain a prediction of the buffet onset angle of attack and related shock-oscillation frequencies. Finally, URANS is performed on the same configuration to compare the predicted buffet frequencies to those obtained from both the ILES and GSA approaches.

\subsubsection{Comparison between RANS and ILES at pre- and post-onset \label{sec:RANS}}
\begin{figure}[h]
\begin{center}
  \includegraphics[width=0.497\columnwidth]{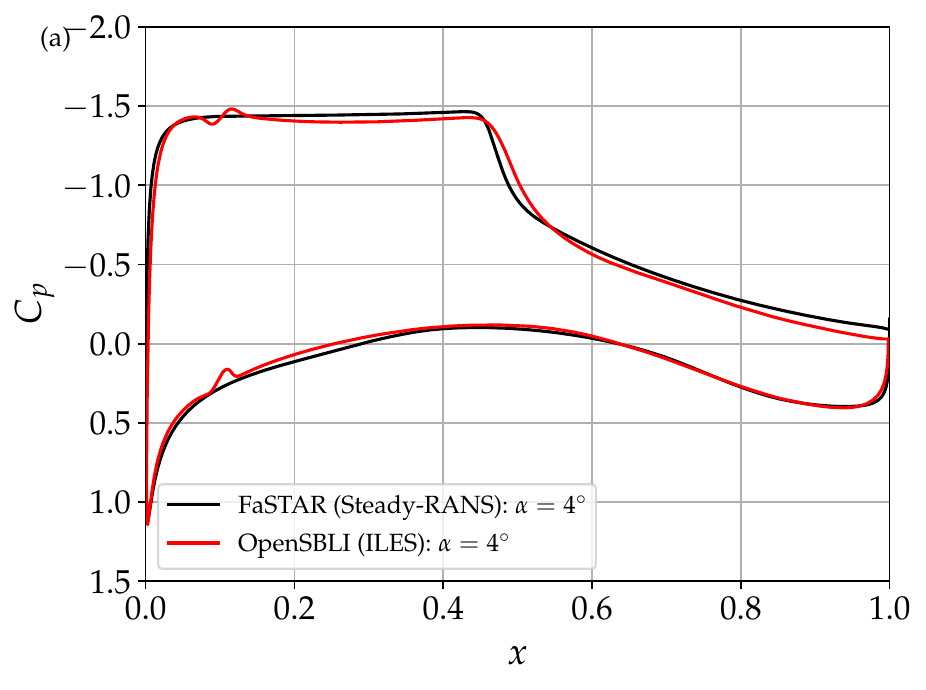}
  \includegraphics[width=0.497\columnwidth]{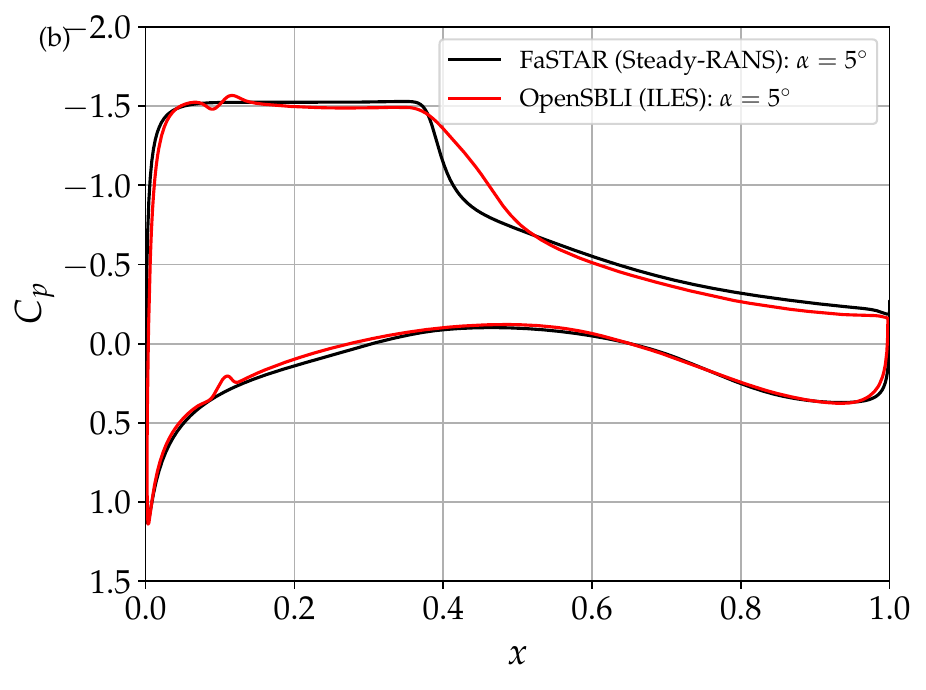}
  \caption{OpenSBLI (ILES, time-averaged) to FaSTAR (Steady-RANS) comparison of pressure distribution for (a) $\alpha=4^{\circ}$ (pre-onset) and (b) $\alpha=5^{\circ}$ (post-onset) conditions.}
\label{fig:FaSTAR_RANS_2D_lines}
\end{center}
\end{figure}

Figure~\ref{fig:FaSTAR_RANS_2D_lines} shows RANS solutions compared to the time- and span-averaged $C_p$ distributions for the ILES cases in section \ref{sec:finding_buffet}. At pre-onset conditions ($\alpha=4^{\circ}$) (Figure~\ref{fig:FaSTAR_RANS_2D_lines}-a), the steady-RANS solution matches the ILES data very well, despite the two simulations differing on several aspects. As well as the obvious differences in the numerical approaches, domain topology, grid resolution and trailing edge geometry, one should be reminded that the RANS solver assumes the flow to be fully turbulent, whereas the ILES data is initially laminar and then tripped to turbulence (the tripping in the ILES simulations is indicated by the wiggles at 10\% chord on both pressure and suction sides). As shown in the previous section, the ILES results at $\alpha=5^{\circ}$ show the occurrence of the 2D buffet instability. Due to the steady nature of the calculation, the RANS solution is incapable of reproducing the unsteadiness and the pressure gradient around the shock is steeper with respect to the ILES one (figure~\ref{fig:FaSTAR_RANS_2D_lines}-b), that appears smeared over a wider $x$-range because of the time-averaging. For the rest of the airfoil, the agreement between RANS and ILES is again excellent.

\subsubsection{GSA-based 2D buffet onset prediction\label{sec:GSA}}

Global stability analysis is performed in the $\alpha = \left[4, 5^{\circ}\right]$ range to determine the onset angle of attack and shock oscillation frequencies with a tolerance of $\Delta \alpha = 0.1^{\circ}$. For each angle of attack investigated, a base flow solution is obtained by sufficiently converging the RANS solver. The linearized URANS solver is then used for the stability calculation and advanced in time with an integration step of $\Delta t = 0.005$ (corresponding to a dimensional time step $\Delta t^*= 1.48 \times 10^{-3}\: s$) and 40 sub-iterations for the pseudo-time integration of the dual time stepping method. The Krylov base used for the stability modal decomposition is constructed by collecting 128 perturbation fields every $\Delta T_{kryl} = 450 \times \Delta t$ (Krylov time step). The GSA calculations were restarted from the last perturbation field of a preliminary simulation and all unstable modes presented in this section have residuals at least below $10^{-5}$. 

\begin{figure}[htb]
\begin{center}
  \includegraphics[width=.9\columnwidth]{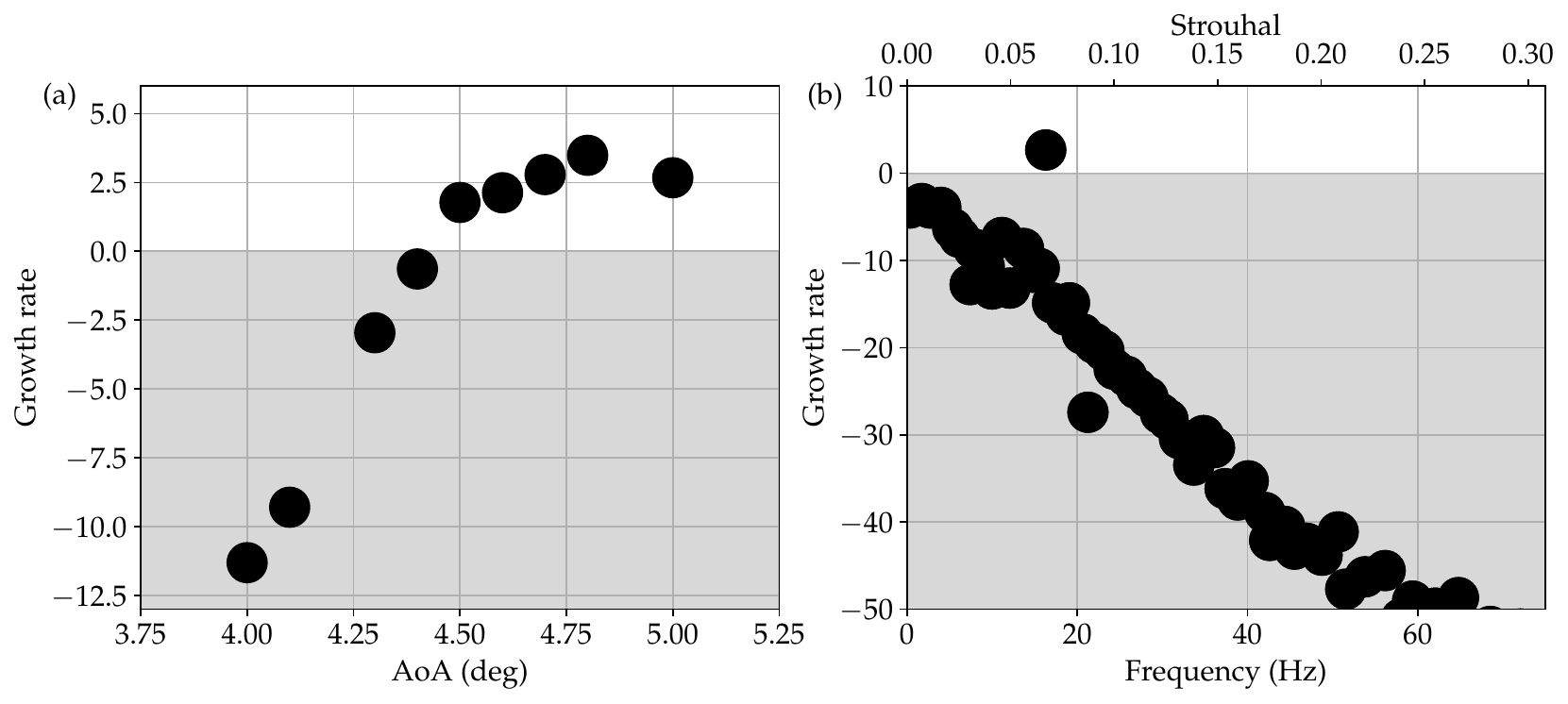}
  \caption{Summary of the GSA results indicating (a) the evolution of the growth rate with the angle of attack and (b) the predicted eigen-spectra (with frequencies in both dimensional and non-dimensional forms) for the $\alpha=5^{\circ}$ case. The shaded region indicates stable solutions.}
\label{fig:FaSTARGSA_spectra}
\end{center}
\end{figure}

\begin{figure}[htb]
\begin{center}
  \begin{subfigure}[t]{0.03\textwidth}
    (a)
  \end{subfigure}
  \begin{subfigure}[t]{0.45\textwidth}
    \includegraphics[width=\linewidth, valign=t]{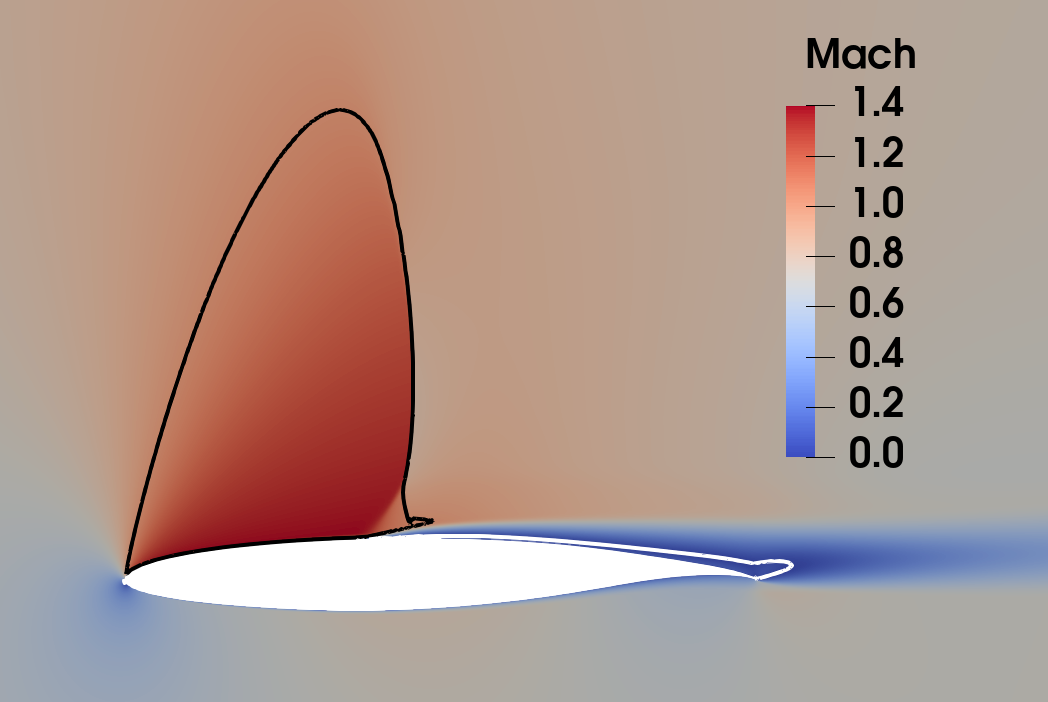}
  \end{subfigure}\hfill
  \begin{subfigure}[t]{0.03\textwidth}
    (b)
  \end{subfigure}
  \begin{subfigure}[t]{0.45\textwidth}
    \includegraphics[width=\linewidth, valign=t]{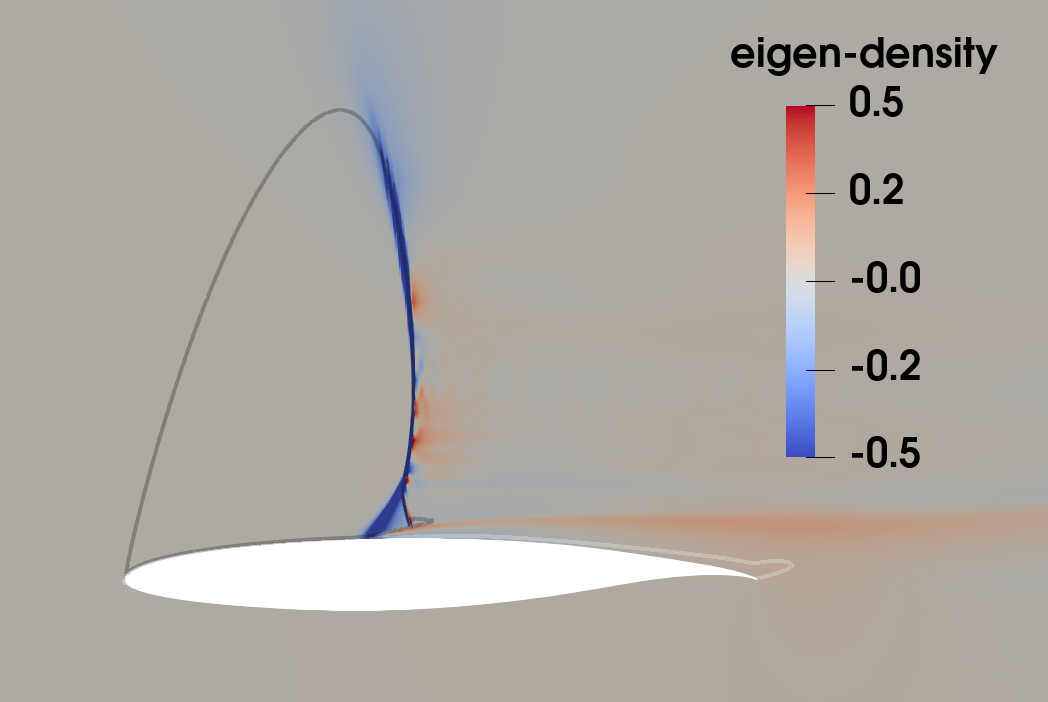}
  \end{subfigure}
  \caption{The (a) RANS/base flow solution and (b) density eigen-function corresponding to the temporally amplified mode at $St=0.068$ are plotted for the $\alpha=5^{\circ}$ case.}
\label{fig:FaSTARGSA_baseflow_mode}
\end{center}
\end{figure}

Figure \ref{fig:FaSTARGSA_spectra}(a) shows the evolution of the least temporally damped / most temporally amplified mode for each angle of attack investigated. The first angle of attack with a temporally amplified mode (or with a positive growth rate) is found at $\alpha=4.5^{\circ}$, indicating 2D buffet onset. For the case at $\alpha=5^{\circ}$ previously discussed, the full spectra is reported in Figure \ref{fig:FaSTARGSA_spectra}(b). A single temporally amplified mode is present at $St=0.068$ ($f=16.53 \: Hz$), confirming that the $\alpha=5^{\circ}$ case is unstable and shock oscillations must be present, as shown by the ILES calculations. The sonic (black) and the zero-streamwise velocity (white) isolines are superimposed to the Mach number contours in Figure \ref{fig:FaSTARGSA_baseflow_mode}(a) for the (unstable) base flow solution at $\alpha=5^{\circ}$. The density eigen-function corresponding to the temporally amplified mode at $St=0.068$ is plotted in Figure \ref{fig:FaSTARGSA_baseflow_mode}(b) and shows the typical perturbation distribution responsible for the 2D buffet shock instability, as seen in the literature \cite{CGMT2009,SMS2015}.

\subsubsection{Comparison between URANS, GSA and ILES at post-onset conditions \label{sec:URANS}}

\begin{figure}[h]
\begin{center}
  \includegraphics[width=1.\columnwidth]{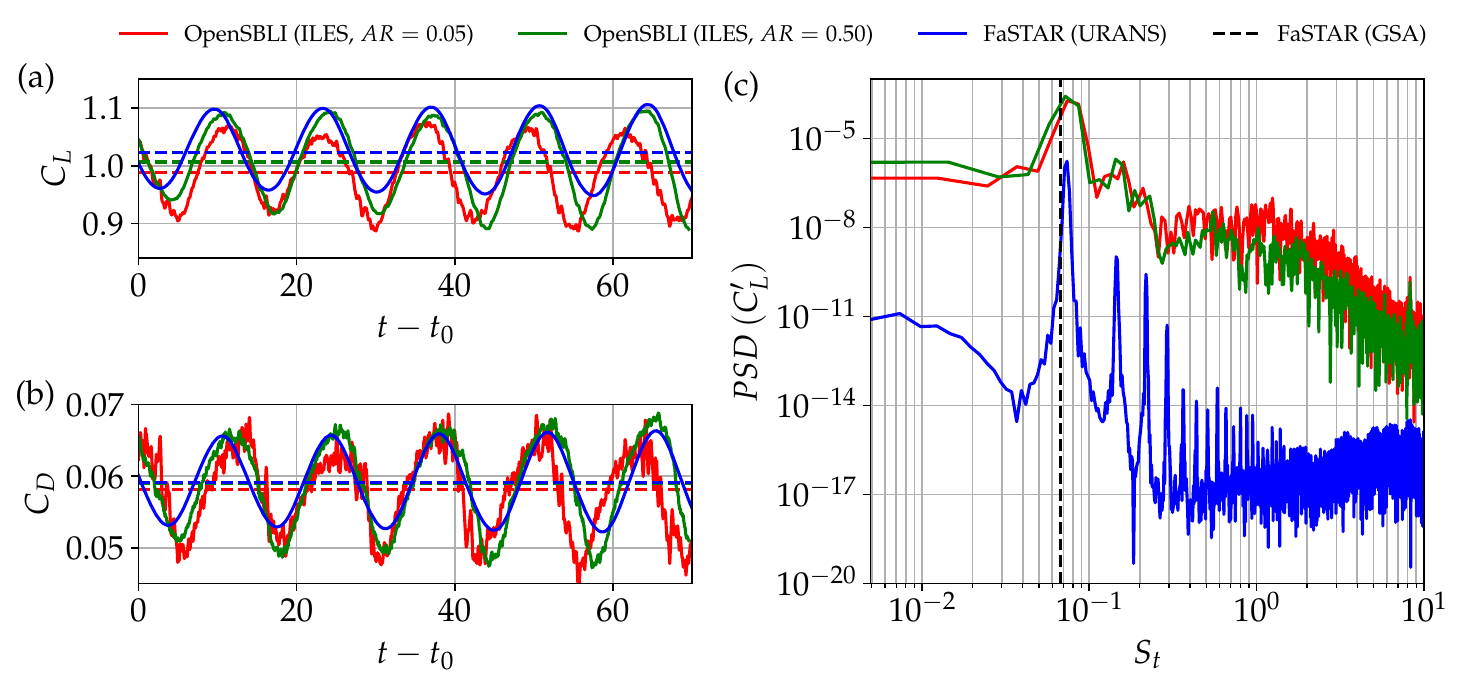}
  \caption{Unsteady OpenSBLI (ILES) data for $\alpha = 5^{\circ}$ at $AR=0.05$ and $AR=0.50$, compared to URANS and GSA results from FaSTAR. Showing (a) lift coefficient, (b) drag coefficient, and (c) PSD of lift fluctuations. The horizontal dashed lines in the lift and drag coefficient plots indicate time-averaged values.}
\label{fig:FaSTAR_URANS_2D_lines}
\end{center}
\end{figure}

The GSA-based onset prediction determined unstable flow conditions for angles of attack above $\alpha=4.5^{\circ}$. To verify the stability results, URANS calculations are carried out at post-onset conditions for the $\alpha = 5^{\circ}$ case. The same time-integration parameters used for the GSA calculations are here selected. In Figure~\ref{fig:FaSTAR_URANS_2D_lines}, the URANS result is compared to ILES data from both the narrow-span ($AR=0.05$) simulation in the previous section, and one computed on a wider airfoil with $AR=0.50$. The time-histories of (a) lift coefficient, (b) drag coefficient, and (c) PSD of lift fluctuations show excellent agreement between the low- and high-fidelity methods, especially for the wider aspect ratio airfoil (as it will be shown later, the $AR=0.50$ case is found to be domain span-width independent). Despite some differences in the mean aerodynamic force values and oscillation amplitudes, the low-fidelity methods can accurately reproduce the dominant frequencies at $St\approx0.077$. As expected, the GSA-predicted frequency is slightly under-estimated, due to the differences between the (steady) base flow solution used for the stability analysis and the URANS time-averaged field far from the onset.

\subsubsection{Discussion on the comparison between low- and high-fidelity methods \label{sec:low-v-high_discussion}}

While the low-fidelity methods showed remarkable agreement, it should be emphasized that the angles of attack here examined fall in a narrow range around the instability onset. As it will be seen later, when the angle of attack is further increased, the 2D URANS solutions are no longer representative of the physics at play and show significant differences to the ILES results and 3D URANS solutions. Because of the larger separation and stronger unsteadiness, the quality of the RANS-based results deteriorates when moving towards higher angles of incidence. Due to the difficulty of obtaining well converged and representative steady solutions, stability methods also suffer when applied far from the onset \cite{SH2023}. Last but not least, it should be emphasized that although the low-fidelity simulations were carefully designed and set up based on general best practices, potential sensitivity to domain topology, grid resolution, time integration parameters and turbulence modeling cannot be entirely ruled out. In this sense, high-fidelity simulations are an important reference and represent an essential tool when investigating complex flow phenomena such as turbulent transonic buffet and the range of scales present. The design of high-fidelity simulations in general is however not exempt from sources of numerical errors. Sensitivity analyses for the high-fidelity simulations in this work are presented in the following sections, in an attempt to provide high-quality data for the characterization of 2D turbulent transonic buffet.

\section{Sensitivity of turbulent transonic buffet to numerical tripping strength}\label{sec:tripping_parametric}

The baseline narrow case (AR005-AoA5) at buffet conditions is selected to test the sensitivity of the solution to different strengths of the tripping, as defined in equation \eqref{eq:tripping_eqn}. The aim of this section is to investigate how sensitive the buffet characteristics are to the upstream boundary-layer state. In particular, whether or not there are substantial changes in buffet frequency, aerodynamic coefficients, and shock-structure, as the tripping amplitude is reduced and the boundary-layer becomes more laminar/transitional. For fully turbulent transonic buffet on supercritical airfoils in the range of Mach numbers considered here, a single lambda shock-wave oscillates across the suction side of the airfoil with low-frequency (Figure~\ref{fig:OpenSBLI_onset_side_view}), as commonly reported in the literature \citep{JMDMS2009, FK2018, SLKHR2022}. In `laminar-' or `free-transitional-buffet' \citep{dandois_mary_brion_2018,MarkusPRF_2020,moise_zauner_sandham_2022,Moise2023_AIAAJ,LongWong2024_Laminar_buffet}, differences have been reported in the dominant frequencies, flow structure, and shock topology. Free-transition buffet cases also typically have far larger regions of flow separation than fully turbulent configurations, as commonly found in studies of SBLI on different flow configurations with differing upstream boundary-layer states \citep{Sandham2014_Transition_SBLI, Sansica_PoF_2014,Sansica_JFM_2016,Lusher2020_FTAC}. \citet{MarkusPRF_2020} and \citet{moise_zauner_sandham_2022,Moise2023_AIAAJ} have reported laminar buffet cases at $Re=5\times 10^5$ (equal to the $Re$ in the present study) which are comprised of a system of multiple shock-wave/expansions which oscillate on the suction side of the airfoil and interact with one another, which deviates from the single shock-wave configuration from turbulent buffet. However, the same authors reported in \cite{moise_zauner_sandham_2022} that the number of observed shock-waves decreased as the Reynolds number was increased. They suggested that there could be a critical Reynolds number at which the number of shock-waves reduces to one, as observed in turbulent buffet. Indeed, this would be consistent with other laminar buffet simulations \citep{dandois_mary_brion_2018} and experiments \citep{Brion2017_Laminar_Experiment,Brion2020_Laminar_Experiment} in the recent literature, which showed laminar buffet with only a single shock-wave at $Re=3 \times 10^6$. As we investigate turbulent buffet at moderate $Re$ in this work, it is of interest to assess how sensitive the configuration is to a modified upstream boundary-layer state. Table~\ref{tab:trip_cases} shows a summary of the cases simulated with six different trip amplitudes, ranging from $A = 0.5\%$ to $A = 10\%$ (the amplitude is expressed in percentage of the free-stream velocity $U_{\infty}$). The baseline AR005-AoA5 case refers to case AR005-AoA5-$A$7.5 in this table, with an amplitude of $A=7.5\%$.

\begin{table}[h]
  \begin{center}
\def~{\hphantom{0}}
  \begin{tabular}{cccccccccc}
      Case                    & AR     & $\alpha$       & Trip Amplitude   & Interaction                   & $\overline{C_L}$ & $\overline{C_{Dp}}$ & $\overline{C_{Df}}$ & $\overline{C_D}$ & $St$\\ \hline
      AR005-AoA5-$A$10   (a)   & 0.05   & $5^{\circ}$    &   $10.0 \% \cdot U_{\infty}$ &  Turbulent &  0.986     &       0.0498             &         0.0080           &        0.0578        &      0.0778 \\
      AR005-AoA5-$A$7.5  (b)        & 0.05   & $5^{\circ}$    &   $7.5 \% \cdot U_{\infty}$  & Turbulent &   0.991     &       0.0501             &         0.0079           &        0.0581         &      0.0778 \\
      AR005-AoA5-$A$5    (c)     & 0.05   & $5^{\circ}$    &   $5.0 \% \cdot U_{\infty}$  & Turbulent &   1.000     &       0.0507             &         0.0077           &        0.0584         &      0.0778 \\
      AR005-AoA5-$A$2.5  (d)         & 0.05   & $5^{\circ}$    &   $2.5 \% \cdot U_{\infty}$  & Transitional &   1.125     &       0.0584             &         0.0066           &        0.0650         &      0.0667 \\
      AR005-AoA5-$A$1    (e)      & 0.05   & $5^{\circ}$    &   $1.0 \% \cdot U_{\infty}$  & Laminar &   1.050     &       0.0532             &         0.0057           &        0.0588         &      0.0778 \\
      AR005-AoA5-$A$0.5  (f)         & 0.05   & $5^{\circ}$    &   $0.5 \% \cdot U_{\infty}$  & Laminar &   1.041     &       0.0527             &         0.0056           &        0.0583         &      0.0778 \\
  \end{tabular}
  \caption{Summary of narrow-span $AR=0.05$ cases at different tripping strengths. For each case, Reynolds and Mach numbers are fixed at $Re=5 \times 10^{5}$ and $M_{\infty}=0.72$. Letters refer to the labels used in figure~\ref{fig:Amplitude_XT} figure~\ref{fig:Amplitude_side_contours}.}
  \label{tab:trip_cases}
  \end{center}
\end{table}

\begin{figure}[h]
\begin{center}
  \includegraphics[width=1\columnwidth]{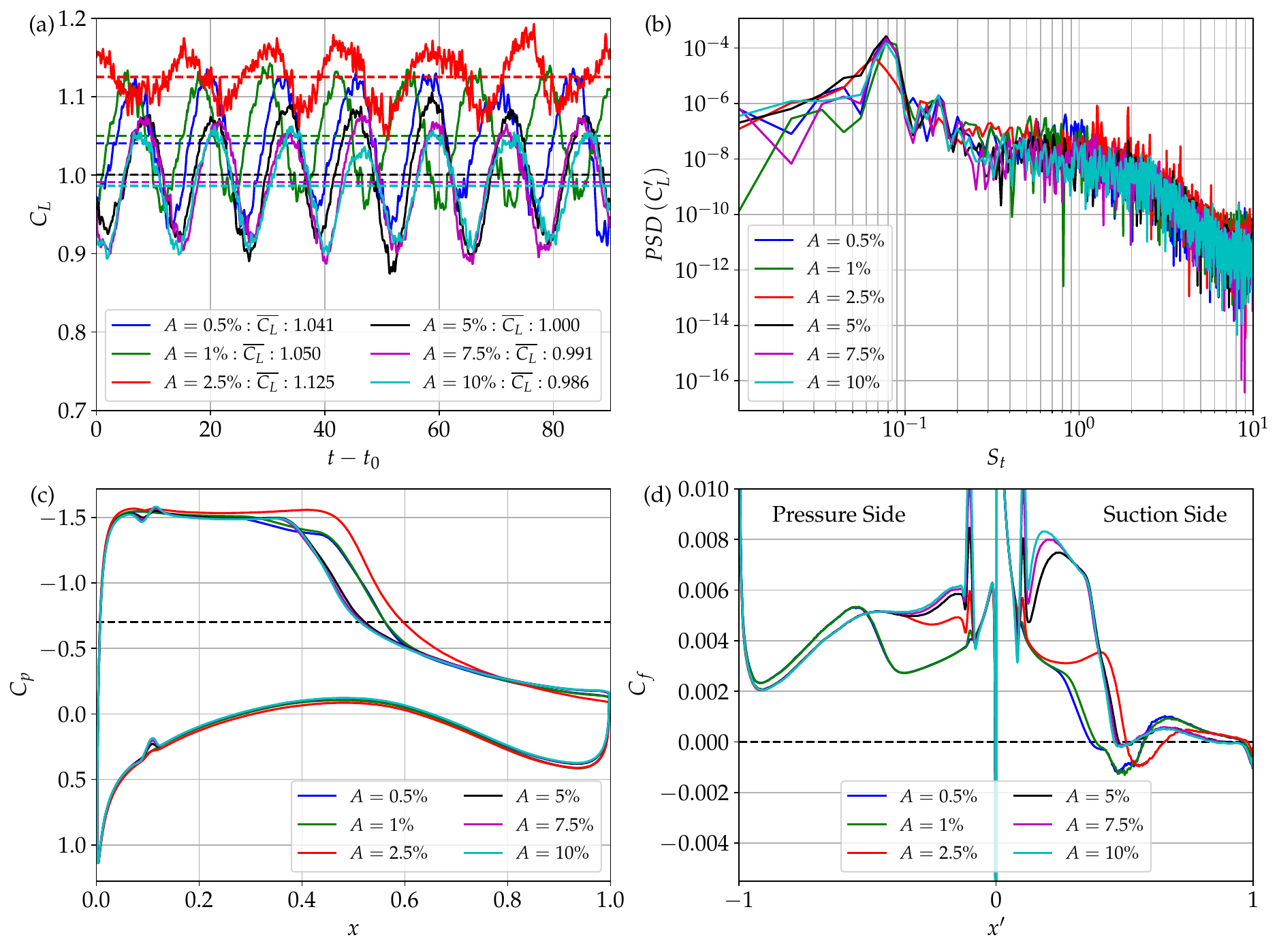}
  \caption{Sensitivity to tripping amplitude for the baseline case of $M=0.72$, $Re=500,000$, $AR=0.05$, $\alpha = 5^{\circ}$. Showing (a) lift coefficient history (b) PSD of fluctuating lift component (c) time-averaged pressure coefficient and (d) time-averaged skin-friction.}
\label{fig:amp_study_lines}
\end{center}
\end{figure}

Figure~\ref{fig:amp_study_lines} shows (a) lift coefficient history, (b) PSD of lift coefficient fluctuations, (c) time-averaged pressure distribution and (d) time-averaged skin-friction. In Figures~\ref{fig:amp_study_lines} (c,d), the variation of trip strength can be seen in height of the peaks in the pressure coefficient and skin-friction distributions around $x=0.1$, as the tripping amplitude is varied by a factor of twenty. Figure~\ref{fig:amp_study_lines} (a) shows the effect this has on the variations of lift coefficient. The first thing to note is that for the three strongest amplitudes ($A=5\%, 7.5\%, 10\%$), the buffet characteristics are very similar and only minor changes on the order of $1\%$ are observed in $\overline{C_L}$. Once the boundary-layer has been tripped to turbulence, further increases in forcing show a minimal impact on aerodynamic coefficients and the skin-friction distributions in Figure~\ref{fig:amp_study_lines} (d) collapse upstream of the mean shock location ($x \approx 0.4$). Based on these findings, we select the $A= 7.5 \%$ tripping strength throughout the later sections of this work as representative fully-turbulent buffet conditions. It should be noted that another confirmation of the turbulent nature of the interaction at this trip amplitude was previously given in section~\ref{sec:RANS_compare} when showing the agreement between the ILES results and the RANS/URANS simulations, which assumed the boundary-layer to be fully-turbulent over the entire airfoil.

While the three strongest tripping amplitudes agree well with minimal differences in aerodynamic coefficients, large variations are observed as the tripping strength is reduced. At $A=2.5\%$, we observe a 15\% increase in mean $C_L$ and a modified low-frequency peak in the PSD given in Figure~\ref{fig:amp_study_lines} (b). Additionally, there are higher frequency peaks visible around $St=1.3$ and $St=1.9$, not present in the fully-tripped cases. In the pressure distribution and skin-friction plots in Figures~\ref{fig:amp_study_lines} (c,d), there is a large downstream shift in the mean shock position and a significant region of time-averaged flow separation around $x=0.6$ on the suction side of the airfoil, distinct from the higher forcing cases. However, further decreases in tripping amplitude do not continue a monotonic response in the lift coefficient, with the weaker cases $(A = 0.5\%, 1\%)$ having mean lift values and buffet frequencies closer to the fully turbulent ones. From the skin-friction plot in Figure~\ref{fig:amp_study_lines} (d), the cases can be categorized based on the $C_f$ upstream of the shock as laminar ($A=0.5\%, 1\%)$, transitional ($A=2.5\%$), and fully-turbulent ($A=5\%, 7.5\%, 10\%$). This categorization is used in rest of this work.

\begin{figure}
\begin{center}
  \includegraphics[width=0.497\columnwidth]{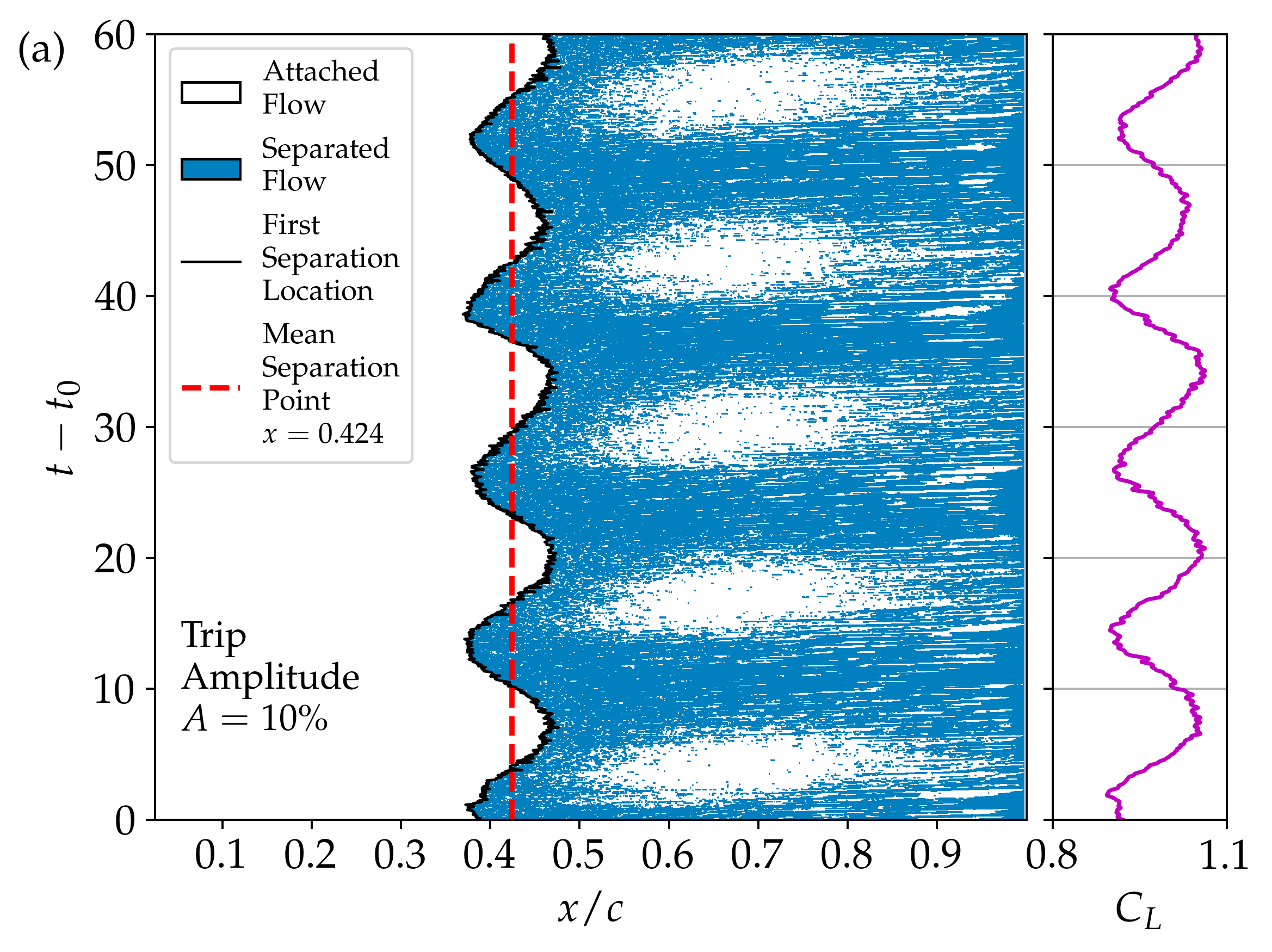}
  \includegraphics[width=0.497\columnwidth]{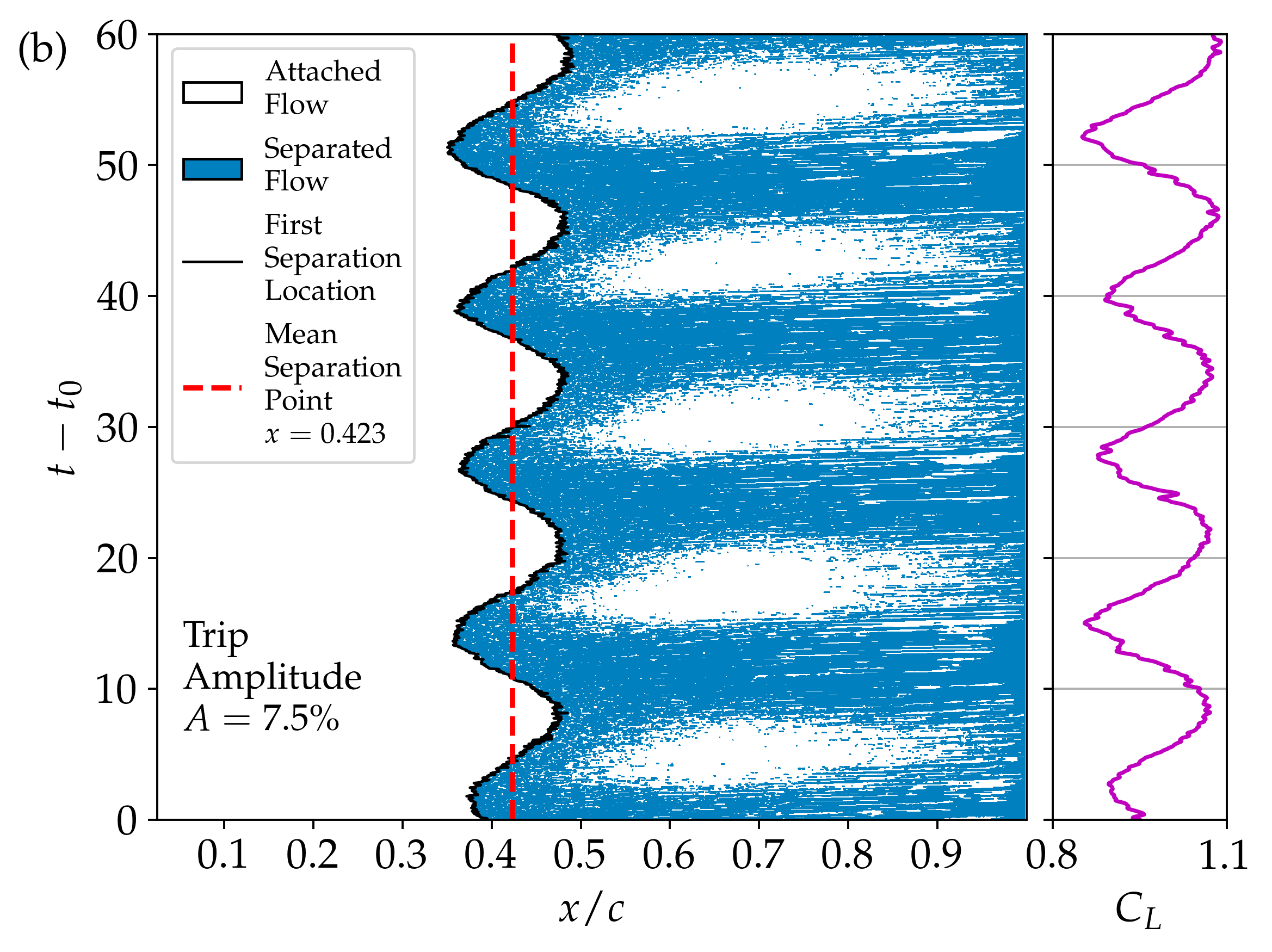}\\
  \includegraphics[width=0.497\columnwidth]{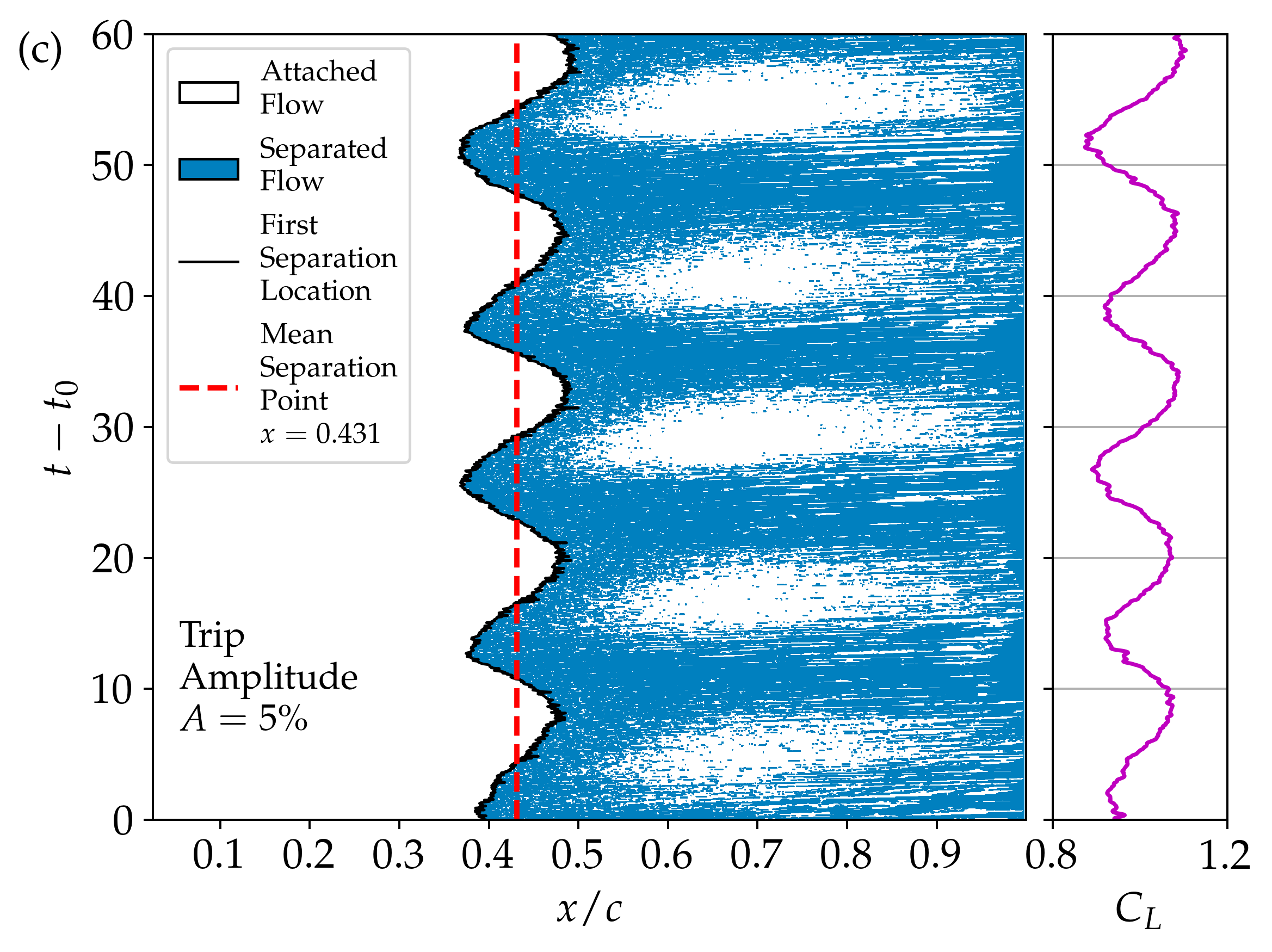}
  \includegraphics[width=0.497\columnwidth]{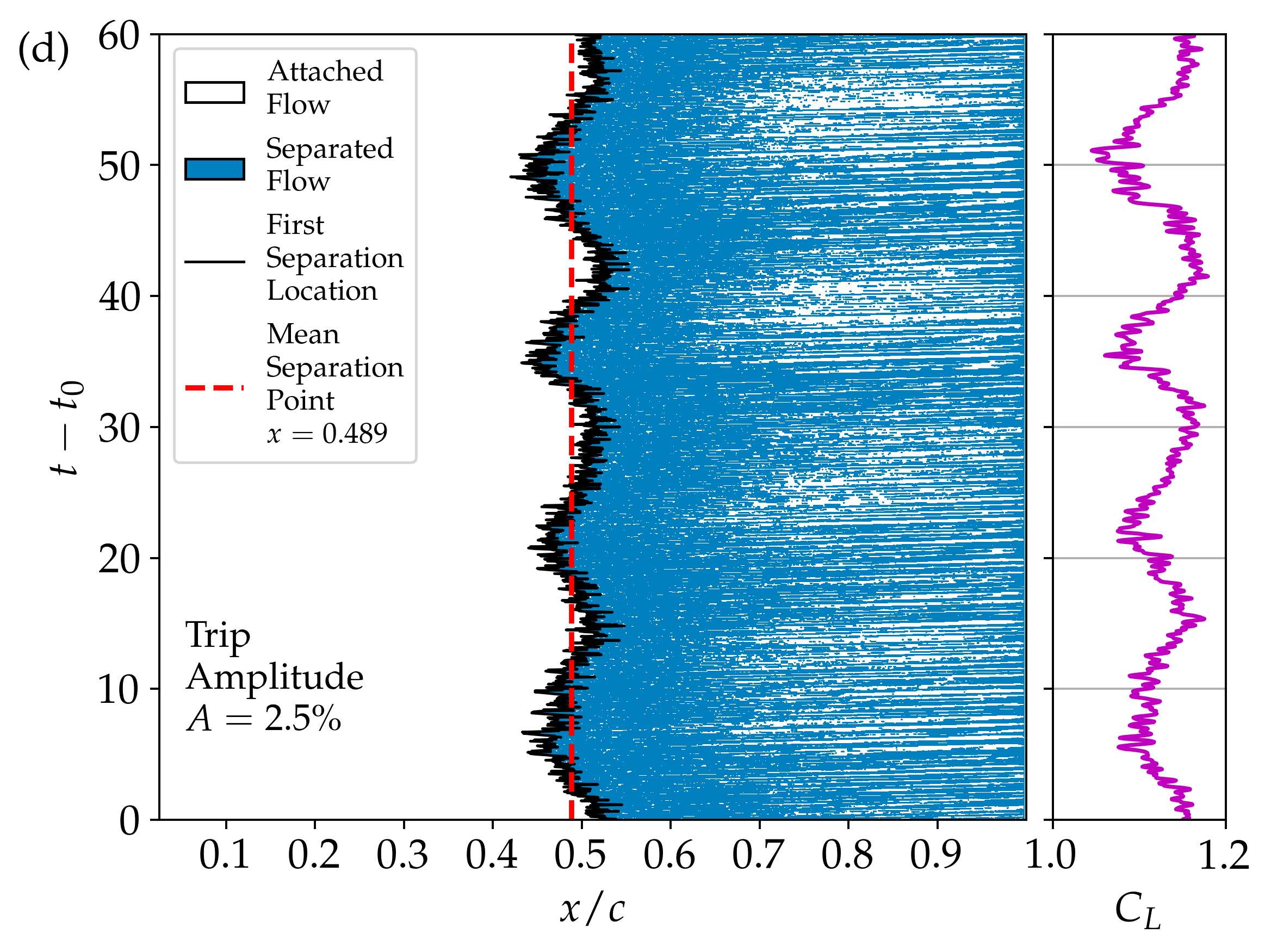}\\
  \includegraphics[width=0.497\columnwidth]{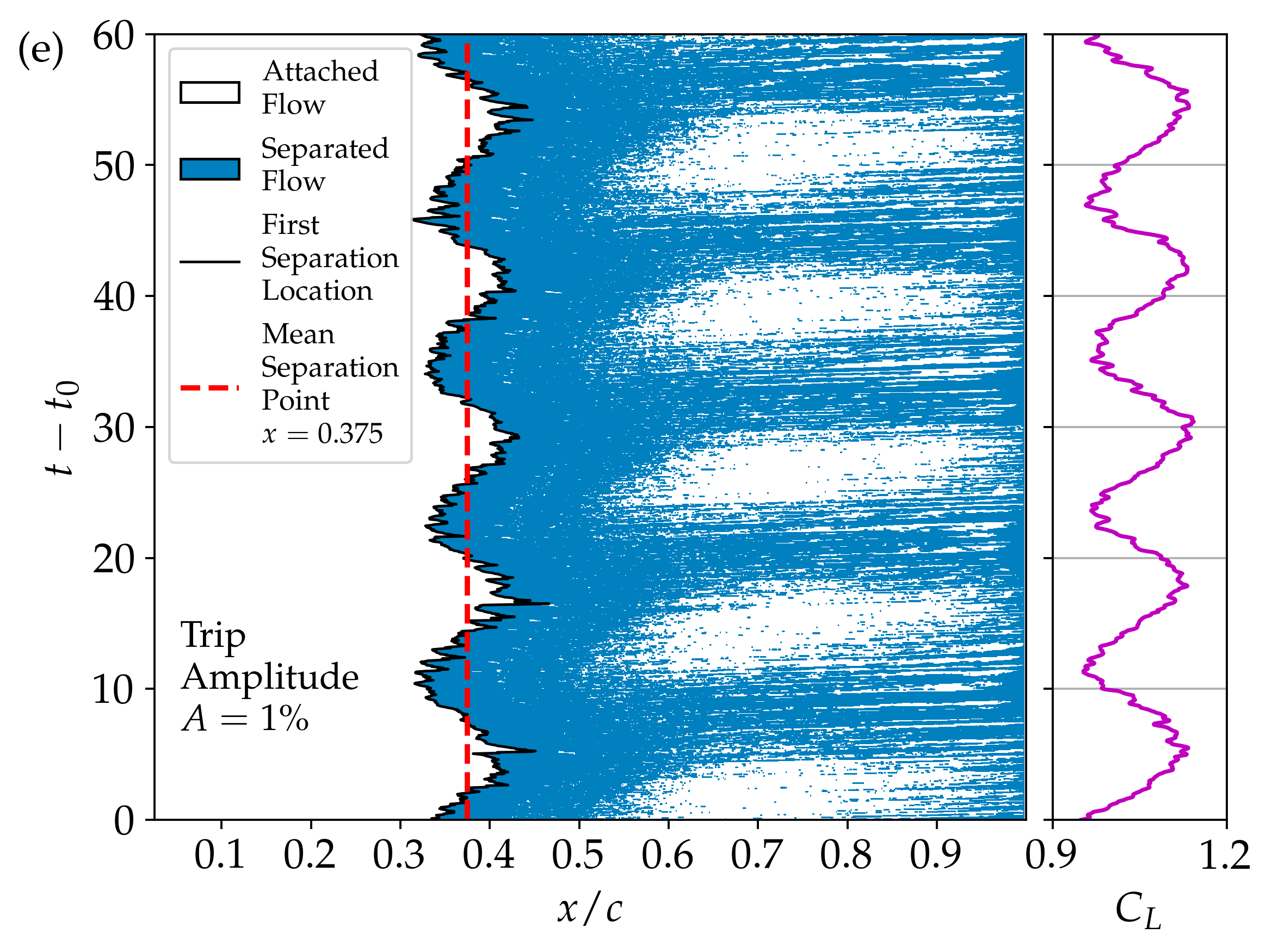}
  \includegraphics[width=0.497\columnwidth]{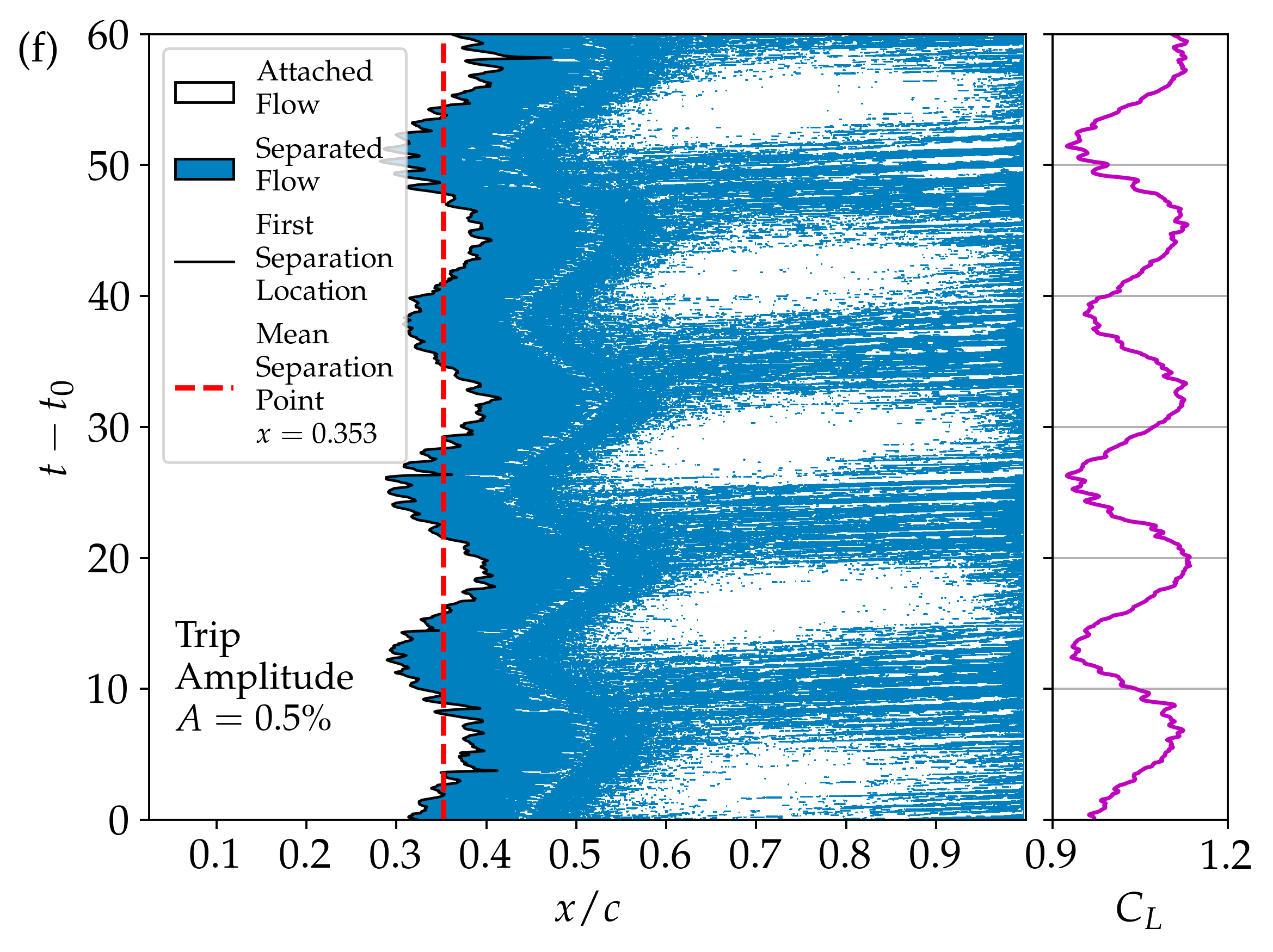}
  \caption{Skin-friction $\left(x-t\right)$ diagrams of attached/separated flow for varying tripping strength cases in Table \ref{tab:trip_cases}. The corresponding time history of lift coefficient is shown on the side plot.}
\label{fig:Amplitude_XT}
\end{center}
\end{figure}

Figure~\ref{fig:Amplitude_XT} shows $x-t$ spatio-temporal plots of how the regions of attached and separated flow vary in time for each of the six tripping strengths. In each case the corresponding lift coefficient is plotted on the side of the figure. The black line represents the first streamwise location of flow separation $(C_f < 0)$ at each given time instance. The three strongest forcing fully-turbulent cases (Figures~\ref{fig:Amplitude_XT} (a)-(c)) are all very similar. All three have periodic low-frequency separation and reattachment downstream of the main interaction ($0.5 < x < 0.9$). The mean separation point is within $1.5\%$ for each of these cases. The transitional case (Figure~\ref{fig:Amplitude_XT} (d)), however, shows a strong deviation from the turbulent topology, with the same high frequency modulation of the lift coefficient observed on the first separation location. This suggests that the additional frequencies observed in the PSD in the $1 < S_t < 5$ range (Figure~\ref{fig:amp_study_lines} (b)) are due to the response of the separation point to the incoming boundary-layer state. As the tripping amplitude is reduced further to $A=0.5\%, 1\%$, clear regions of separation and attachment are restored. As the boundary-layer becomes more laminar upstream of the shock-wave, the separation point oscillates with an intermediate frequency in range $1 < St < 5$ (Figure~\ref{fig:amp_study_lines} (b)). In contrast to the turbulent cases, the weakest trip case also shows a reattachment line around $x=0.45$ which oscillates at the same low-frequency as the buffet cycle. The increased values of $\overline{C_L}$ for the laminar and transitional cases relative to the fully turbulent configuration are consistent with the RANS/GSA findings of \citet{Garbaruk2021}. 

\begin{figure}
\begin{center}
  \includegraphics[width=0.485\columnwidth]{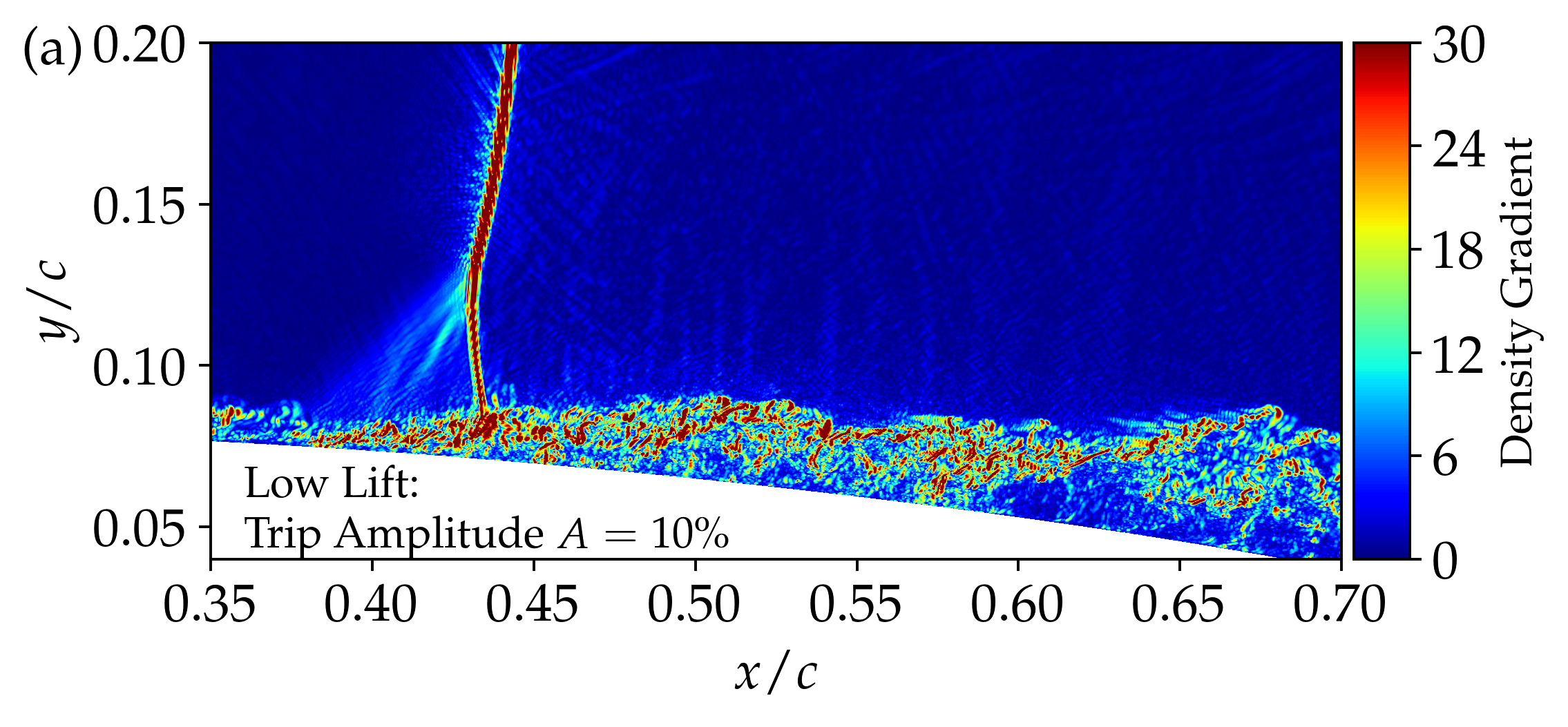}
  \includegraphics[width=0.485\columnwidth]{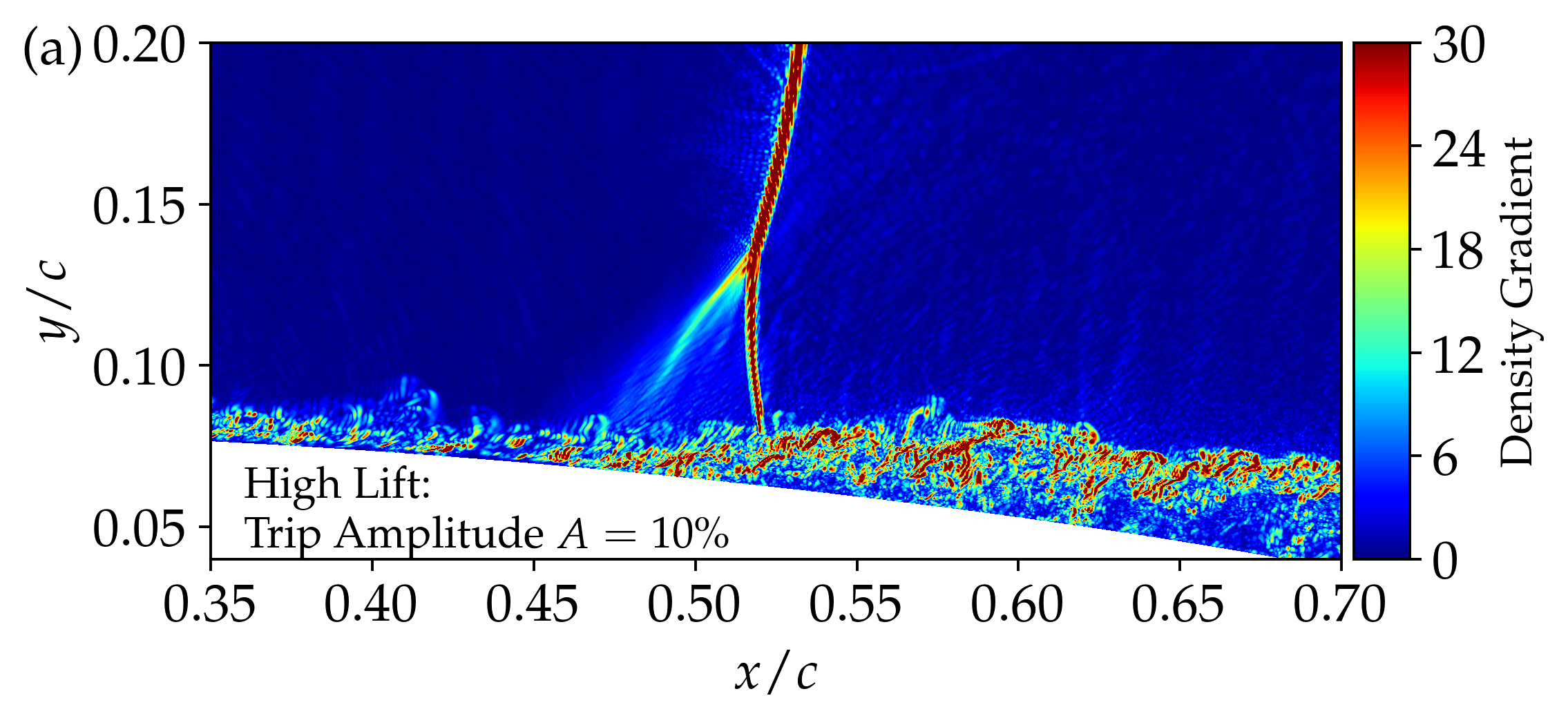}\\
   \includegraphics[width=0.485\columnwidth]{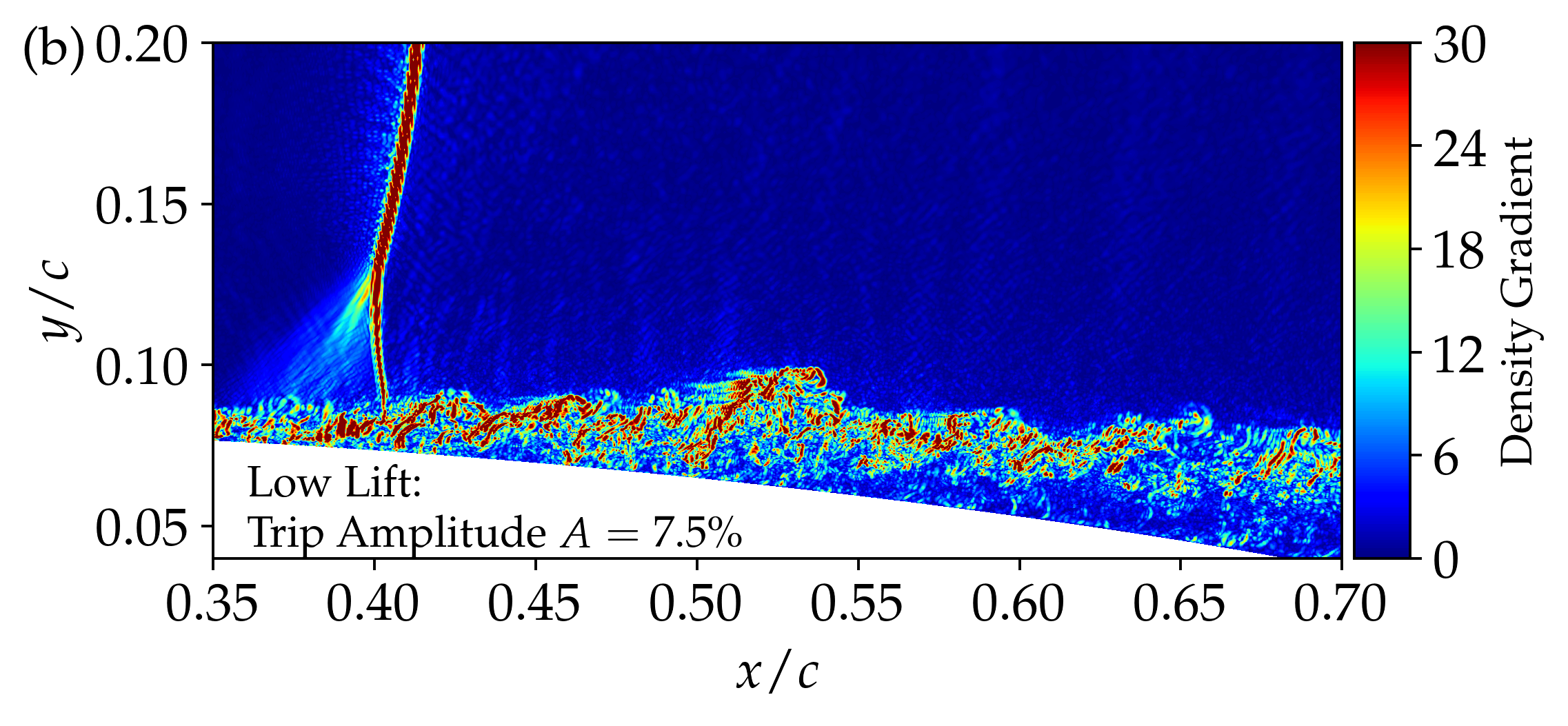}
  \includegraphics[width=0.485\columnwidth]{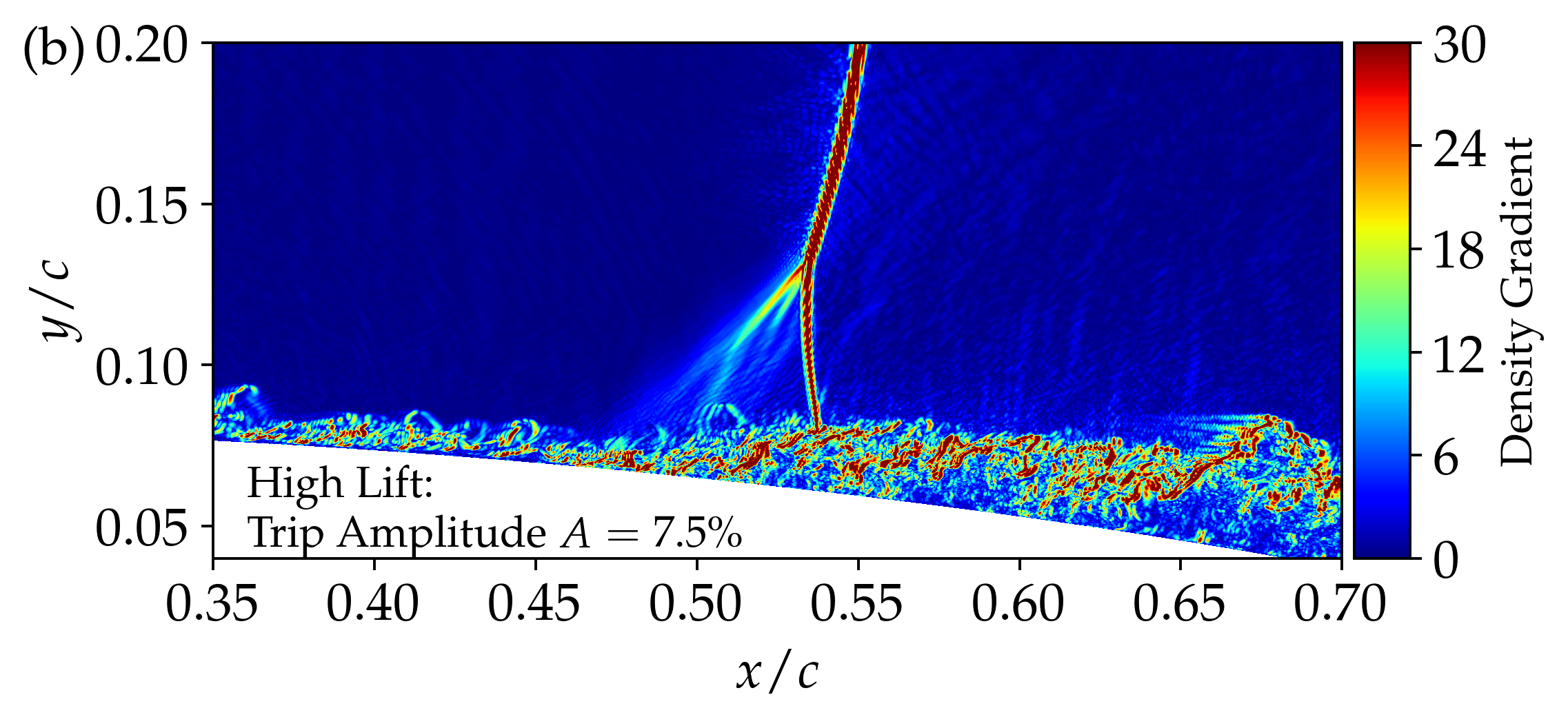}\\
   \includegraphics[width=0.485\columnwidth]{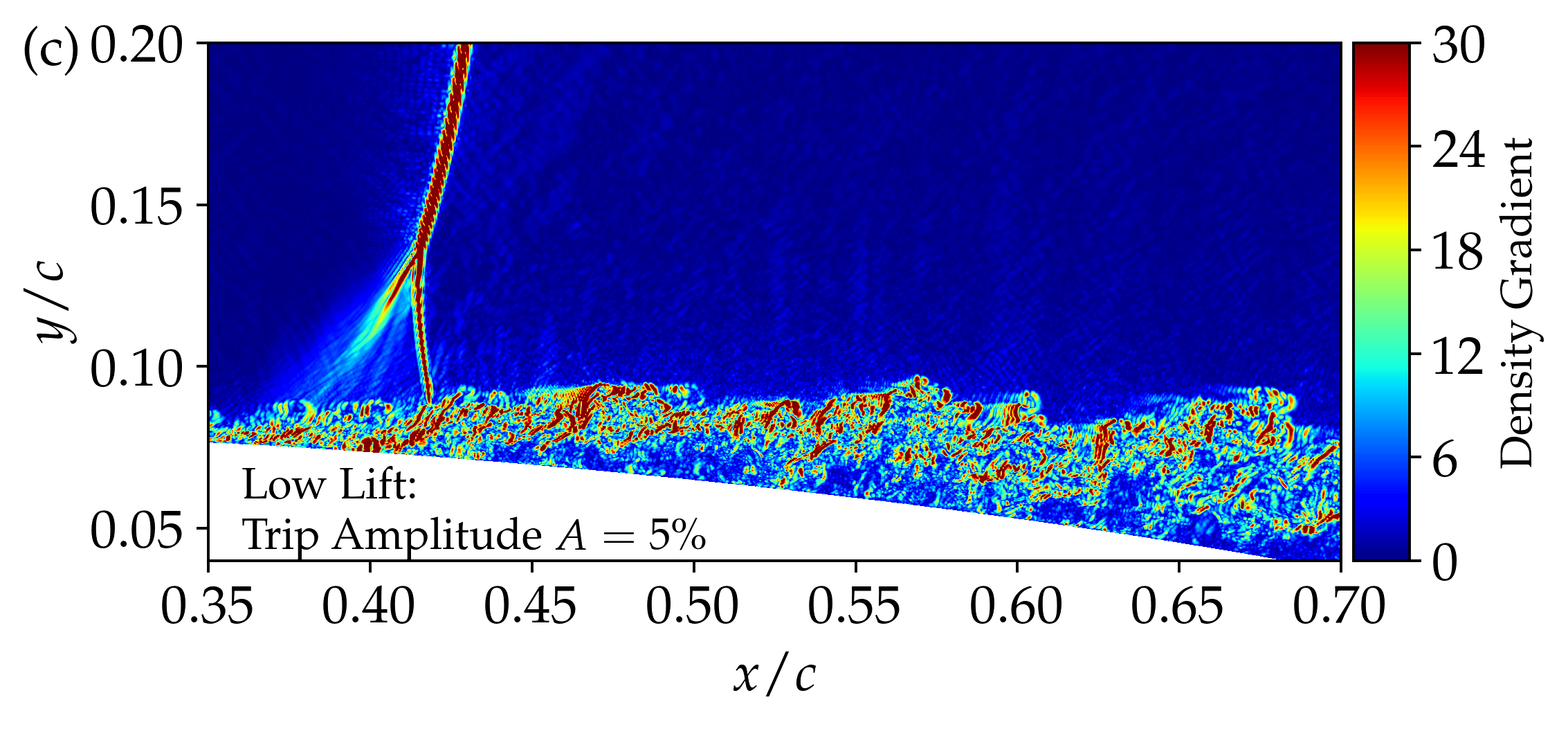}
  \includegraphics[width=0.485\columnwidth]{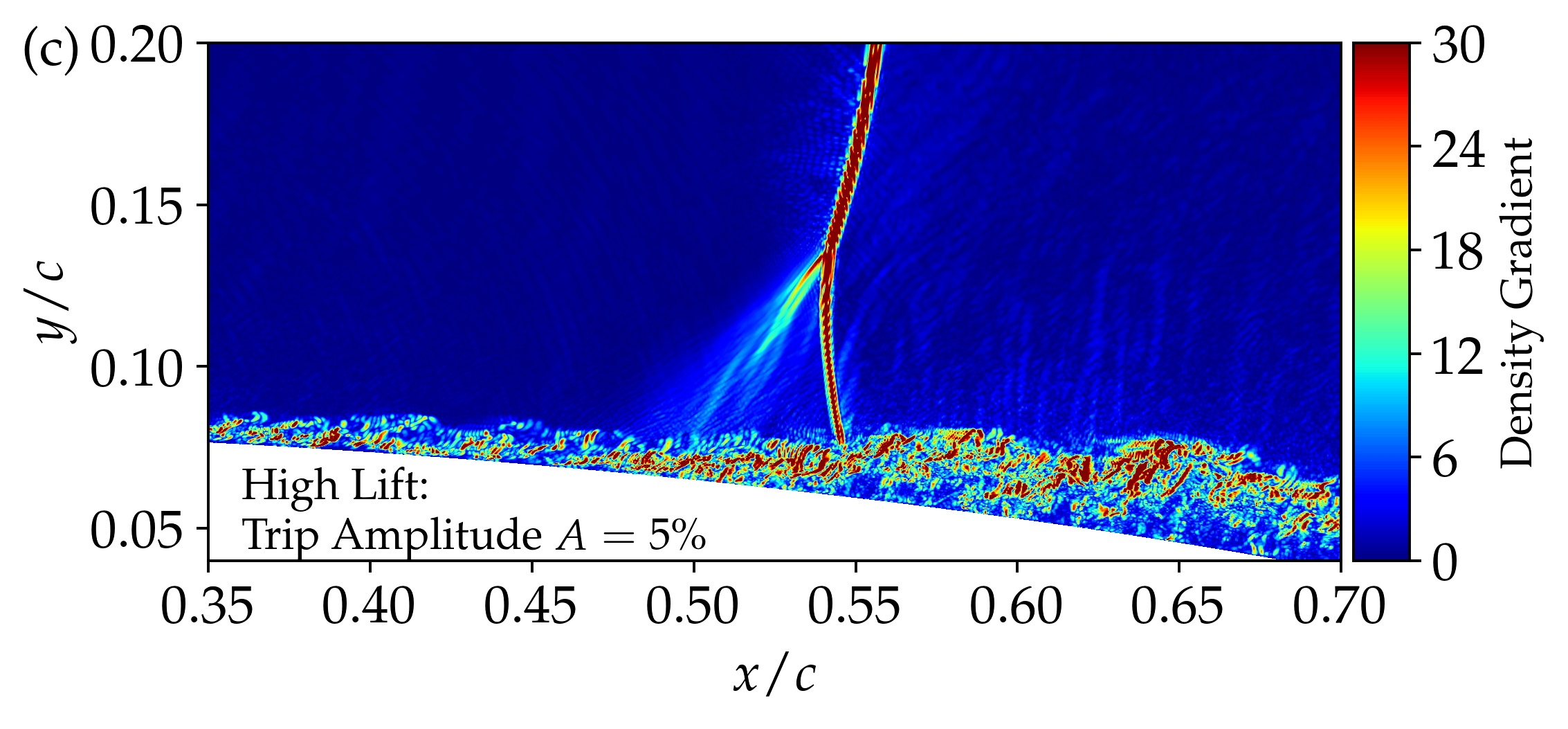}\\
  \includegraphics[width=0.485\columnwidth]{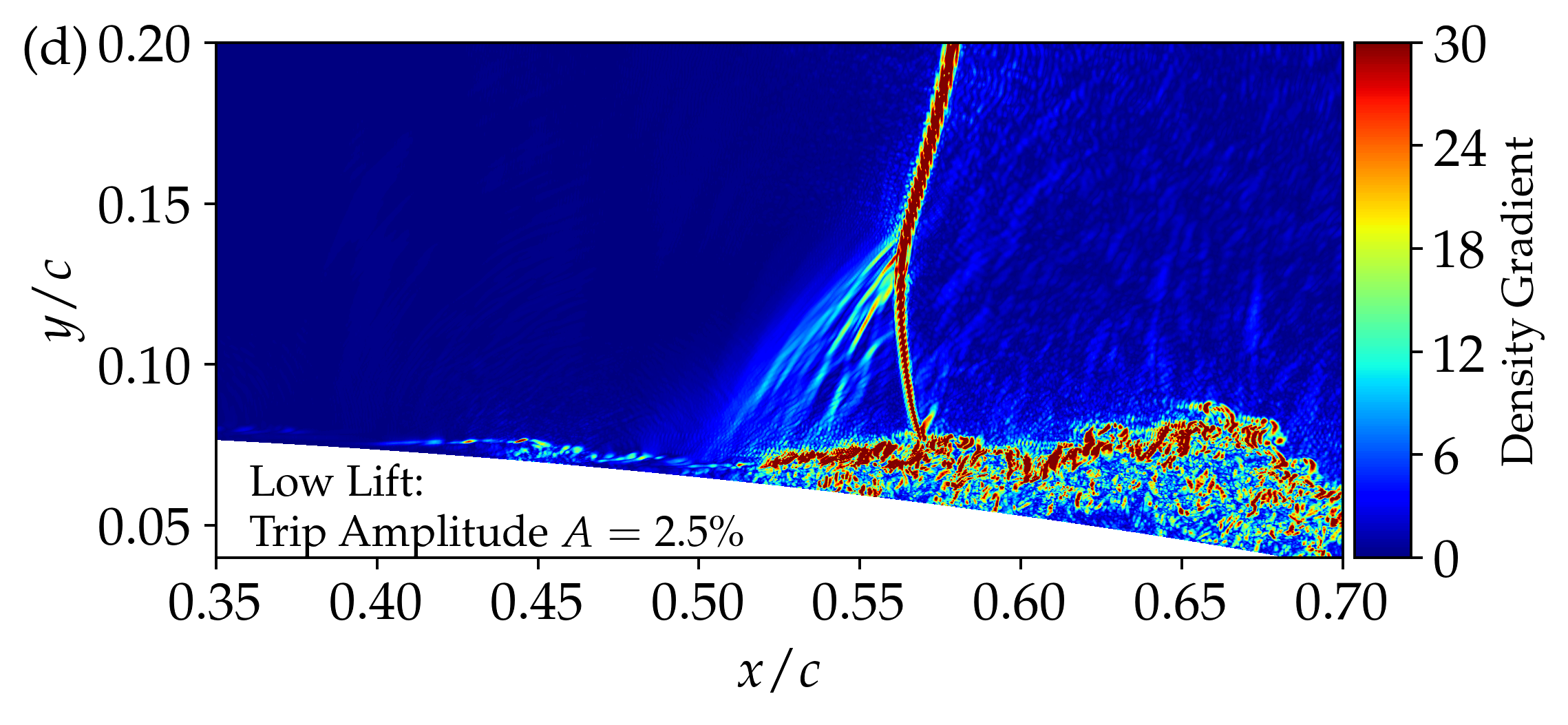}
  \includegraphics[width=0.485\columnwidth]{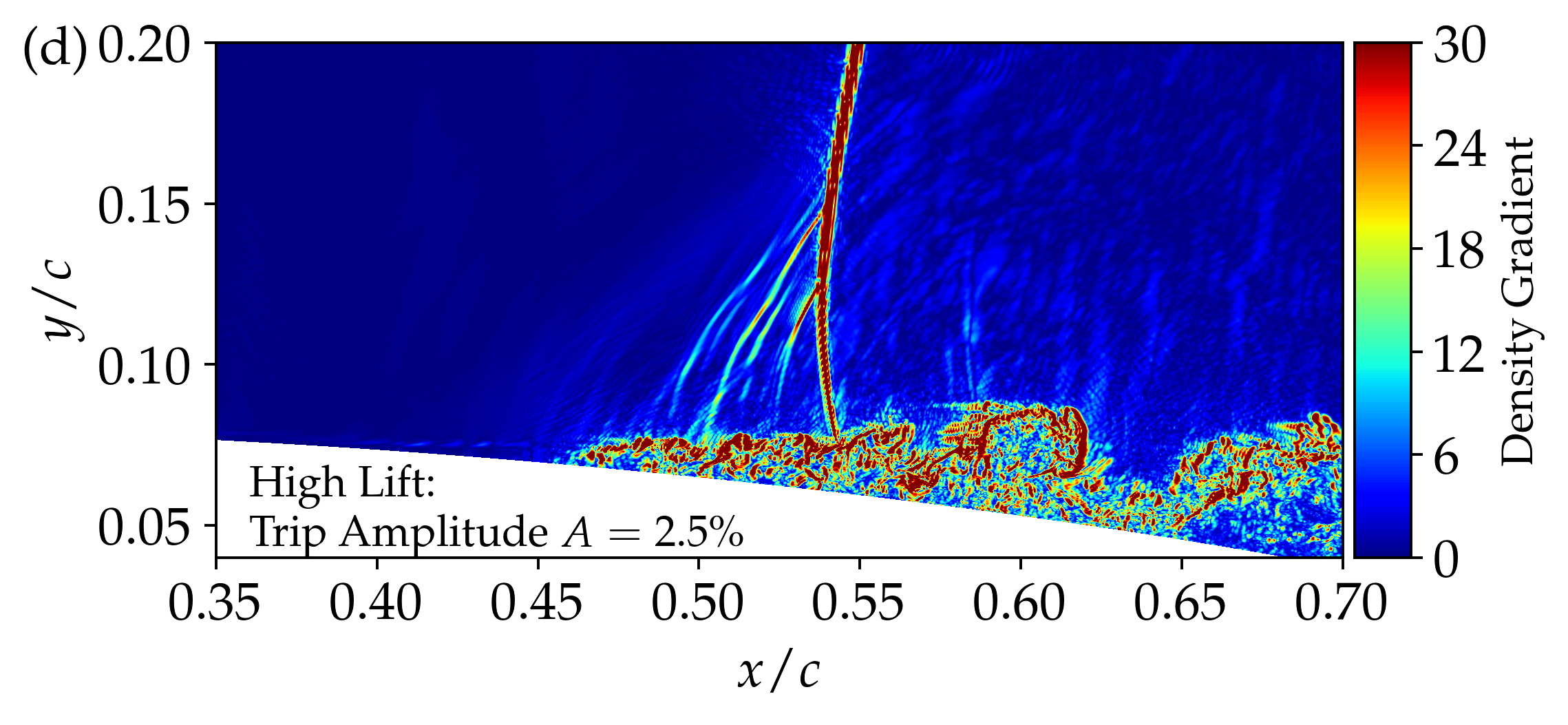}\\    
  \includegraphics[width=0.485\columnwidth]{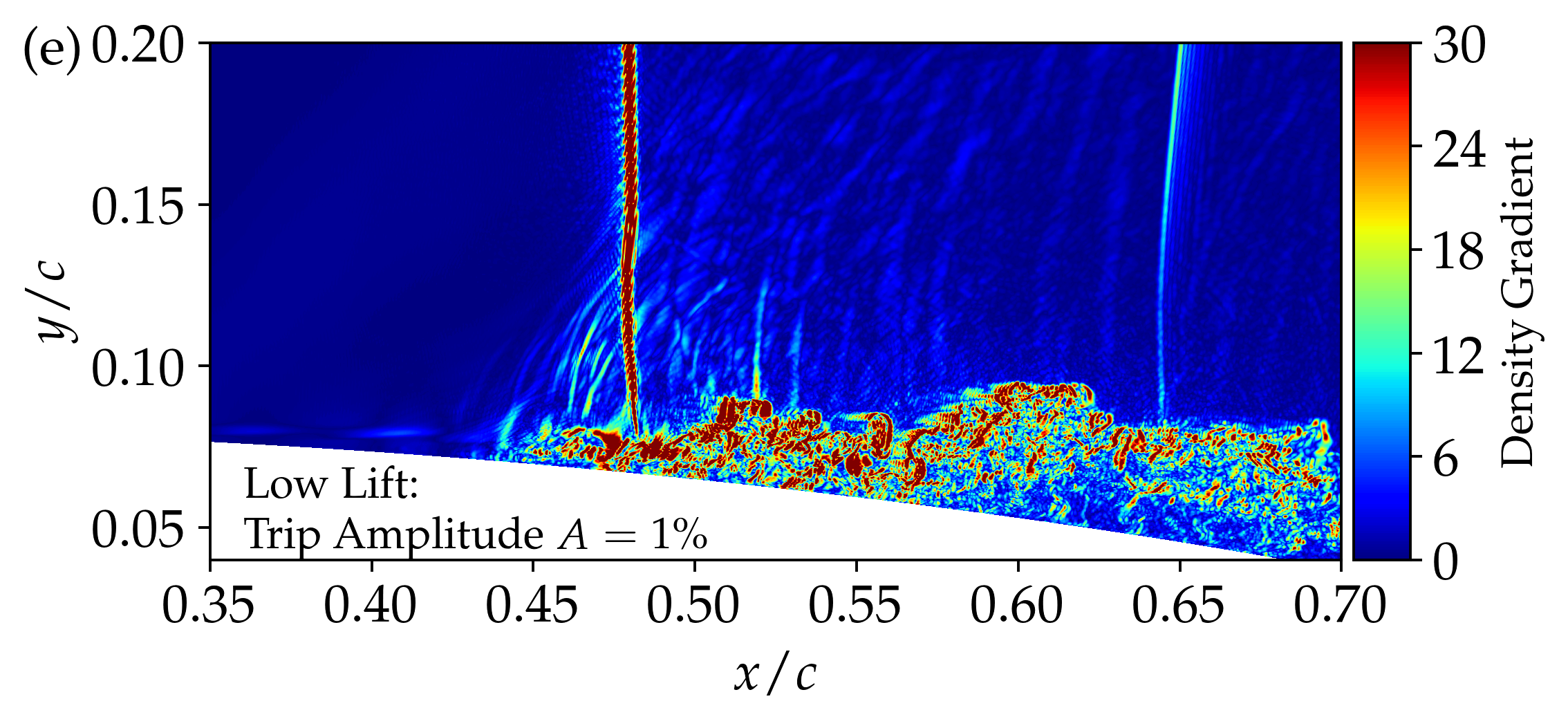}
  \includegraphics[width=0.485\columnwidth]{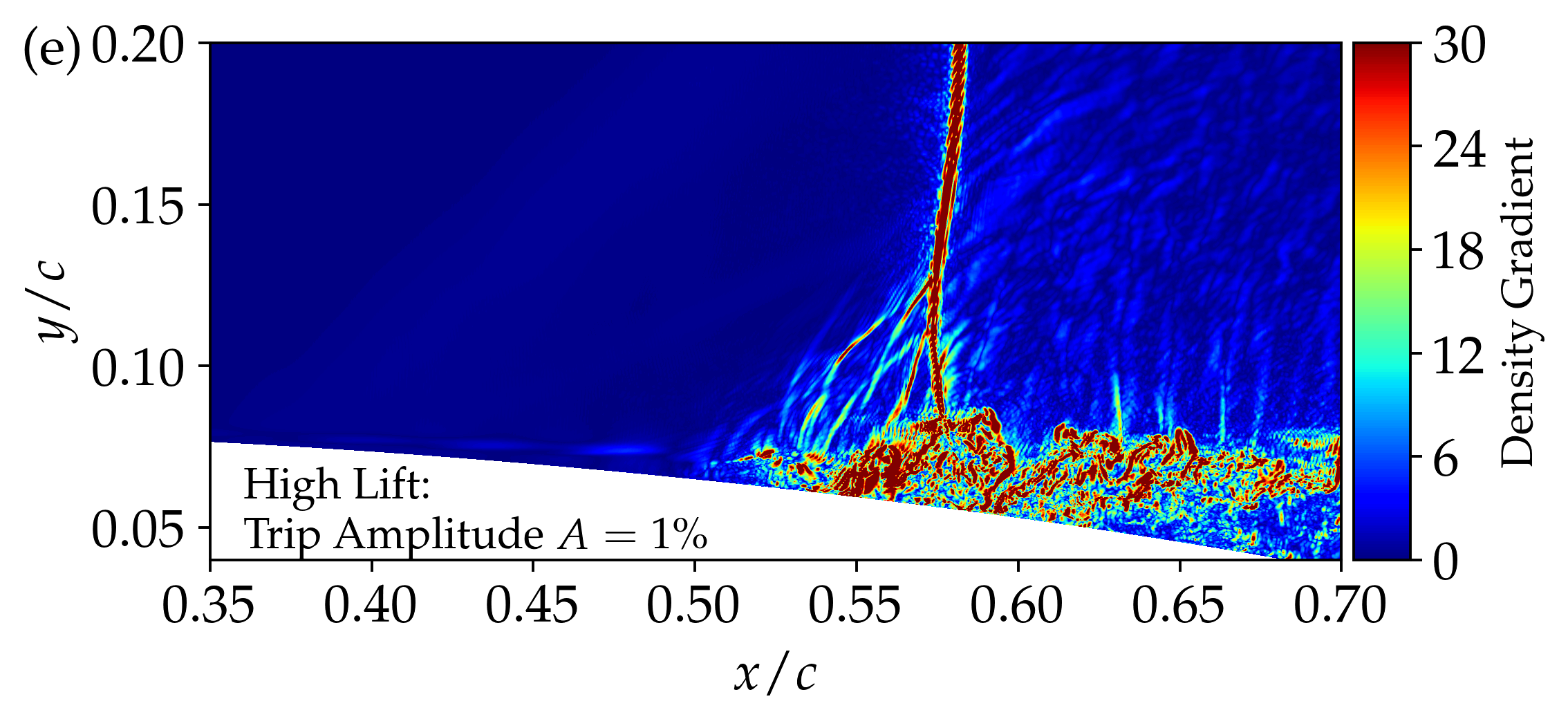}\\
  \includegraphics[width=0.485\columnwidth]{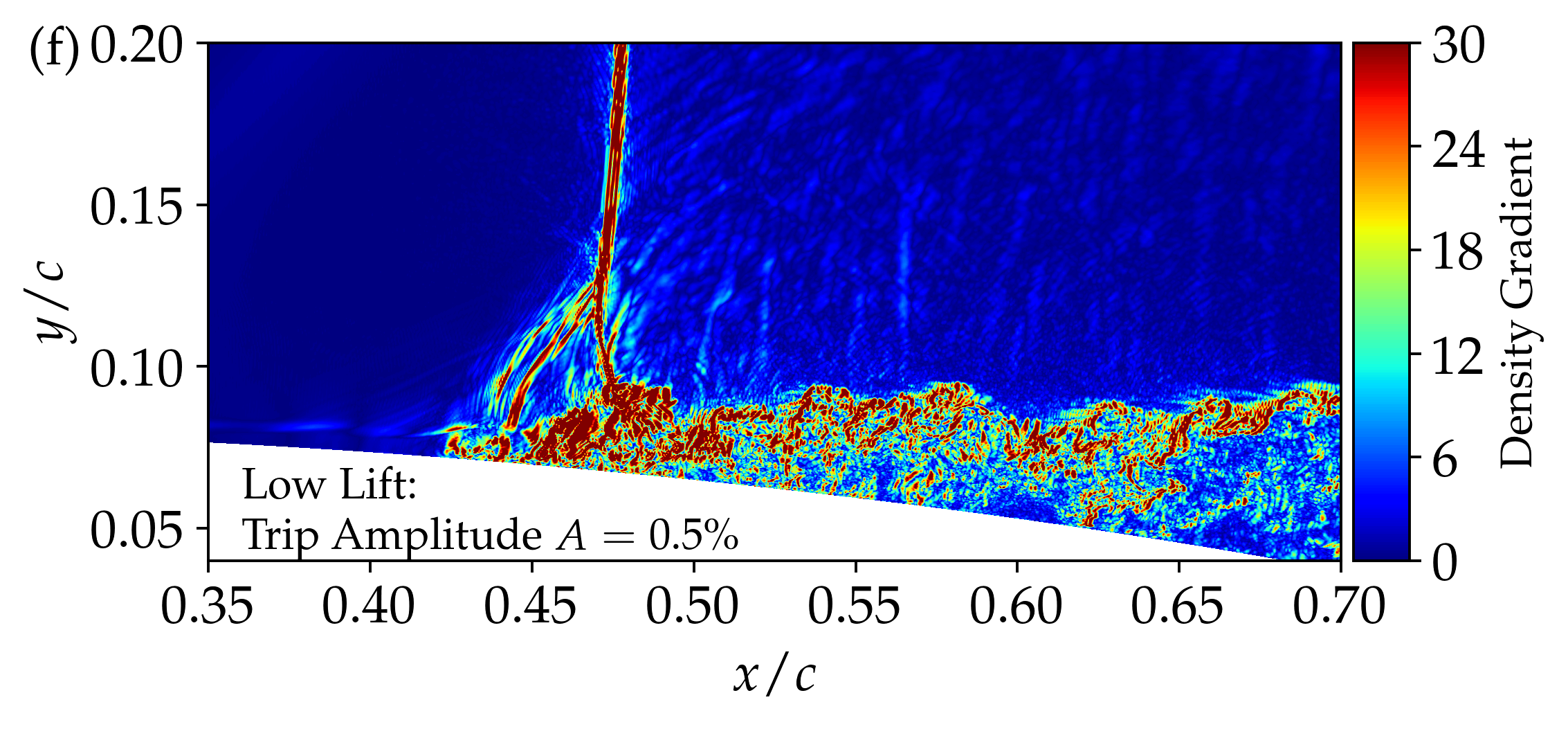}
  \includegraphics[width=0.485\columnwidth]{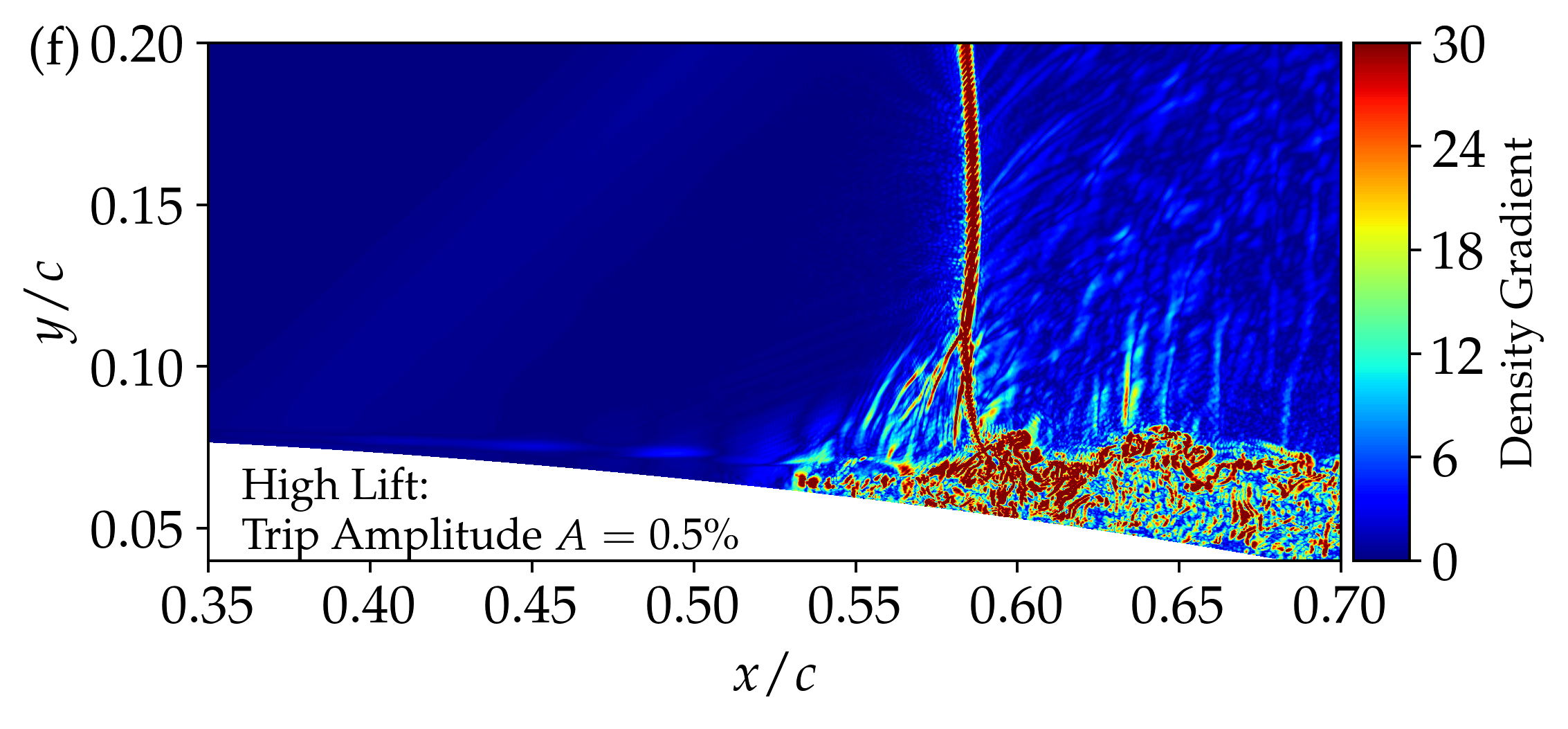}\\
  \caption{Instantaneous streamwise density gradient showing boundary-layer state and shock-wave structure on the suction side of the airfoil at (left) low- and (right) high-lift phases for different tripping amplitudes of $A=\left[0.5\%, 10\%\right]$.}
\label{fig:Amplitude_side_contours}
\end{center}
\end{figure}

Figure~\ref{fig:Amplitude_side_contours} shows instantaneous contours of density gradient at the main shock-wave location for the six tripping amplitudes in (a)-(f). The left and right hand columns of this figure show the low- and high-lift phases of the buffet cycle, respectively. The three strongest cases (Figures~\ref{fig:Amplitude_side_contours} (a)-(c)) all consist of a clear lambda shock-wave on top of the turbulent boundary-layers. The upstream and downstream shock-wave positions at the low- and high-lift phases are very similar for the three strongest tripping amplitudes. As the tripping strength is reduced in Figures~\ref{fig:Amplitude_side_contours} (d)-(f), the lambda shock structure is seen to partially breakdown. In these cases the boundary-layer upstream of the main SBLI is no longer turbulent, with the flow observed to rapidly transition around the shock position. While the turbulent cases (Figures~\ref{fig:Amplitude_side_contours} (a)-(c)) all possess a terminating shock-wave approximately perpendicular to the surface curvature, the laminar cases have a shock-wave that is far more vertical. Importantly, we note that despite the very low forcing amplitude ($A=0.5\%)$ and essentially laminar boundary-layer upstream of the SBLI, the interaction still contains only a single main shock-wave, as typically attributed to fully-turbulent airfoil buffet \citep{JMDMS2009, FK2018, SLKHR2022}. 

The configuration of multiple travelling shock-wave/expansions attributed in some (but not all) works as representative of `free-transitional' or `laminar' buffet \citep{MarkusPRF_2020,moise_zauner_sandham_2022,Moise2023_AIAAJ} is not observed, even when the upstream boundary-layer in the present cases is far from being in a turbulent state (Figures~\ref{fig:Amplitude_side_contours} (e,f)). Laminar buffet with a single main shock-wave has been reported both computationally \citep{dandois_mary_brion_2018} and experimentally \citep{Brion2017_Laminar_Experiment,Brion2020_Laminar_Experiment}, showing shock topologies closer to those observed for the laminar cases in the present work. Furthermore, we note the presence of sharp density gradients oriented vertically above the boundary-layer for the almost laminar cases (Figures~\ref{fig:Amplitude_side_contours} (e,f)), which are not observed once the interaction is fully turbulent (Figures~\ref{fig:Amplitude_side_contours} (a-c). This feature is most apparent in Figure~\ref{fig:Amplitude_side_contours} (e) at low-lift, where a secondary weak shock is visible at $x=0.65$. Recent laminar buffet simulations by \citet{LongWong2024_Laminar_buffet} reported a similar small intermittently-present terminating shock downstream of the main-shock-wave, for a fully-untripped OALT25 profile at a higher Reynolds number of $Re=1\times 10^{6}$ with $AR=0.25$. Finally, we note that the distance over which the shock oscillates between $x_\textrm{High}$ and $x_\textrm{Low}$ is also sensitive to the change in upstream boundary-layer state. For the turbulent cases (Figures~\ref{fig:Amplitude_side_contours} (a)-(c)), the main shock-wave traverses over a region of $\approx 0.13 c$ between low- and high-lift phases, compared to the laminar cases with $\approx 0.1 c$. The transitional case shows the least streamwise shock movement of only $\approx 0.02 c$. This lack of shock movement for the transitional case could partially explain the lower amplitudes observed for the low-frequency lift coefficient (Figure~\ref{fig:amp_study_lines} (a)), relative to the other tripping amplitudes. The observed oscillations of the separation point and reduced overall shock-wave movement is similar to the laminar buffet cases of \cite{dandois_mary_brion_2018} at the higher Reynolds number of $Re=3 \times 10 ^6$. The authors of that work found lower shock oscillation amplitudes compared to turbulent buffet and argued that oscillations of the shock-foot played a more prominent role in the instability. Modal analysis in future work will investigate the transitional cases in more detail, to understand the role of the shock-foot and separation point in laminar/transitional buffet.

\section{Sensitivity of turbulent transonic buffet to span-width}
\label{sec:spanwidth_sensitivity}

\begin{figure}[h]
\begin{center}
  \includegraphics[width=1\columnwidth]{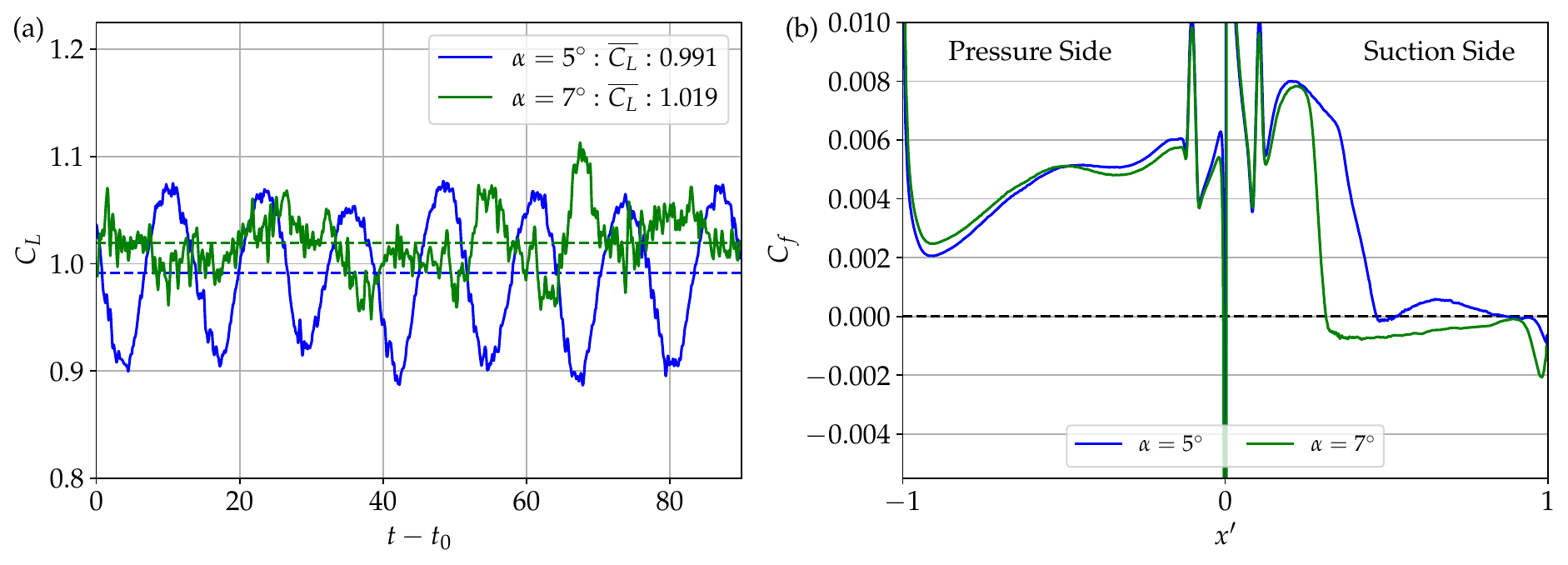}
  \caption{Moderate and largely separated cases for the baseline configuration ($AR=0.05$ and $A=7.5\%$) at angles of attack $\alpha = 5$ and $7^{\circ}$. Showing (a) lift coefficient history and (b) time-averaged skin-friction.}
\label{fig:5vs7}
\end{center}
\end{figure}

In this section, the sensitivity of transonic buffet to span-wise domain width is investigated for aspect ratios in the range of $AR=\left[0.025, 0.5\right]$. At the upper end, this is significantly wider than the domains typically used ($AR=0.05 - 0.065$) in previous high-fidelity studies of transonic buffet \citep{FK2018,Nguyen2022,Moise2023_AIAAJ}. Two post-onset buffet conditions are selected at $\alpha = 5$ and $7^{\circ}$. For both angles of attack, the flow is unsteady, and the shock oscillates and causes fluctuations in the lift-coefficient histories (Figure~\ref{fig:5vs7} (a)). While the differences in dominant frequencies and flow separation characteristics at different aspect ratios will be discussed later in this section, it is important to first note Figure~\ref{fig:5vs7} (b), that while the boundary-layer is (in a time-averaged sense) marginally separated at $\alpha = 5^{\circ}$, the separation is far more extensive due to a stronger shock and increased inclination at $\alpha = 7^{\circ}$. A summary of the cases simulated is shown in Table \ref{tab:domain_cases}, with their main aerodynamic coefficients shown. 

\begin{table}[h]
  \begin{center}
\def~{\hphantom{0}}
  \begin{tabular}{ccccccccc}
      Case       & Method    & AR     & $\alpha$       & $\overline{C_L}$ & $\overline{C_{Dp}}$ & $\overline{C_{Df}}$ & $\overline{C_D}$ & $St_{2D}$\\ \hline

      AR0025-AoA5 & ILES    & 0.025 & $5^{\circ}$  &         0.968      &       0.0505             &         0.0078           &        0.0583         &      0.0776 \\
      AR005-AoA5 & ILES    & 0.05 & $5^{\circ}$  &          0.993       &       0.0504             &         0.0079           &        0.0583         &      0.0778 \\
      AR010-AoA5 & ILES    & 0.10 & $5^{\circ}$  &          1.004       &       0.0505             &         0.0079           &        0.0584         &      0.0778 \\
      AR025-AoA5 & ILES    & 0.25 & $5^{\circ}$  &          1.015       &       0.0509            &         0.0079           &        0.0588         &      0.0778 \\
      AR050-AoA5 & ILES    & 0.50 & $5^{\circ}$  &          1.007       &       0.0506             &         0.0079           &        0.0586         &      0.0778 \\\hline

      AR0025-AoA7 & ILES    & 0.025 & $7^{\circ}$  &          Unstable       &       Unstable             &         Unstable          &        Unstable         &      Unstable \\
      AR005-AoA7 & ILES    & 0.05 & $7^{\circ}$  &          1.019       &       0.0984             &         0.0066           &        0.1049         &      0.0778 \\
      AR010-AoA7 & ILES    & 0.10 & $7^{\circ}$  &          0.978       &       0.0870             &         0.0066           &        0.0936         &      0.0889 \\
      AR025-AoA7 & ILES    & 0.25 & $7^{\circ}$  &          0.988      &        0.0865             &         0.0066           &        0.0931         &      0.1001 \\
      AR050-AoA7 & ILES    & 0.50 & $7^{\circ}$  &          0.988       &       0.0868             &         0.0066           &        0.0934         &      0.1112 \\\hline

  \end{tabular}
  \caption{Summary of the span-width sensitivity cases at moderate AoA (buffet onset, $5^{\circ}$), and high AoA (buffet, $7^{\circ}$) conditions. For each case, the tripping amplitude is fixed to $A=7.5\%$ of freestream velocity.}
  \label{tab:domain_cases}
  \end{center}
\end{table}

\subsection{Span-width sensitivity for moderately separated flows}\label{sec:spanwidth_moderate}

\begin{figure}
\begin{center}
  \includegraphics[width=1\columnwidth]{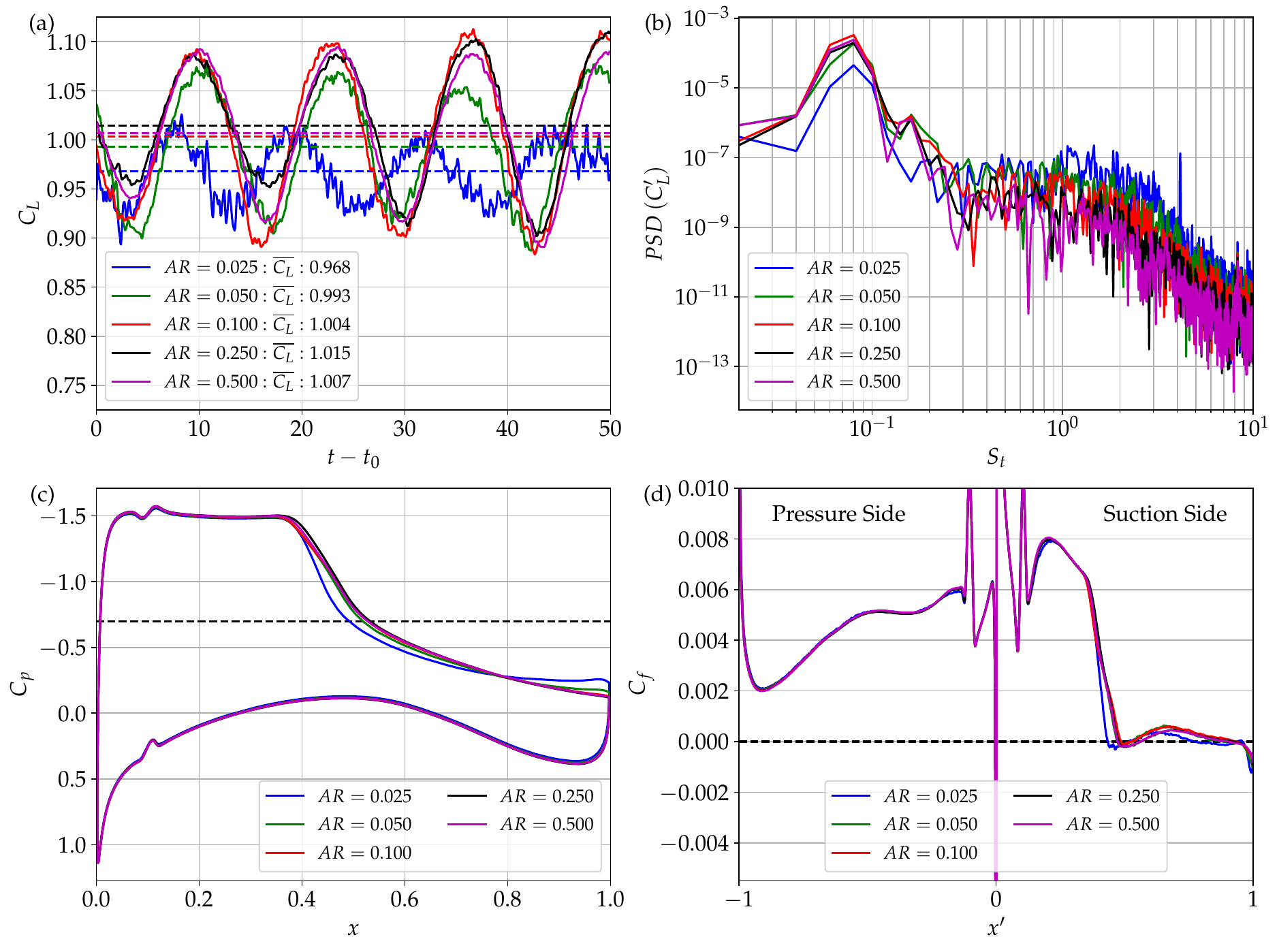}
  \caption{Span-width sensitivity at $\textrm{AoA} = 5^{\circ}$. Showing (a) lift coefficient history (b) PSD of fluctuating lift component (c) time-averaged pressure coefficient and (d) time-averaged skin-friction.}
\label{fig:domain_deg5_lines}
\end{center}
\end{figure}

Figure~\ref{fig:domain_deg5_lines} shows the aerodynamic quantities near buffet onset conditions ($\alpha = 5^{\circ}$) for  $AR=0.025, 0.05, 0.1, 0.25, 0.5$. The narrowest domain of $AR=0.025$ under predicts $\overline{C_L}$, and there are large deviations in distributions of $C_p$ and $C_f$. While the pressure side of the airfoil and transition region upstream of the SBLI are essentially insensitive to domain width, the narrow domain is seen to be insufficient at the shock location and near the trailing edge. At this AoA, there is a clear sensitivity to domain width near the trailing edge in both the $C_p$ and $C_f$ distributions in Figure~\ref{fig:domain_deg5_lines} (c) and (d). There is still a slight over-prediction of the separated region near the TE at $AR=0.05$, whereas for $AR=0.1$ and above, the time-averaged distributions converge exactly. Despite the under-prediction of $\overline{C_L}$ for the narrowest two domains, all five aspect ratios reproduce the same low-frequency buffet peak in the PSD in figure~\ref{fig:domain_deg5_lines} (b) of $S_t \approx 0.078$. The narrowest domain exhibits a spurious peak at $S_t = 4.1$ compared to the other aspect ratios where it is not present. For these buffet cases within $\alpha = 0.5^{\circ}$ of onset and moderate regions of time-averaged flow separation, an aspect ratio of $AR=0.1$ is required to achieve insensitivity to domain width. While the $AR=0.05$ case generally agrees well for the low-frequency buffet, it does slightly over-predict separation near the trailing edge as the flow is artificially constrained to be more two-dimensional than in the wider aspect ratios.

\begin{figure}
\begin{center}
  \includegraphics[width=0.497\columnwidth]{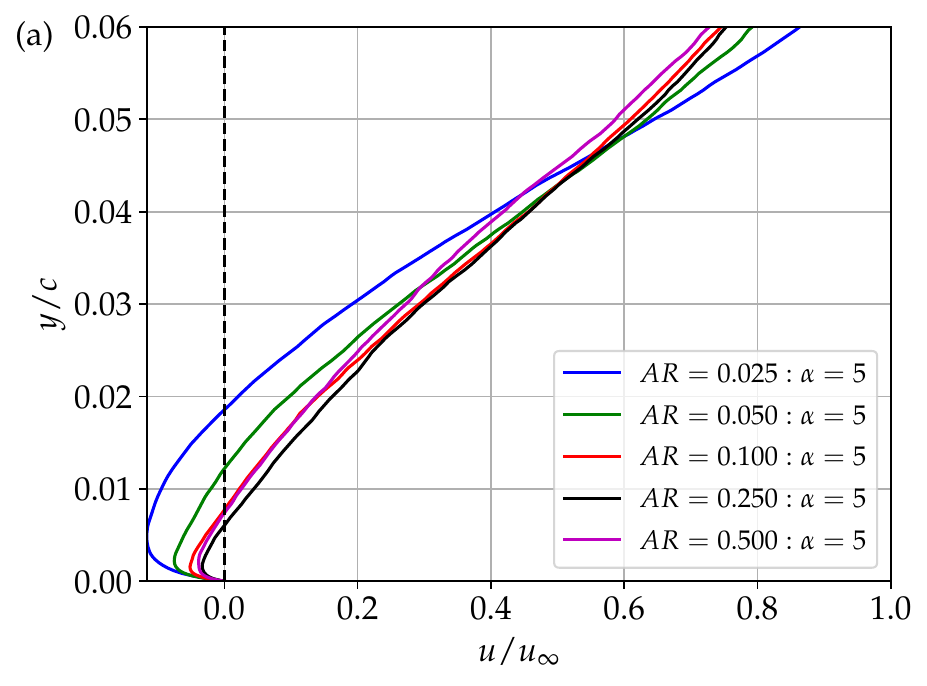}
  \includegraphics[width=0.497\columnwidth]{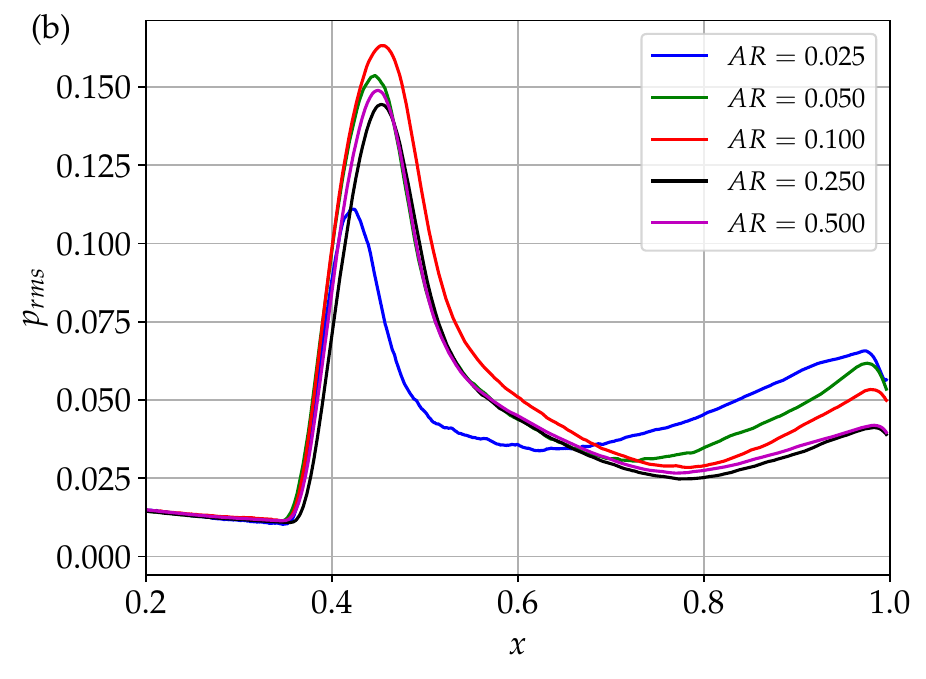}\\
  \caption{Spanwise domain width sensitivity for (a) trailing edge streamwise velocity boundary-layer profiles and (b) RMS of pressure for $\alpha = 5^{\circ}$.}
\label{fig:deg5_TE_BL}
\end{center}
\end{figure}

Figure~\ref{fig:deg5_TE_BL} shows the (a) mean streamwise velocity boundary-layer profiles at $97.5\%$ of chord length and (b) RMS distributions of pressure along the suction side of the airfoil, for aspect ratios of $AR=\left[0.025, 0.05, 0.1, 0.25, 0.5\right]$. Data to the left of the vertical dashed $u=0$ line represents regions of reverse flow near the trailing edge. The wider aspect ratios exhibit far weaker separations compared to the narrow ($AR=0.025, 0.05$) cases, both in magnitude and vertical extent. The time-averaged reverse flow region extends to a distance of $\approx 0.007c$ above the airfoil surface at $AR=0.1$ and above, compared to $\approx 0.012c$ and $\approx 0.018c$ for the narrower $AR=0.05$ and $AR=0.025$. The narrow domains artificially constrain the flow separation to be overly two-dimensional, leading to the domain-dependent artefacts seen near the trailing edge in the pressure and skin-friction distributions of Figures~\ref{fig:domain_deg5_lines} (c,d). At $AR=0.1$ and above the profiles converge and the domain sensitivity is removed. 

Figure~\ref{fig:deg5_TE_BL} (b) shows the RMS of pressure fluctuations across the suction side of the airfoil for each of the aspect ratios. The pressure fluctuations upstream of the shock-wave $0.2 < x < 0.35$ collapse for all aspect ratios considered, indicating that the tripped transition method is insensitive to the domain width. The pressure rise at the shocked region is similar for all aspect ratios except the narrowest case ($AR=0.025$), which appears overly narrow to reproduce the SBLI dynamics captured by the wider simulations. The main differences appear close to the trailing edge as previously seen. The three narrowest aspect ratios all overestimate the pressure fluctuations at the trailing edge compared to the $AR=0.25$ and $AR=0.5$ cases. This is similar to the findings of \citet{GD2010}, who observed a significant reduction in trailing edge pressure fluctuations when increasing the domain with from $AR=0.0365$ to $AR=0.073$. The simulations of \citet{GD2010} were later reviewed by \citet{Giannelis_buffet_review}, who commented that the wider domain better captured unsteady pressures at the trailing edge, as three-dimensional turbulent structures were allowed to develop with a reduced intensity relative to the two-dimensional structures from the narrow domain. However, they went on to comment that even at $AR=0.073$ the pressure fluctuations were still overestimated compared to experiment. This is consistent with the present cases, which require between $AR=0.1$ and $AR=0.25$ to remove domain sensitivity at the trailing edge. Given the critical role that separation height has been argued to have for driving the underlying self-sustaining buffet mechanism \citep{KAWAI_RESOLVENT2023}, it is essential that sufficiently wide domains are used even for predictions of the two-dimensional part of the buffet phenomenon. The next section investigates whether similar domain sensitivity trends are observed at higher AoA ($\alpha = 7^{\circ}$) when flow separation is even larger.

\subsection{Span-width sensitivity for largely separated flows}\label{sec:spanwidth_large}

\begin{figure}
\begin{center}
  \includegraphics[width=1\columnwidth]{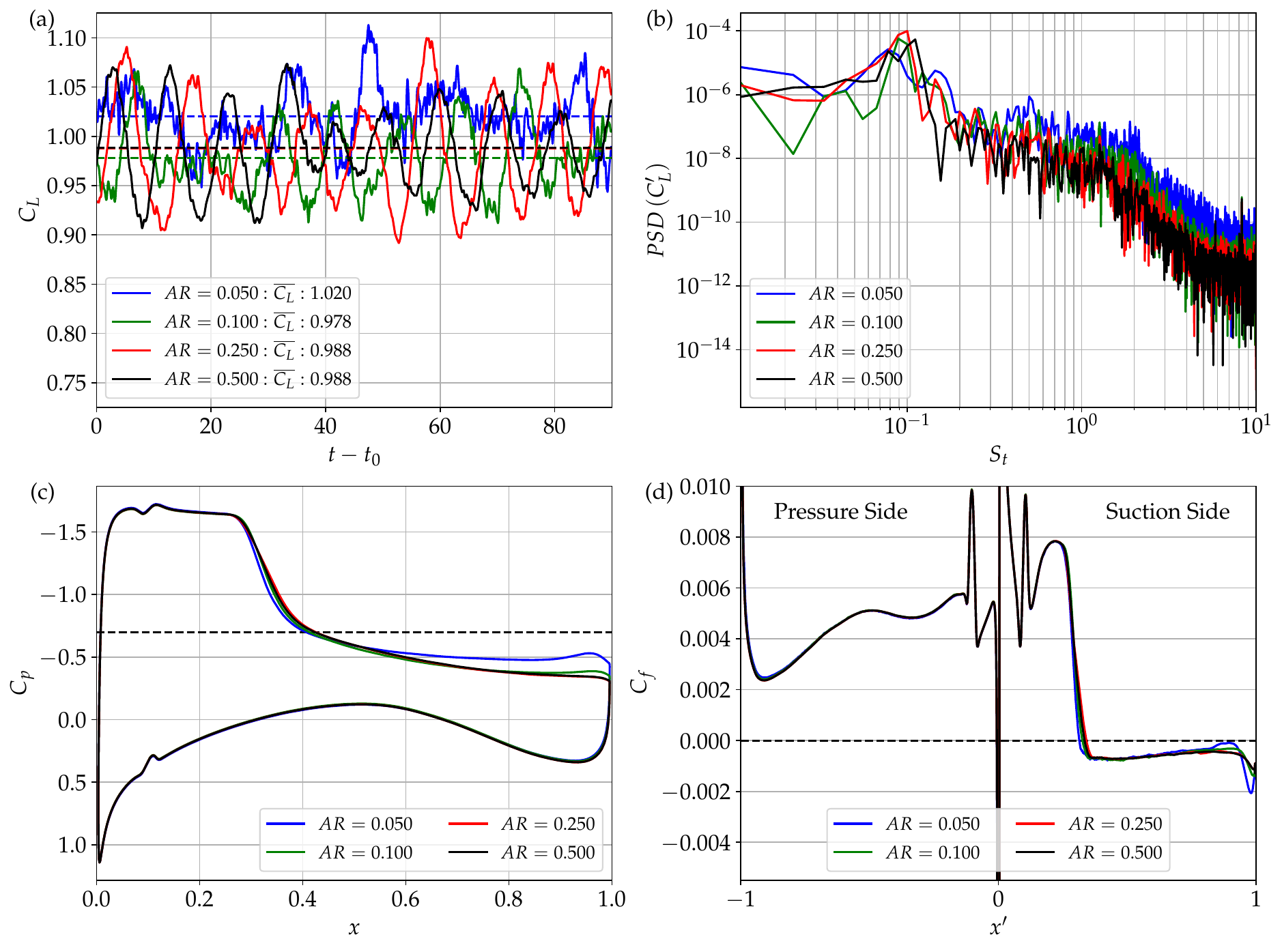}

  \caption{Span-width sensitivity at $\textrm{AoA} = 7^{\circ}$. Showing (a) lift coefficient history (b) PSD of fluctuating lift component (c) time-averaged pressure coefficient and (d) time-averaged skin-friction.}
\label{fig:domain_deg7_lines}
\end{center}
\end{figure}

\begin{figure}
\begin{center}
  \includegraphics[width=0.497\columnwidth]{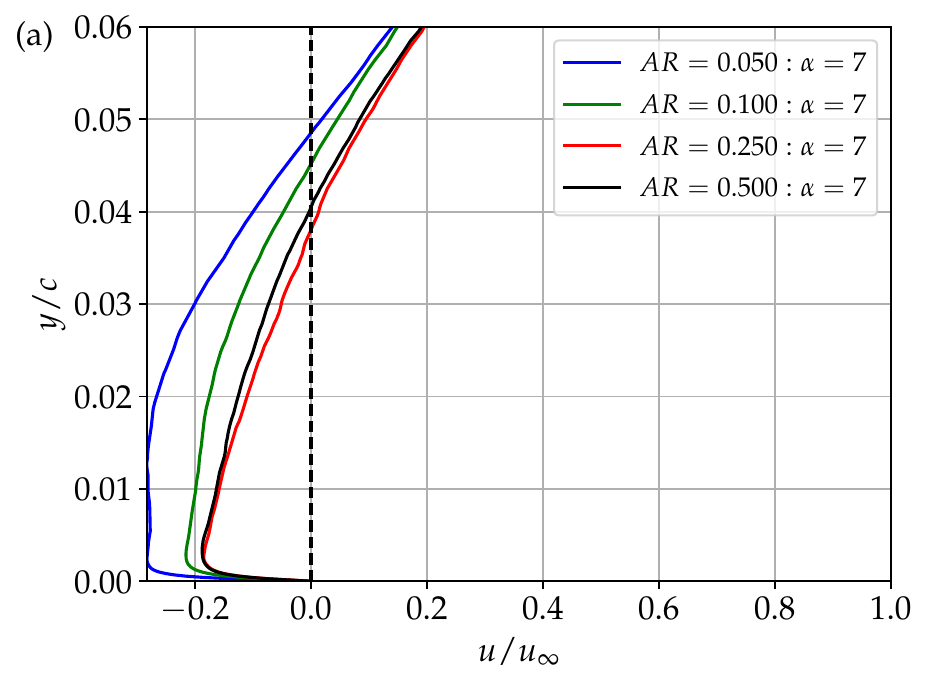}
  \includegraphics[width=0.497\columnwidth]{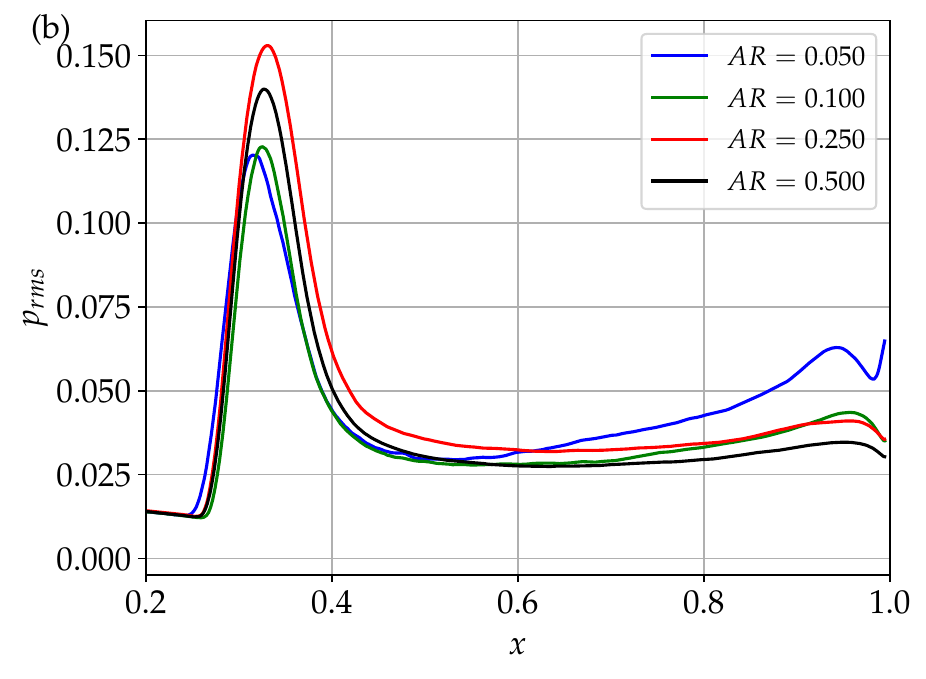}\\
  \caption{Spanwise domain width sensitivity for (a) trailing edge streamwise velocity boundary-layer profiles and (b) RMS of pressure for $\alpha = 7^{\circ}$}
\label{fig:deg7_TE_BL}
\end{center}
\end{figure}

In this section the influence of span-width is assessed at a higher angle of attack of $\alpha = 7^{\circ}$, to investigate whether similar domain sensitivity trends found at $\alpha = 5^{\circ}$ are present for buffet in the presence of more extensive flow separation. As shown in Table~\ref{tab:domain_cases}, the narrowest domain ($AR=0.025$) proved to be unstable at the higher AoA and was unsuitable to even provide a solution. The domain width study at $\alpha = 7^{\circ}$ was instead performed at $AR=0.05, 0.1, 0.25, 0.5$. Figure~\ref{fig:domain_deg7_lines} shows the aerodynamic coefficients for (a) lift, (b) PSD of lift fluctuations, (c) pressure coefficient, and (d) skin-friction distribution at each of the four aspect ratios. With respect to the previous moderate angle of attack, at $\alpha = 7^{\circ}$ the shock-wave is stronger, moves upstream, and causes the flow to largely separate. Unlike the lift-coefficient convergence with increasing aspect ratio observed at moderate AoA post-onset (Figure~\ref{fig:domain_deg5_lines} (a)), at $\alpha=7^{\circ}$ no clear convergence is found. However, the time-averaged $\overline{C_L}$ does agree to 3 decimal points at $AR=0.25$ and $AR=0.5$ despite irregularities between the individual buffet periods. Similar to the trends observed for moderate AoA buffet cases (Figures~\ref{fig:FaSTAR_URANS_2D_lines} and ~\ref{fig:domain_deg5_lines}), increasing the aspect ratio leads to increased regularity of the low-frequency buffet oscillations and reduction in the higher frequency content. While the results presented here are span-averaged, the same behaviour was observed with increasing aspect ratio when the aerodynamic coefficients were instead evaluated over a single spanwise grid location. This suggests that the trend of more regular low-frequency oscillations is intrinsic to the dynamics on the wider airfoils and not just an averaging effect.

In the PSD of lift fluctuations in Figure~\ref{fig:domain_deg7_lines} (b), there are low-frequency peaks visible in the 2D buffet range for each aspect ratio despite the higher-frequency noise observed in the lift histories (Figure~\ref{fig:domain_deg7_lines} (a)). However, unlike the lower-AoA buffet case (Figure~\ref{fig:domain_deg5_lines} (b)), the buffet frequency does not converge exactly with increasing aspect ratio. This suggests that the buffet dynamics are potentially different between $\alpha=5^{\circ}$ and $\alpha=7^{\circ}$, beyond just a simple shift in the buffet frequency. In Figure~\ref{fig:domain_deg7_lines} (c), we see that the narrowest domain of $AR=0.05$ which is similar to the values commonly used in previous buffet studies (e.g. $AR=0.0365 - 0.073$ \cite{GD2010}, $AR=0.05$ \cite{moise_zauner_sandham_2022,Moise2023_AIAAJ}, and $AR=0.065$ \cite{FK2018,Nguyen2022}), vastly over-predicts the pressure distribution from mid-chord to TE ($0.5 < x < 1$) at this high AoA. Elsewhere on the airfoil, all aspect ratios collapse and no domain-width sensitivity is observed. The $AR=0.1$ case shows a strong improvement, and is much closer to the wide-span results which converge in the trailing edge region at $AR=0.25$ and above. Similarly, the skin-friction distributions in Figure~\ref{fig:domain_deg7_lines} (d) show the inadequacy of the $AR=0.05$ domain at high AoA. The narrowest two both overestimate the separation size at the trailing edge, most notably at $AR=0.05$. The $AR=0.05$ case actually reattaches around $x=0.9$ before separating again, a feature not observed on the wider aspect ratios. This highlights that insufficient spanwise domain width can, in some cases, lead to opposing flow states, with flow features and energy content that is only present as a result of modelling error when selecting the airfoil width.

Figure~\ref{fig:deg7_TE_BL} again shows the boundary-layer profiles and RMS of pressure at the trailing edge, at the higher AoA of $7^{\circ}$. The same, albeit stronger, trends from the previous cases (Figure~\ref{fig:deg5_TE_BL}) are observed once more. The narrowest $AR=0.05$ case greatly overestimates the size of the separated region at the aft position of the airfoil both in magnitude and spatial extent, in contrast to the convergence seen for the wider domains. Finally, the RMS of pressure fluctuations in Figure~\ref{fig:deg7_TE_BL} (b) shows a similar trend of sensitivity at the trailing edge for the narrowest domains. In addition to the trailing edge, at $\alpha = 7^{\circ}$ we also observe some sensitivity at the shock location which will be later clarified. 

\subsection{Spanwise domain width selection criterion}\label{sec:domain_width_criterion}

\begin{figure}[h]
\begin{center}
  \includegraphics[width=1.\columnwidth]{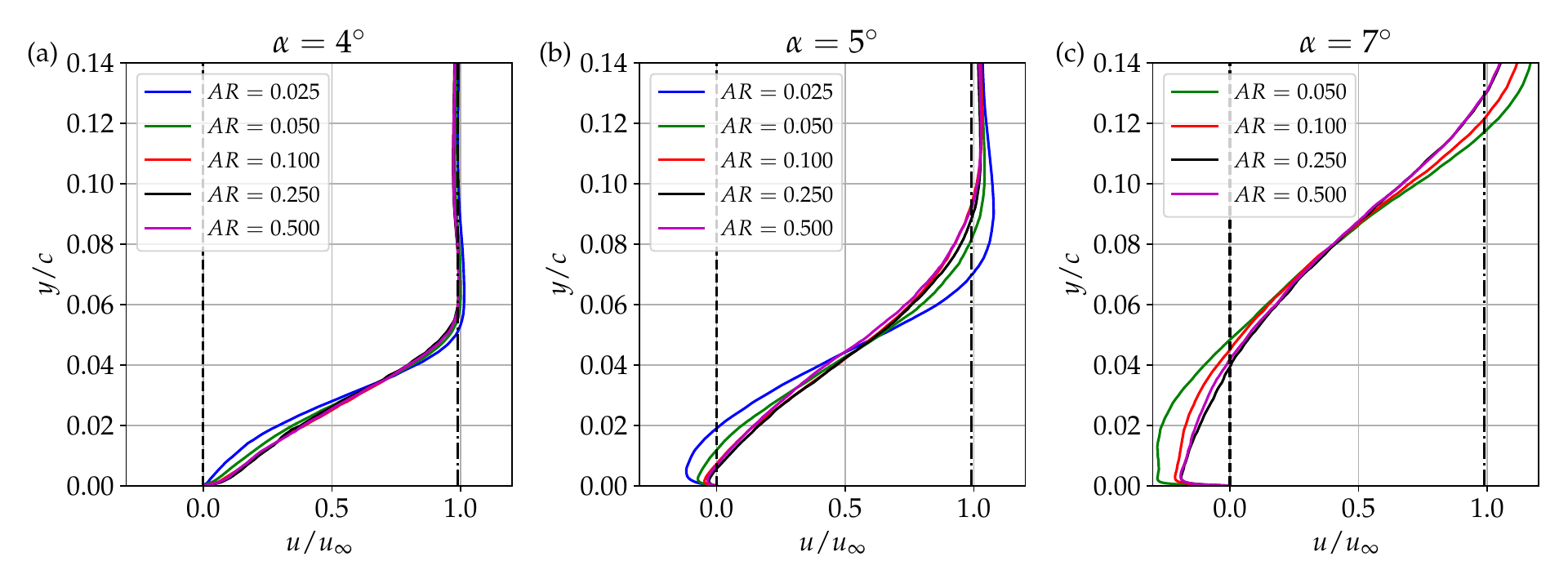}
  \caption{Spanwise domain width sensitivity for trailing edge streamwise velocity boundary-layer profiles at angles of attack of (a) $\alpha = 4^{\circ}$, (b) $\alpha = 5^{\circ}$, and (c) $\alpha = 7^{\circ}$. The ashed line represents zero crossing for reverse flow region, the dashed-dot line represents $99\%$ of freestream velocity.}
\label{fig:deg4-5-7_TE_BL}
\end{center}
\end{figure}

\begin{table}[h]
  \begin{center}
\def~{\hphantom{0}}
  \begin{tabular}{cccc|cccc|cccc}
      $\alpha$  & AR & $\overline{\delta}_{99}^{TE}$ & $AR/\overline{\delta}_{99}^{TE}$ & $\alpha$  & AR & $\overline{\delta}_{99}^{TE}$ & $AR/\overline{\delta}_{99}^{TE}$ & $\alpha$  & AR & $\overline{\delta}_{99}^{TE}$ & $AR/\overline{\delta}_{99}^{TE}$ \\ 
      \hline
$4^\circ$ & 0.025 & 0.050 & 0.500  & $5^\circ$ & 0.025 & 0.069 & 0.362 & $7^\circ$ & 0.025 & N/A   & N/A \\
$4^\circ$ & 0.05  & 0.054 & 0.926  & $5^\circ$ & 0.05  & 0.081 & 0.617 & $7^\circ$ & 0.05  & 0.117 & 0.427 \\
$4^\circ$ & 0.10  & 0.058 & 1.724  & $5^\circ$ & 0.10  & 0.092 & 1.087 & $7^\circ$ & 0.10  & 0.122 & 0.820 \\
$4^\circ$ & 0.25  & 0.058 & 4.310  & $5^\circ$ & 0.25  & 0.087 & 2.874 & $7^\circ$ & 0.25  & 0.129 & 1.938 \\
$4^\circ$ & 0.50  & 0.057 & 8.772 & $5^\circ$  & 0.50  & 0.091 & 5.495 & $7^\circ$ & 0.50  & 0.129 & 3.876 \\ \hline
  \end{tabular}
  \caption{Summary of the span-width selection criterion for cases at pre-onset ($\alpha = 4^{\circ}$), post-onset moderately separated ($\alpha = 5^{\circ}$), and post-onset largely separated ($\alpha = 7^{\circ}$). For each case, the tripping amplitude is fixed to $A=7.5\%$ of freestream velocity.}
  \label{tab:domain_width_criterion}
  \end{center}
\end{table}

The span-width sensitivity studies are summarized here in the attempt to determine quantitative criteria for appropriate selection of the spanwise domain width. Along with the results presented above for moderately and largely separated flows at post-onset conditions ($\alpha = 5^{\circ}$ and $\alpha = 7^{\circ}$, respectively), the pre-onset case at $\alpha = 4^{\circ}$ described in section~\ref{sec:finding_buffet} is also reported. Figure~\ref{fig:deg4-5-7_TE_BL} shows the time-averaged boundary-layer profiles near the trailing edge ($x/c=0.975$) for the three angles of attack and all investigated aspect ratios. As expected, the time-averaged boundary-layer thickens for increasing angles of attack, and by inspecting the reverse flow near the wall it is possible to observe the flow switching from fully-attached to moderately and largely separated. While in the fully-attached steady pre-onset and moderately separated unsteady post-onset cases the profiles seem to converge for $AR=0.1$, for the largely separated case an aspect ratio of at least $AR=0.25$ is needed. The $99\%$ boundary-layer thickness at the trailing edge ($\overline{\delta_{99}^{TE}}$) for all investigated cases is summarized in Table~\ref{tab:domain_width_criterion}, along with the $AR/\overline{\delta_{99}^{TE}}$ ratio. This study seems to suggest that to obtain a spanwise domain width independent solution, the span-width must be at least greater than the boundary-layer thickness at the trailing edge region, i.e. $AR/\overline{\delta_{99}^{TE}} > 1$. Previous buffet simulations \citep{FK2018} have used a domain-width set equal to experimental time-averaged boundary-layer thickness values measured at $75\%$ chord. This is in general agreement with our findings, although we set our criteria based on boundary-layer thickness at $97.5\%$ chord where the boundary-layer is thicker. While the spanwise domain width guidelines given here should be considered appropriate for the present geometry and flow conditions, there seems to be a general trend which may also apply to other configurations.

\subsection{Comparison to 2D and 3D URANS at $AR=1$}\label{sec:high_AoA_URANS}

\begin{figure}[h]
\begin{center}
  \includegraphics[width=1.\columnwidth]{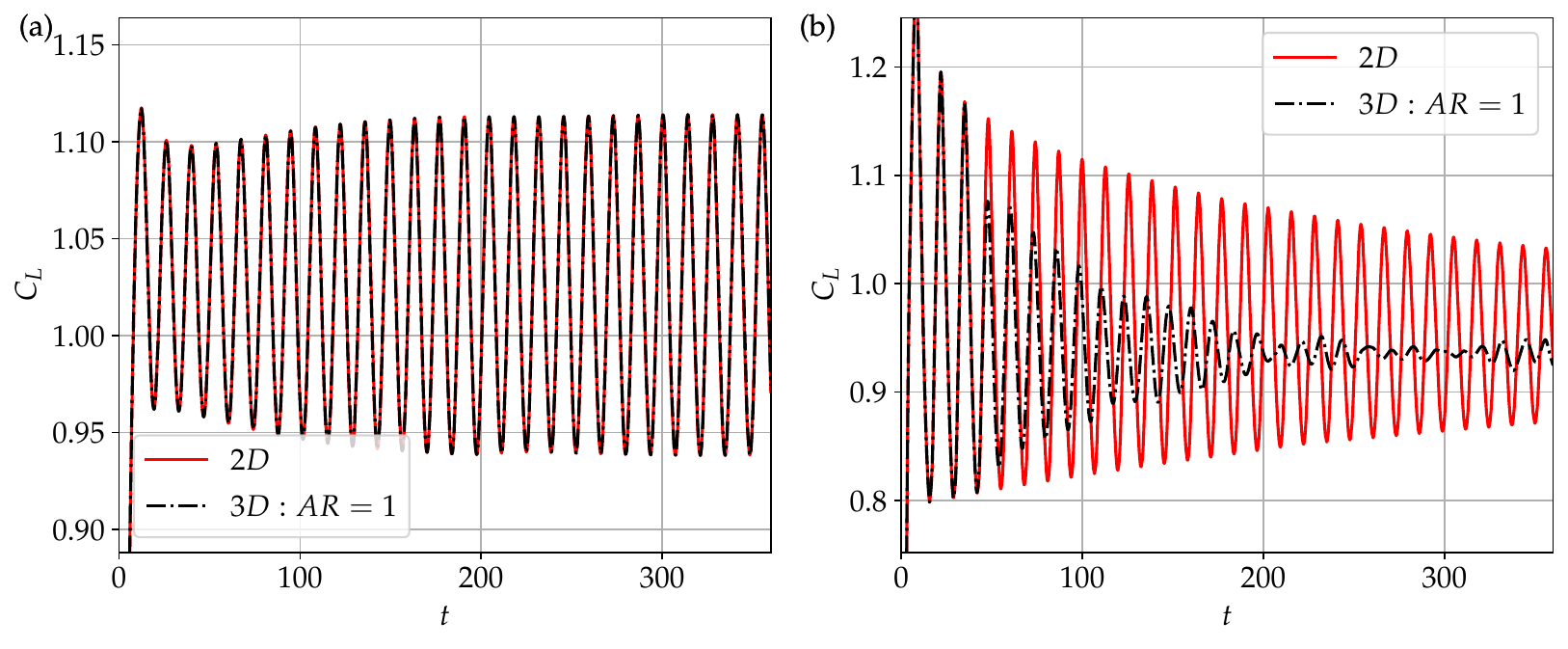}
  \caption{Lift coefficient history comparisons for URANS cases performed on 2D and 3D ($AR=1$) configurations. Showing angles of attack of (a) $\alpha = 5^{\circ}$, and (b) $\alpha = 7^{\circ}$.}
\label{fig:FaSTAR_URANS_AR_sensitvity_histories}
\end{center}
\end{figure}

\begin{figure}
\begin{center}
 \includegraphics[width=1.\columnwidth]{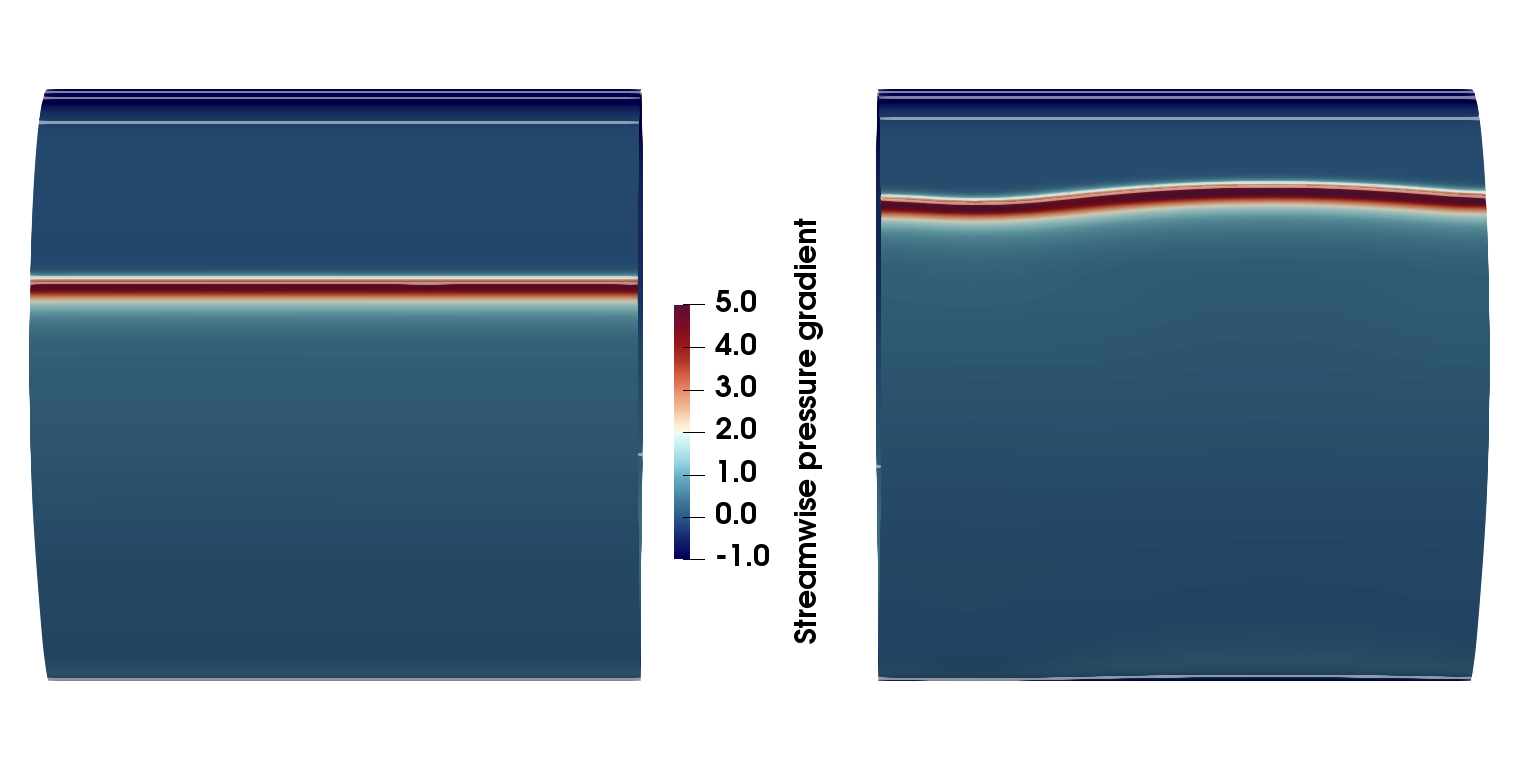}
  \caption{Instantaneous streamwise pressure gradient on the airfoil surface for URANS cases at $AR=1$. Showing angles of attack of (left) $\alpha = 5^{\circ}$, and (right) $\alpha = 7^{\circ}$. The white lines show contours of skin-friction in both cases.}
\label{fig:FaSTAR_URANS_AR_sensitvity_top-view}
\end{center}
\end{figure}

The domain-width sensitivity studies have shown that while the pressure side of supercritical airfoils for type II buffet is largely insensitive to domain width, both the SBLI and separated regions on the suction side show strong sensitivity to domain width within the range of values typically used in the literature. At moderate AoA close to buffet onset $(\alpha_{\textrm{onset}} +0.5^{\circ})$ with minimal flow separation, similar frequencies are predicted, but narrow-span simulations ($AR=0.05$) under-predict lift, show less regularity of the buffet cycles, and over-predict trailing edge separations which are un-physically constrained. At higher AoA with more extensive flow separation narrow-span simulations are shown to be highly inadequate. While high-fidelity simulations offer improvements over low-fidelity methods by reduced modelling and explicit resolution of turbulence, care must be taken not to artificially constrain the problem with overly-narrow domains. It is also clear that domain width sensitivity studies should be performed at representative flow configurations (at the highest AoA considered in the work), where the span-wise width requirements are at their most critical. 

In this final section we aim to address the disparity in behavior between moderately separated buffet ($\alpha = 5^{\circ}$) which converges with increasing span width (Figure~\ref{fig:domain_deg5_lines}), and the higher AoA case with extensive flow separation which does not ($\alpha = 7^{\circ}$, Figure~\ref{fig:domain_deg7_lines}). To do this, we extend the 2D URANS simulations from Section~{\ref{sec:URANS}} to 3D with an even wider aspect ratio of $AR=1$. The 2D URANS grid is extruded in the spanwise direction and discretized using 21 cells. Similarly to the 2D unsteady calculations, the 3D URANS are started from uniform flow conditions. Figure~\ref{fig:FaSTAR_URANS_AR_sensitvity_histories} shows the $C_L$ histories for 2D and 3D ($AR=1$) simulations at (a) $\alpha = 5^{\circ}$ and (b) $\alpha = 7^{\circ}$). For the case with moderate flow separation, the predictions from the 2D URANS and 3D URANS are identical. In each case the 2D shock-oscillation buffet mode is reproduced and the extension to a 3D simulation and a wide domain has no influence during the initial transient or when the nonlinear dynamics fully saturates and buffet is fully developed. However, at $\alpha = 7^{\circ}$, despite having a similar buffet frequency and amplitude during the initial transient, the 2D and 3D URANS calculations diverge. The amplitude of the 3D simulations drops significantly, and additional frequencies are observed beyond the 2D case.

\begin{figure}
\begin{center}
 \includegraphics[width=0.495\columnwidth]{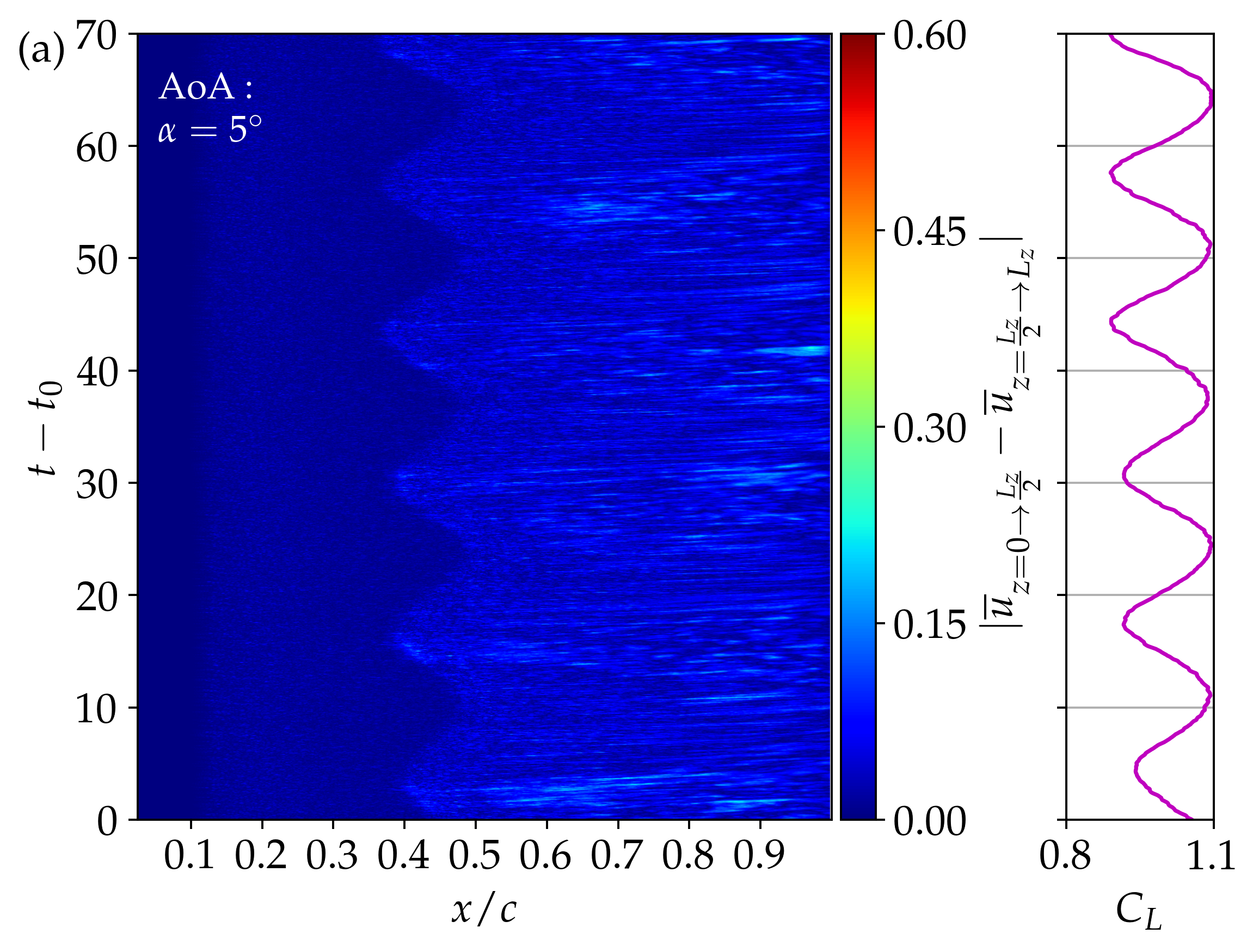}
 \includegraphics[width=0.495\columnwidth]{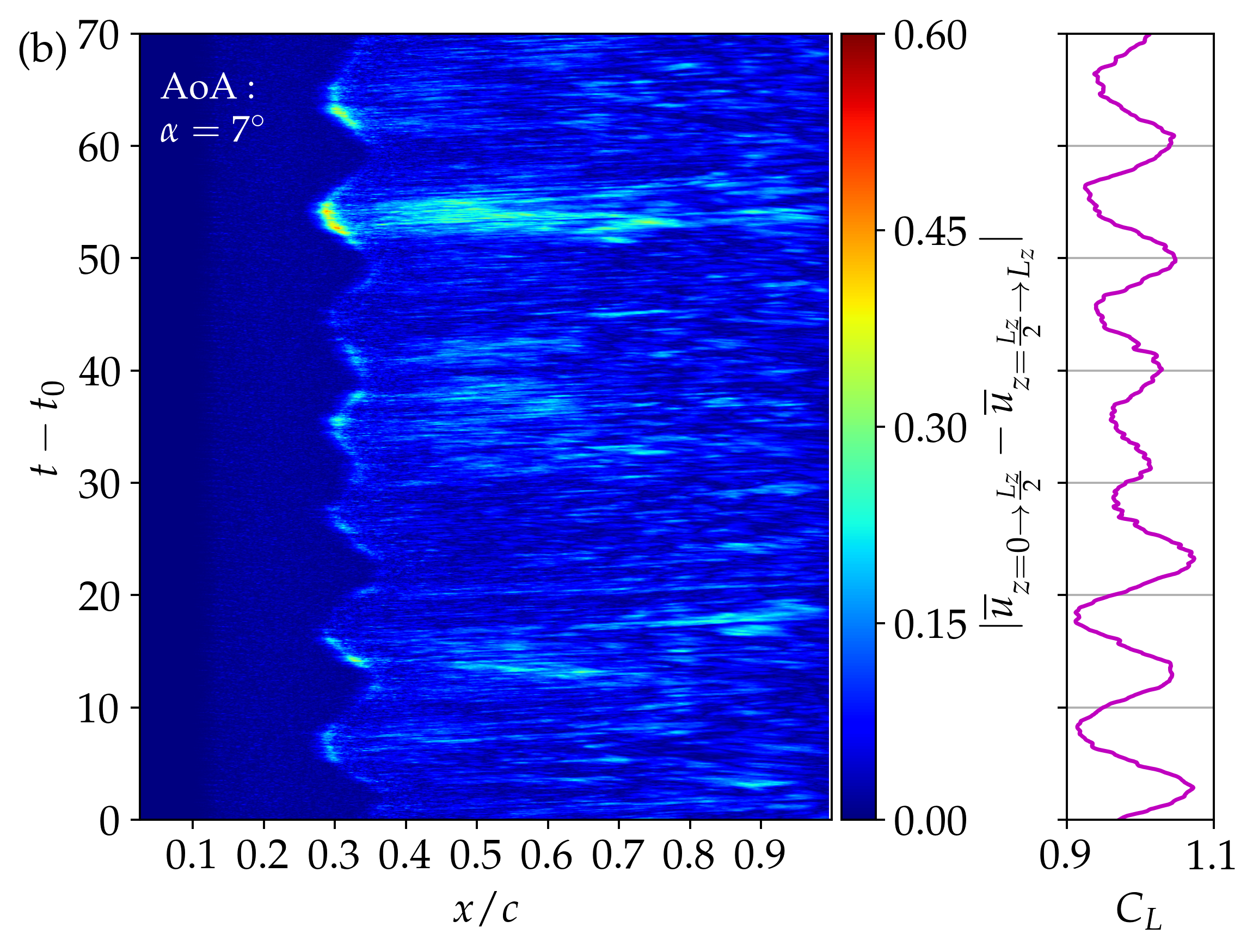}
  \caption{Three-dimensional variations across the span for the ILES data at $AR=0.5$. Showing the absolute difference in the streamwise velocity when averaged over the first and second half of the span individually. Showing the evolution in time for angles of attack of (a) $\alpha = 5^{\circ}$, and (b) $\alpha = 7^{\circ}$.}
\label{fig:OpenSBLI_wide_top-view}
\end{center}
\end{figure}

Figure~\ref{fig:FaSTAR_URANS_AR_sensitvity_top-view} shows instantaneous streamwise pressure gradient plots on the suction side of the airfoil for the URANS at $AR=1$, at (left) $\alpha = 5^{\circ}$ and (right) $\alpha = 7^{\circ}$. While the lower AoA case has a shock with an entirely normal orientation to the oncoming flow regardless of the spanwise position, the high AoA case shows modulation of the shock-front across the span and is no longer quasi-2D across the periodic direction. This indicates that the two-dimensionality of the buffet phenomenon has been lost, and, at this higher AoA, three-dimensional effects such as buffet/stall-cells (commonly seen with aspect ratio wavelengths around unity and above \citep{Giannelis_buffet_review,PDL2020, PDBLS2021}), may be starting to affect the role that the spanwise domain width has. Figure~\ref{fig:OpenSBLI_wide_top-view} shows $x$-$t$ diagrams for the ILES data on the widest domain of $AR=0.5$, for the cases of (a) $\alpha = 5^{\circ}$, and (b) $\alpha = 7^{\circ}$. At each time instance, the quantity plotted is the absolute difference between the $u$-velocity component inside the boundary-layer when averaged over the first and second halves of the span-wise width ($z = 0 \to \frac{L_z}{2}$ and $z = \frac{L_z}{2} \to L_z$). For a purely two-dimensional phenomenon with no significant variation across the spanwise direction, both halves of the span should have the same mean value. In the moderate AoA case ($\alpha = 5^{\circ}$) this is largely true, with only small variations seen between the mean value in each of the two halves of the airfoil. However, at the higher AoA this is no longer the case. Significant variations are now observed between the two halves of the airfoil span. These are most notable around mid-chord and at the main SBLI. Interestingly, the peaks of the variation are strongest during the low-lift phases of the buffet cycle when the shock-wave is at its farthest upstream location, with reduced activity at the point of maximum lift.

\section{Conclusions}\label{sec:conclusions}

Turbulent transonic buffet has been investigated with high-fidelity scale-resolving (ILES) simulations on periodic unswept wing configurations of the NASA-CRM geometry. Data from the baseline ILES cases at pre- and post-onset buffet conditions was cross-validated to low-fidelity (RANS/URANS) and global stability (GSA) methods. Excellent agreement was observed between the methods, in terms of the average pressure distributions (ILES/RANAS), onset criteria (ILES, GSA), plus unsteady histories of lift and drag and their low-frequency oscillations (ILES/URANS).

After validating the initial results, the sensitivity of buffet characteristics to both upstream boundary-layer state (via wall-tripping of varying strength) and span-wise domain width were then investigated via ILES. The role of the boundary-layer state upstream of the main shock-wave for transonic airfoil buffet was investigated by varying the tripping amplitude over the range of $0.5\%$ to $10\%$ of freestream velocity. It was found that buffet with only a single main shock-wave can persist for both laminar and turbulent boundary-layers with similar low-frequency oscillations and periodic flow separation and reattachment. In addition to the two limiting regimes (laminar/turbulent), we show there also exists a transitional interaction for which the main shock-wave is more stationary and higher frequencies are present. The laminar and transitional cases exhibit far larger mean flow separation compared to the turbulent case, but the transitional case does not share the same separation bubble shape as the laminar ones. The transitional case exhibited $13.5\%$ and $7\%$ higher mean lift than the turbulent and laminar cases, respectively.

The role of the span-wise domain extent was assessed at post-onset for a fully-turbulent case with moderate flow separation ($\alpha = 5^{\circ}$), and post-onset with large flow separation ($\alpha = 7^{\circ}$). Domains of up to $AR=0.5$ were simulated with ILES, up to an order of magnitude wider than what commonly used in the literature. It was observed that narrow domains ($AR=0.025, 0.05$) commonly used in high-fidelity simulations of buffet were insufficient to provide domain insensitive solutions. At pre- and post-onset with minimal flow separation, although the main low-frequency buffet peak was largely insensitive to domain width, at least $AR=0.1$ was necessary for domain independent amplitudes and pressure/skin-friction distributions. The main discrepancy was found at the SBLI and at the trailing edge. The separation at the trailing edge was found to be overly-constrained on narrow domains, decreasing in extent as the domain was widened, and the flow near the trailing edge had higher RMS pressure fluctuations compared to solutions that converged on wider domains. At higher AoA with more extensive flow separation, even greater span-width sensitivity was observed. Simulations on $AR=0.05$ domains failed to reproduce clear low-frequency buffet peaks, and led to strongly diverged predictions of pressure distributions and trailing edge separation. This indicates that high-fidelity airfoil simulations should assess domain sensitivity on the highest AoA considered, as there is a strong dependence on increasing AoA where flow separation becomes larger. A summary of the cases showed that the span-width must be at least greater than the boundary-layer thickness at the trailing edge to obtain domain-independent solutions. 

Furthermore, unlike the case at moderate AoA close to onset ($\alpha = 5^{\circ}$), for which the shock-oscillations remained quasi-2D even up to $AR=0.5$ (ILES) and $AR=1$ (URANS), additional frequencies and non-2D features were observed at $\alpha = 5^{\circ}$. It's possible that close to onset ($\alpha = 5^{\circ}$) the problem remains essentially 2D, whereas at $\alpha = 7^{\circ}$ 3D buffet/stall-cell phenomena start appearing, which will have a spanwise wavelength determined by the aspect ratio used. Instantaneous surface plots and $z-t$ time histories showed that the moderate AoA case remains essentially 2D while the higher AoA started to show 3D effects across the span. The study of 3D buffet effects on wider aspect ratios will be the topic of a future work.

\section*{Appendix}

\subsection{Sensitivity to spanwise grid resolution}\label{sec:grid_study}

\begin{figure}
\begin{center}
  \includegraphics[width=1\columnwidth]{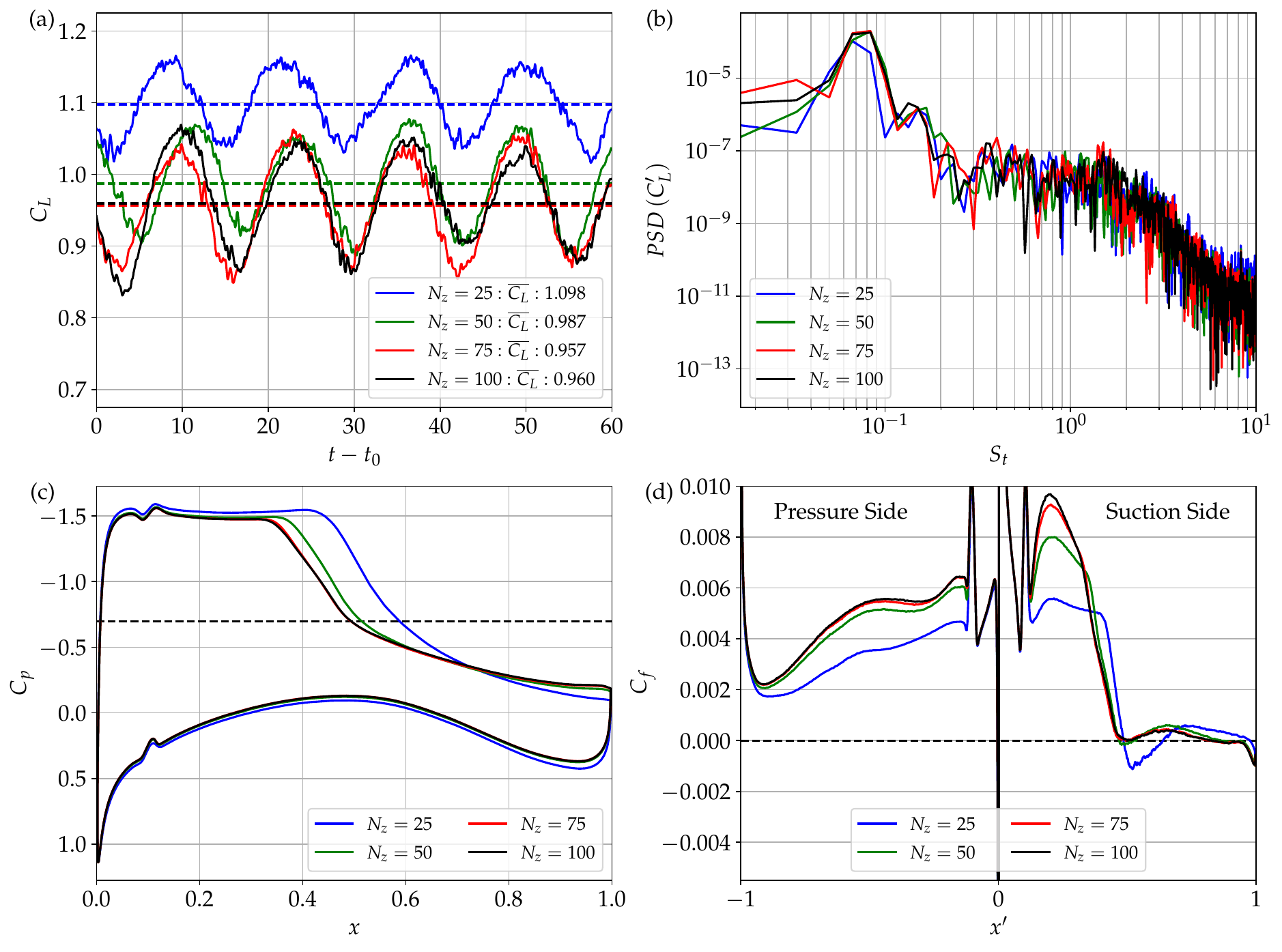}
  \caption{Sensitivity to spanwise grid resolution for the baseline case of $M=0.72$, $Re=500,000$, $AR=0.05$, $\alpha = 5^{\circ}$. Showing (a) lift coefficient history (b) PSD of fluctuating lift component (c) time-averaged pressure coefficient and (d) time-averaged skin-friction.}
\label{fig:grid_study_lines}
\end{center}
\end{figure}

\begin{figure}
\begin{center}
  \includegraphics[width=0.497\columnwidth]{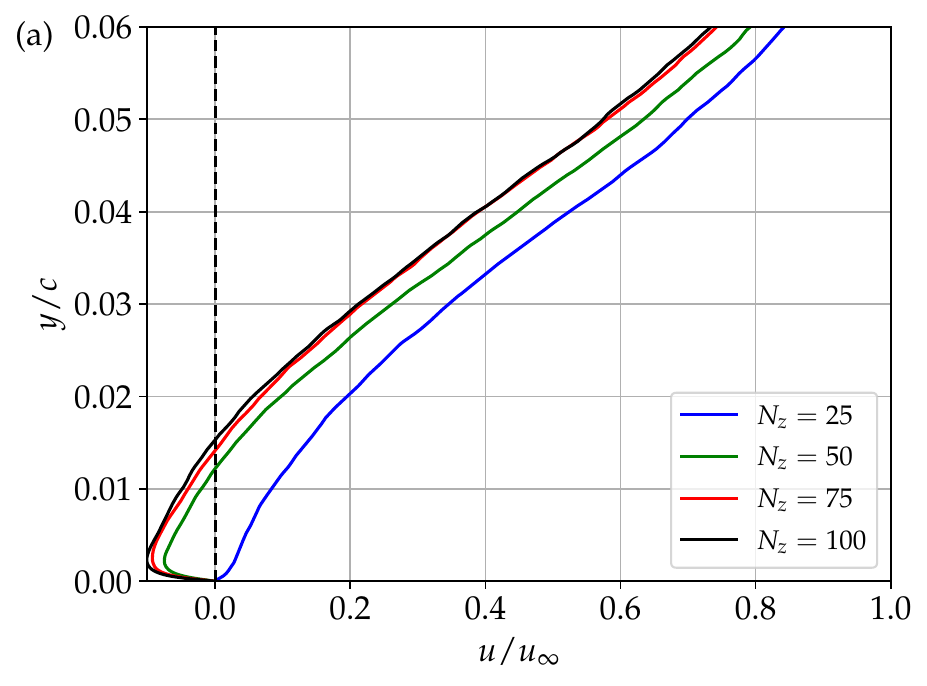}
  \includegraphics[width=0.497\columnwidth]{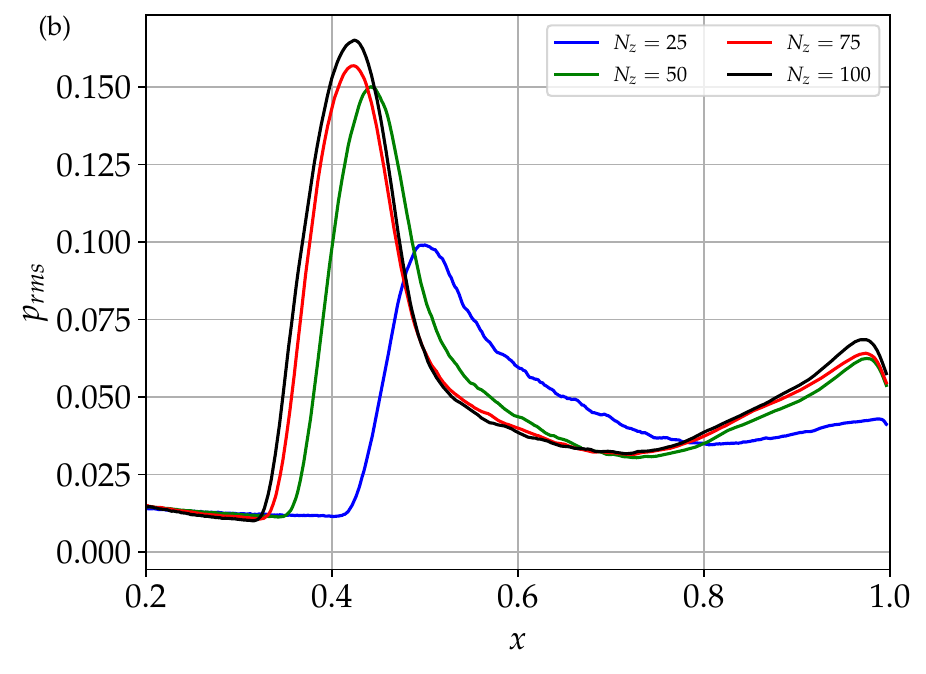}\\
  \caption{Spanwise grid resolution sensitivity for (a) trailing edge streamwise velocity boundary-layer profiles and (b) RMS of pressure fluctuations.}
\label{fig:grid_study_TE_BL}
\end{center}
\end{figure}

Given the large number of cases in the present study with aspect ratios up to an order of magnitude wider than commonly used for high-fidelity buffet simulations (i.e. $AR=0.5$ compared to commonly used $AR=0.05$ \citep{moise_zauner_sandham_2022,Moise2023_AIAAJ} or $AR=0.065$ \citep{FK2018,Nguyen2022}), it was necessary to a identify a moderate spanwise grid resolution that could reproduce the main buffet dynamics while still remaining computationally feasible for the wider aspect ratios. This is of particular relevance in the case of buffet, where long time integration is required to capture multiple cycles of the low-frequency phenomenon. With this in mind, Figure~\ref{fig:grid_study_lines} shows aerodynamic coefficients for the $AR=0.05$ baseline case at $\alpha=5^{\circ}$, with four spanwise grid resolutions of $N_z = \left[25, 50, 75, 100\right]$. Despite the factor of four difference in spanwise resolution, all resolutions tested reproduce the low-frequency buffet oscillation, with peaks in the same Strouhal number range (figure~\ref{fig:grid_study_lines} (b)). The main difference in the lift is an over-prediction of $\overline{C_L}$ for the coarsest $N_z = 25$ grid. The medium resolution ($N_z=50$) follows the finer meshes well, with a slight over-prediction of $\overline{C_L}$ by $2.8\%$. Similarly, the pressure coefficient in Figure~\ref{fig:grid_study_lines} (c) shows a large downstream shift of pressure gradient  near mid-chord at $N_z=25$, and a much smaller one at $N_z=50$, close to the finer mesh results. While the coarsest grid also diverges from the finer solutions on the pressure side, transition location, and near the trailing edge, the medium resolution agrees well in these areas. In a similar vein, the $N_z=50$ resolution matches the finer mesh skin-friction distributions very well in Figure~\ref{fig:grid_study_lines} (d), in terms of the shock location, SBLI separation and reattachment locations, and trailing edge separation. Figure~\ref{fig:grid_study_TE_BL} shows the same cases for the (a) mean streamwise velocity boundary-layer profile at $97.5\%$ chord, and (b) RMS of pressure fluctuations on the suction side of the airfoil. In both cases the coarsest $N_z=25$ case shows to be insufficient, prediction attached flow in the region that is separated at the finer resolutions, and large disparity in $p_{rms}$. In contrast, the medium $N_z=50$ resolution matches the behaviour well both in terms of boundary-layer profile and $p_{rms}$ compared to the finer meshes. Importantly, the separation magnitude and size in figure~\ref{fig:grid_study_TE_BL} (a) are actually slightly larger on the refined grids, suggesting that the separation based domain sensitivity conclusions reported in section~\ref{sec:moderate_separation} and section~\ref{sec:large_separation} are likely to be more prevalent on finer mesh solutions closer to a full DNS. For these reasons, the $N_z=50$ medium resolution is selected to perform the wide-aspect ratio sensitivity tests in this work.

\section*{Funding Sources}

Dr. David J. Lusher is funded by the Japan Society for the Promotion of Science (JSPS), on a postdoctoral fellowship awarded to the JAXA Chofu Aerospace Center. Additional funding was provided by a JSPS KAKENHI grant award (22F22059). The numerical work on global stability in FaSTAR was partially supported by the JSPS KAKENHI Grant-in-Aid for Early-Career Scientists 20K14953 awarded to Dr. Andrea Sansica.

\section*{Acknowledgments}
Computational time was provided by the JAXA JSS3 supercomputing facility and associated support staff, and the Fugaku supercomputer at RIKEN on projects hp220195, hp220226. The authors thank Dr. Markus Zauner for the help with initial testing on the URANS setup.

\bibliography{sample}

\begin{thebibliography}{77}
\newcommand{\enquote}[1]{``#1''}
\providecommand{\natexlab}[1]{#1}
\providecommand{\url}[1]{\texttt{#1}}
\providecommand{\urlprefix}{URL }
\expandafter\ifx\csname urlstyle\endcsname\relax
  \providecommand{\doi}[1]{\discretionary{}{}{}https://doi.org/#1}\else
  \providecommand{\doi}[1]{\discretionary{}{}{}\urlstyle{rm}\url{https://doi.org/#1}}\fi

\bibitem[{Dolling(2001)}]{D2001}
Dolling, D., \enquote{Fifty years of shock-wave/boundary-layer interaction
  research: what next?} \emph{AIAA Journal}, Vol. 39(8), 2001, pp. 1517--1531.

\bibitem[{Lee(1990)}]{L1990}
Lee, B. H.~K., \enquote{Oscillatory shock motion caused by transonic shock
  boundary-layer interaction,} \emph{AIAA Journal}, Vol. 28(5), 1990, pp.
  942--944.

\bibitem[{Lee(2001)}]{L2001}
Lee, B. H.~K., \enquote{Self-sustained shock oscillations on airfoils at
  transonic speeds,} \emph{Progress in Aerospace Sciences}, Vol.~37, 2001, pp.
  147--196.

\bibitem[{Giannelis et~al.(2017)Giannelis, Vio, and
  Levinski}]{Giannelis_buffet_review}
Giannelis, N.~F., Vio, G.~A., and Levinski, O., \enquote{A review of recent
  developments in the understanding of transonic shock buffet,} \emph{Progress
  in Aerospace Sciences}, Vol.~92, 2017, pp. 39--84.
\newblock \doi{https://doi.org/10.1016/j.paerosci.2017.05.004}.

\bibitem[{Plante(2020)}]{plante2020towards}
Plante, F., \enquote{Towards understanding stall cells and transonic buffet
  cells,} Ph.D. thesis, Ecole Polytechnique, Montreal (Canada), 2020.

\bibitem[{Crouch et~al.(2009)Crouch, Garbaruk, Magidov, and Travin}]{CGMT2009}
Crouch, J.~D., Garbaruk, A., Magidov, D., and Travin, A., \enquote{Origin of
  Transonic Buffet on Aerofoils,} \emph{Journal of Fluid Mechanics}, Vol. 628,
  2009, pp. 357--369.

\bibitem[{Crouch et~al.(2019)Crouch, Garbaruk, and Strelets}]{CGS2019}
Crouch, J.~D., Garbaruk, A., and Strelets, M., \enquote{Global instability in
  the onset of transonic-wing buffet,} \emph{Journal of Fluid Mechanics}, Vol.
  881, 2019, pp. 3--22.

\bibitem[{Paladini et~al.(2019)Paladini, Beneddine, Dandois, Sipp, and
  Robinet}]{PBDSR2019}
Paladini, E., Beneddine, S., Dandois, J., Sipp, D., and Robinet, J.-C.,
  \enquote{Transonic buffet instability: From two-dimensional airfoils to
  three-dimensional swept wings,} \emph{Phys. Rev. Fluids}, Vol.~4, 2019, p.
  103906.
\newblock \doi{10.1103/PhysRevFluids.4.103906},
  \urlprefix\url{https://link.aps.org/doi/10.1103/PhysRevFluids.4.103906}.

\bibitem[{Jacquin et~al.(2009)Jacquin, Molton, Deck, Maury, and
  Soulevant}]{JMDMS2009}
Jacquin, L., Molton, P., Deck, S., Maury, B., and Soulevant, D.,
  \enquote{Experimental study of shock oscillation over a transonic
  supercritical profile,} \emph{AIAA Journal}, Vol.~47, 2009, pp. 1985--1994.

\bibitem[{Aihara and Kawai(2023)}]{AK2023}
Aihara, A., and Kawai, S., \enquote{Effects of Spanwise Domain Size on
  LES-Predicted Aerodynamics of Stalled Airfoil,} \emph{AIAA Journal}, Vol.~61,
  No.~3, 2023, pp. 1440--1446.
\newblock \doi{10.2514/1.J062375}.

\bibitem[{Iovnovich and Raveh(2015)}]{IR2015}
Iovnovich, M., and Raveh, D.~E., \enquote{Numerical study of shock buffet on
  three-dimensional wings,} \emph{AIAA Journal}, Vol. 53(2), 2015, pp.
  449--463.

\bibitem[{Plante et~al.(2020)Plante, Dandois, and Laurendeau}]{PDL2020}
Plante, F., Dandois, J., and Laurendeau, E., \enquote{Similarities between
  cellular patterns occurring in transonic buffet and subsonic stall,}
  \emph{AIAA Journal}, Vol.~58, No.~1, 2020, pp. 71--84.
\newblock \doi{10.2514/1.J058555}.

\bibitem[{Plante et~al.(2021)Plante, Dandois, Beneddine, Laurendeau, and
  Sipp}]{PDBLS2021}
Plante, F., Dandois, J., Beneddine, S., Laurendeau, E., and Sipp, D.,
  \enquote{Link between subsonic stall and transonic buffet on swept and
  unswept wings: from global stability analysis to nonlinear dynamics,}
  \emph{Journal of Fluid Mechanics}, Vol. 908, 2021.
\newblock \doi{10.1017/jfm.2020.848}.

\bibitem[{Sansica et~al.(2022{\natexlab{a}})Sansica, Hashimoto, Koike, and
  Kouchi}]{SHKK2022}
Sansica, A., Hashimoto, A., Koike, S., and Kouchi, T., \enquote{Side-Wall
  Effects on the Global Stability of Swept and Unswept Supercritical Wings at
  Buffet Conditions,} \emph{AIAA SCITECH 2022 Forum, AIAA Paper 2022-1972},
  2022{\natexlab{a}}.
\newblock \doi{10.2514/6.2022-1972}.

\bibitem[{Ohmichi et~al.(2018)Ohmichi, Ishida, and Hashimoto}]{OIH2018}
Ohmichi, Y., Ishida, T., and Hashimoto, A., \enquote{Modal decomposition
  analysis of three-dimensional transonic buffet phenomenon on a swept wing,}
  \emph{AIAA Journal}, Vol.~56, No.~10, 2018, pp. 3938--3950.
\newblock \doi{10.2514/1.J056855}.

\bibitem[{Masini et~al.(2020)Masini, Timme, and Peace}]{MTP2020}
Masini, L., Timme, S., and Peace, A.~J., \enquote{Analysis of a civil aircraft
  wing transonic shock buffet experiment,} \emph{Journal of Fluid Mechanics},
  Vol. 884, 2020.

\bibitem[{Timme(2020)}]{T2020}
Timme, S., \enquote{Global instability of wing shock-buffet onset,}
  \emph{Journal of Fluid Mechanics}, Vol. 885, 2020.

\bibitem[{Sugioka et~al.(2021)Sugioka, Nakakita, Koike, Nakajima, Nonomura, and
  Asai}]{SNKNNA2021}
Sugioka, Y., Nakakita, K., Koike, S., Nakajima, T., Nonomura, T., and Asai, K.,
  \enquote{Characteristic unsteady pressure field on a civil aircraft wing
  related to the onset of transonic buffet,} \emph{Experiments in Fluids}, ,
  No. 62:20, 2021.

\bibitem[{Sansica and Hashimoto(2023)}]{SH2023}
Sansica, A., and Hashimoto, A., \enquote{Global Stability Analysis of
  Full-Aircraft Transonic Buffet at Flight Reynolds Numbers,} \emph{AIAA
  Journal}, Vol.~61, No.~10, 2023, pp. 4437--4455.
\newblock \doi{10.2514/1.J062808}.

\bibitem[{Tinoco(2019)}]{T2019dpw}
Tinoco, E.~N., \enquote{An Evaluation and Recommendations for Further CFD
  Research Based on the NASA Common Research Model (CRM) Analysis from the AIAA
  Drag Prediction Workshop (DPW) Series,} \emph{NASA/CR-2019-220284}, 2019, pp.
  13--17.

\bibitem[{Tinoco et~al.(2018)Tinoco, Brodersen, Keye, Laflin, Feltrop, Wahls,
  Morrison, Vassberg, Mani, Rider, Hue, Roy, Mavriplis, and
  Murayama}]{dpw-summ1}
Tinoco, E.~N., Brodersen, O.~P., Keye, S., Laflin, K.~R., Feltrop, E., Wahls,
  R.~A., Morrison, J.~H., Vassberg, J.~C., Mani, M., Rider, B., Hue, D., Roy,
  C.~J., Mavriplis, D.~J., and Murayama, M., \enquote{Summary of Data from the
  Sixth AIAA CFD Drag Prediction Workshop: CRM Cases 2 to 5,} \emph{Journal of
  Aircraft}, Vol.~55, No.~4, 2018, pp. 1352--1379.
\newblock \doi{10.2514/1.C034409}.

\bibitem[{Sartor et~al.(2015)Sartor, Mettot, and Sipp}]{SMS2015}
Sartor, F., Mettot, C., and Sipp, D., \enquote{Stability, receptivity, and
  sensitivity analyses of buffeting transonic flow over a profile,} \emph{AIAA
  Journal}, Vol. 53(7), 2015, pp. 1980--1933.

\bibitem[{Deck(2005)}]{D2005}
Deck, S., \enquote{Numerical simulation of transonic buffet over a
  supercritical airfoil,} \emph{AIAA Journal}, Vol. 43(7), 2005, pp.
  1556--1566.

\bibitem[{Hartmann et~al.(2013)Hartmann, Feldhusen, and Shroder}]{HFS2013}
Hartmann, A., Feldhusen, A., and Shroder, W., \enquote{On the interaction of
  shock waves and sound waves in transonic buffet flow,} \emph{AIAA Journal},
  Vol.~28, No. 942, 2013.
\newblock \doi{10.1063/1.4791603}.

\bibitem[{Iwatani et~al.(2023)Iwatani, Asada, Yeh, Taira, and
  Kawai}]{KAWAI_RESOLVENT2023}
Iwatani, Y., Asada, H., Yeh, C.-A., Taira, K., and Kawai, S.,
  \enquote{Identifying the Self-Sustaining Mechanisms of Transonic Airfoil
  Buffet with Resolvent Analysis,} \emph{AIAA Journal}, Vol.~61, No.~6, 2023,
  pp. 2400--2411.
\newblock \doi{10.2514/1.J062294}.

\bibitem[{Thiery and Coustols(2006)}]{TC2006}
Thiery, M., and Coustols, E., \enquote{Numerical prediction of shock induced
  oscillations over a {2D} airfoil: Influence of turbulence modelling and test
  section walls,} \emph{International Journal of Heat and Fluid Flow}, Vol.~27,
  2006, pp. 661--670.

\bibitem[{Poplinger et~al.(2019)Poplinger, Raveh, and Dowell}]{PRD2019}
Poplinger, L., Raveh, D.~E., and Dowell, E.~H., \enquote{Modal analysis of
  transonic shock buffet on 2D airfoil,} \emph{AIAA Journal}, Vol.~57, No.~7,
  2019.

\bibitem[{Sansica et~al.(2022{\natexlab{b}})Sansica, Loiseau, Kanamori, and
  Robinet}]{SLKHR2022}
Sansica, A., Loiseau, J.-C., Kanamori, A., M.~Hashimoto, and Robinet, J.-C.,
  \enquote{System Identification of Two-Dimensional Transonic Buffet,}
  \emph{AIAA Journal}, Vol.~60, No.~5, 2022{\natexlab{b}}.

\bibitem[{Grossi et~al.(2014)Grossi, Braza, and Hoarau}]{GBH2014}
Grossi, F., Braza, M., and Hoarau, Y., \enquote{Prediction of Transonic Buffet
  by Delayed Detached-Eddy Simulation,} \emph{AIAA Journal}, Vol.~52, 2014, pp.
  2300--2312.

\bibitem[{Fukushima and Kawai(2018)}]{FK2018}
Fukushima, Y., and Kawai, S., \enquote{Wall-modeled Large-Eddy Simulation of
  transonic airfoil buffet at high Reynolds number,} \emph{AIAA Journal}, Vol.
  56(6), 2018, pp. 1--18.

\bibitem[{Memmolo et~al.(2018)Memmolo, Bernardini, and Pirozzoli}]{MBP2018}
Memmolo, A., Bernardini, M., and Pirozzoli, S., \enquote{Scrutiny of buffet
  mechanisms in transonic flow,} \emph{International Journal of Numerical
  Methods for Heat and Fluid Flow}, Vol.~28, 2018, pp. 1031--1046.

\bibitem[{Grinstein et~al.(2007)Grinstein, Margolin, and
  Rider}]{grinstein2007implicit}
Grinstein, F.~F., Margolin, L.~G., and Rider, W.~J., \emph{Implicit large eddy
  simulation}, Vol.~10, Cambridge university press Cambridge, 2007.

\bibitem[{Garnier et~al.(1999)Garnier, Mossi, Sagaut, Comte, and
  Deville}]{Garnier_ILES_1999}
Garnier, E., Mossi, M., Sagaut, P., Comte, P., and Deville, M., \enquote{On the
  Use of Shock-Capturing Schemes for Large-Eddy Simulation,} \emph{Journal of
  Computational Physics}, Vol. 153, No.~2, 1999, pp. 273--311.
\newblock \doi{https://doi.org/10.1006/jcph.1999.6268}.

\bibitem[{Fu(2023)}]{fu2023review}
Fu, L., \enquote{{Review of the high-order TENO schemes for compressible gas
  dynamics and turbulence},} \emph{Archives of Computational Methods in
  Engineering}, Vol.~30, No.~4, 2023, pp. 2493--2526.

\bibitem[{Garnier and Deck(2013)}]{GD2010}
Garnier, E., and Deck, S., \emph{Large-Eddy Simulation of transonic buffet over
  a supercritical airfoil}, Turbulence and Interactions (ed. M. Deville, T.-H.
  Le and P. Sagaut). Springer, Berlin, Heidelberg, 2013.

\bibitem[{Turner and Kim(2020)}]{TJW2020}
Turner, J.~M., and Kim, J.~W., \enquote{{Effect of spanwise domain size on
  direct numerical simulations of airfoil noise during flow separation and
  stall},} \emph{Physics of Fluids}, Vol.~32, No.~6, 2020, p. 065103.
\newblock \doi{10.1063/5.0009664}.

\bibitem[{Nguyen et~al.(2022)Nguyen, Terrana, and Peraire}]{Nguyen2022}
Nguyen, N.~C., Terrana, S., and Peraire, J., \enquote{Large-Eddy Simulation of
  Transonic Buffet Using Matrix-Free Discontinuous Galerkin Method,} \emph{AIAA
  Journal}, Vol.~60, No.~5, 2022, pp. 3060--3077.
\newblock \doi{10.2514/1.J060459}.

\bibitem[{Moise et~al.(2022)Moise, Zauner, and
  Sandham}]{moise_zauner_sandham_2022}
Moise, P., Zauner, M., and Sandham, N.~D., \enquote{Large-eddy simulations and
  modal reconstruction of laminar transonic buffet,} \emph{Journal of Fluid
  Mechanics}, Vol. 944, 2022, p. A16.
\newblock \doi{10.1017/jfm.2022.471}.

\bibitem[{Moise et~al.(2023)Moise, Zauner, Sandham, Timme, and
  He}]{Moise2023_AIAAJ}
Moise, P., Zauner, M., Sandham, N.~D., Timme, S., and He, W.,
  \enquote{Transonic Buffet Characteristics Under Conditions of Free and Forced
  Transition,} \emph{AIAA Journal}, Vol.~61, No.~3, 2023, pp. 1061--1076.
\newblock \doi{10.2514/1.J062362}.

\bibitem[{Song et~al.()Song, Wong, Ghate, and
  Lele}]{LongWong2024_Laminar_buffet}
Song, H., Wong, M.~L., Ghate, A.~S., and Lele, S.~K., \emph{Numerical study of
  transonic laminar shock buffet on the OALT25 airfoil}, ????, Chaps. AIAA
  2024-2148.
\newblock \doi{10.2514/6.2024-2148}.

\bibitem[{Zauner and Sandham(2020)}]{MarkusPRF_2020}
Zauner, M., and Sandham, N.~D., \enquote{Wide domain simulations of flow over
  an unswept laminar wing section undergoing transonic buffet,} \emph{Phys.
  Rev. Fluids}, Vol.~5, 2020, p. 083903.
\newblock \doi{10.1103/PhysRevFluids.5.083903},
  \urlprefix\url{https://link.aps.org/doi/10.1103/PhysRevFluids.5.083903}.

\bibitem[{Dandois et~al.(2018)Dandois, Mary, and
  Brion}]{dandois_mary_brion_2018}
Dandois, J., Mary, I., and Brion, V., \enquote{Large-eddy simulation of laminar
  transonic buffet,} \emph{Journal of Fluid Mechanics}, Vol. 850, 2018, p.
  156–178.
\newblock \doi{10.1017/jfm.2018.470}.

\bibitem[{{Brion, V.} et~al.(2017){Brion, V.}, {Dandois, J.}, {Abart, J.-C.},
  and {Paillart, P.}}]{Brion2017_Laminar_Experiment}
{Brion, V.}, {Dandois, J.}, {Abart, J.-C.}, and {Paillart, P.},
  \enquote{Experimental analysis of the shock dynamics on a transonic laminar
  airfoil,} , 2017.
\newblock \doi{10.1051/eucass/2016090365},
  \urlprefix\url{https://doi.org/10.1051/eucass/2016090365}.

\bibitem[{Brion et~al.(2020)Brion, Dandois, Mayer, Reijasse, Lutz, and
  Jacquin}]{Brion2020_Laminar_Experiment}
Brion, V., Dandois, J., Mayer, R., Reijasse, P., Lutz, T., and Jacquin, L.,
  \enquote{Laminar buffet and flow control,} \emph{Proceedings of the
  Institution of Mechanical Engineers, Part G: Journal of Aerospace
  Engineering}, Vol. 234, No.~1, 2020, pp. 124--139.
\newblock \doi{10.1177/0954410018824516}.

\bibitem[{Garbaruk et~al.(2021)Garbaruk, Strelets, and Crouch}]{Garbaruk2021}
Garbaruk, A., Strelets, M., and Crouch, J.~D., \enquote{Effects of Extended
  Laminar Flow on Wing Buffet-Onset Characteristics,} \emph{AIAA Journal},
  Vol.~59, No.~8, 2021, pp. 2848--2854.
\newblock \doi{10.2514/1.J060707}.

\bibitem[{Sugioka et~al.(2018)Sugioka, Koike, Nakakita, Numata, Nonomura, and
  Asai}]{SKNNNA2018}
Sugioka, Y., Koike, S., Nakakita, K., Numata, D., Nonomura, T., and Asai, K.,
  \enquote{Experimental analysis of transonic buffet on a {3D} swept wing using
  fast-response pressure-sensitive paint,} \emph{Experiments in Fluids},
  Vol.~59, No. 108, 2018.

\bibitem[{Sugioka et~al.(2022)Sugioka, Kouchi, and Koike}]{SKK2022}
Sugioka, Y., Kouchi, T., and Koike, S., \enquote{Experimental comparison of
  shock buffet on unswept and 10-deg swept wings,} \emph{Experiments in
  Fluids}, Vol.~63, No. 132, 2022.

\bibitem[{Lusher et~al.(2021)Lusher, Jammy, and Sandham}]{OpenSBLI_2021_CPC}
Lusher, D.~J., Jammy, S.~P., and Sandham, N.~D., \enquote{{OpenSBLI: Automated
  code-generation for heterogeneous computing architectures applied to
  compressible fluid dynamics on structured grids},} \emph{Computer Physics
  Communications}, Vol. 267, 2021, p. 108063.
\newblock \doi{https://doi.org/10.1016/j.cpc.2021.108063}.

\bibitem[{Hashimoto et~al.(2012)Hashimoto, Murakami, Aoyama, Ishiko, Hishida,
  Sakashita, and Lahur}]{FaSTAR_Hashimoto_2012}
Hashimoto, A., Murakami, K., Aoyama, T., Ishiko, K., Hishida, M., Sakashita,
  M., and Lahur, P., \enquote{Toward the Fastest Unstructured CFD Code
  'FaSTAR',} \emph{50th AIAA Aerospace Sciences Meeting including the New
  Horizons Forum and Aerospace Exposition}, , No. AIAA 2012-1075, 2012.
\newblock \doi{10.2514/6.2012-1075}.

\bibitem[{Lusher et~al.(2018)Lusher, Jammy, and Sandham}]{LUSHER201817}
Lusher, D.~J., Jammy, S.~P., and Sandham, N.~D.,
  \enquote{{Shock-wave/boundary-layer interactions in the automatic source-code
  generation framework OpenSBLI},} \emph{Computers \& Fluids}, Vol. 173, 2018,
  pp. 17 -- 21.

\bibitem[{Lusher et~al.(2023)Lusher, Zauner, Sansica, and Hashimoto}]{LZSH2023}
Lusher, D.~J., Zauner, M., Sansica, A., and Hashimoto, A., \enquote{{Automatic
  Code-Generation to Enable High-Fidelity Simulations of Multi-Block Airfoils
  on GPUs},} \emph{AIAA Scitech 2023 Forum, AIAA SciTech Forum}, 2023.

\bibitem[{Reguly et~al.(2014)Reguly, Mudalige, Giles, Curran, and
  McIntosh-Smith}]{Reguly_2014_OPSC}
Reguly, I.~Z., Mudalige, G.~R., Giles, M.~B., Curran, D., and McIntosh-Smith,
  S., \enquote{{The OPS Domain Specific Abstraction for Multi-block Structured
  Grid Computations},} IEEE Press, 2014, pp. 58--67.

\bibitem[{Coppola et~al.(2019)Coppola, Capuano, Pirozzoli, and {de
  Luca}}]{Coppola_CubicSplit_2019}
Coppola, G., Capuano, F., Pirozzoli, S., and {de Luca}, L.,
  \enquote{Numerically stable formulations of convective terms for turbulent
  compressible flows,} \emph{Journal of Computational Physics}, Vol. 382, 2019,
  pp. 86--104.
\newblock \doi{https://doi.org/10.1016/j.jcp.2019.01.007}.

\bibitem[{Carpenter and Kennedy(1994)}]{carpenter_kennedy_1994}
Carpenter, M.~H., and Kennedy, C.~A., \enquote{Fourth-order 2{N}-storage
  {Runge-Kutta} schemes,} \emph{NASA Langley Research Center}, 1994.

\bibitem[{Bogey and Bailly(2004)}]{BOGEY2004194}
Bogey, C., and Bailly, C., \enquote{A family of low dispersive and low
  dissipative explicit schemes for flow and noise computations,} \emph{Journal
  of Computational Physics}, Vol. 194, No.~1, 2004, pp. 194--214.
\newblock \doi{https://doi.org/10.1016/j.jcp.2003.09.003}.

\bibitem[{Borges et~al.(2008)Borges, Carmona, Costa, and Don}]{Borges2008}
Borges, R., Carmona, M., Costa, B., and Don, W.~S., \enquote{{An improved
  weighted essentially non-oscillatory scheme for hyperbolic conservation
  laws},} \emph{Journal of Computational Physics}, Vol. 227, No.~6, 2008, pp.
  3191--3211.

\bibitem[{Lusher and Sandham(2021)}]{lusher2021assessment}
Lusher, D.~J., and Sandham, N.~D., \enquote{Assessment of low-dissipative
  shock-capturing schemes for the compressible Taylor--Green vortex,}
  \emph{AIAA Journal}, Vol.~59, No.~2, 2021, pp. 533--545.

\bibitem[{Hamzehloo et~al.(2021)Hamzehloo, Lusher, Laizet, and
  Sandham}]{hamzehloo2021_IJNMF}
Hamzehloo, A., Lusher, D.~J., Laizet, S., and Sandham, N.~D., \enquote{{On the
  performance of WENO/TENO schemes to resolve turbulence in DNS/LES of
  high-speed compressible flows},} \emph{International Journal for Numerical
  Methods in Fluids}, Vol.~93, No.~1, 2021, pp. 176--196.
\newblock \doi{https://doi.org/10.1002/fld.4879}.

\bibitem[{Yee et~al.(1999)Yee, Sandham, and Djomehri}]{Yee1999_filter_schemes}
Yee, H., Sandham, N., and Djomehri, M., \enquote{Low-Dissipative High-Order
  Shock-Capturing Methods Using Characteristic-Based Filters,} \emph{Journal of
  Computational Physics}, Vol. 150, No.~1, 1999, pp. 199--238.
\newblock \doi{https://doi.org/10.1006/jcph.1998.6177}.

\bibitem[{Yee and Sj{\"{o}}green(2018)}]{Yee2018}
Yee, H.~C., and Sj{\"{o}}green, B., \enquote{{Recent developments in accuracy
  and stability improvement of nonlinear filter methods for DNS and LES of
  compressible flows},} \emph{Computers {\&} Fluids}, Vol. 169, 2018, pp.
  331--348.

\bibitem[{Ducros et~al.(1999)Ducros, Ferrand, Nicoud, Weber, Darracq,
  Gacherieu, and Poinsot}]{DUCROS1999517}
Ducros, F., Ferrand, V., Nicoud, F., Weber, C., Darracq, D., Gacherieu, C., and
  Poinsot, T., \enquote{Large-Eddy Simulation of the Shock/Turbulence
  Interaction,} \emph{Journal of Computational Physics}, Vol. 152, No.~2, 1999,
  pp. 517 -- 549.

\bibitem[{Bhagatwala and Lele(2009)}]{Bhagatwala2009}
Bhagatwala, A., and Lele, S.~K., \enquote{{A modified artificial viscosity
  approach for compressible turbulence simulations},} \emph{Journal of
  Computational Physics}, Vol. 228, No.~14, 2009, pp. 4965--4969.
\newblock \doi{https://doi.org/10.1016/j.jcp.2009.04.009}.

\bibitem[{Ishida et~al.(2016)Ishida, Ishiko, Hashimoto, Aoyama, and
  Takekawa}]{IIHAT2016}
Ishida, T., Ishiko, K., Hashimoto, A., Aoyama, T., and Takekawa, K.,
  \enquote{Transonic buffet simulation over supercritical airfoil by
  unsteady-{FaSTAR} code,} \emph{AIAA Paper 2016-1310}, 2016.

\bibitem[{Obayashi and Guruswamy(1995)}]{O1995}
Obayashi, S., and Guruswamy, G.~P., \enquote{Convergence Acceleration of an
  Aeroelastic {Navier-Stokes} Solver,} \emph{AIAA Journal}, Vol.~33, 1995, pp.
  1134--1141.

\bibitem[{Mavriplis(2003)}]{M2003}
Mavriplis, D.~J., \enquote{Revisiting the Least-Squares Procedure for Gradient
  Reconstruction on Unstructured,} \emph{16th AIAA Computational Fluid Dynamics
  Conference, AIAA paper 2003-3986}, 2003.
\newblock \doi{10.2514/6.2003-3986}.

\bibitem[{Spalart and Allmaras(1992)}]{SA1992}
Spalart, P.~R., and Allmaras, S.~R., \enquote{A One-Equation Turbulence Model
  for Aerodynamic Flows,} \emph{30th Aerospace Sciences Meeting and Exhibit,
  Aerospace Sciences Meetings}, 1992.
\newblock \doi{10.2514/6.1992-439}.

\bibitem[{Dacles-Mariani et~al.(1995)Dacles-Mariani, Zilliac, Chow, and
  Bradshaw}]{DM1995}
Dacles-Mariani, J., Zilliac, G.~G., Chow, J.~S., and Bradshaw, P.,
  \enquote{Numerical/Experimental Study of a Wingtip Vortex in the Near Field,}
  \emph{AIAA Journal}, Vol.~33, 1995, pp. 1561--1568.
\newblock \doi{10.2514/3.12826}.

\bibitem[{Spalart(2000)}]{S2000}
Spalart, P.~R., \enquote{Strategies for Turbulence Modelling and Simulation,}
  \emph{International Journal of Heat and Fluid Flow}, Vol.~21, 2000, pp.
  252--263.
\newblock \doi{10.1016/S0142-727X(00)00007-2}.

\bibitem[{Sharov and Nakahashi(1998)}]{Setail1998}
Sharov, D., and Nakahashi, K., \enquote{Reordering of hybrid unstructured grids
  for lower-upper symmetric {Gauss-Seidel} computations,} \emph{AIAA Journal},
  Vol.~36, 1998, pp. 484--486.
\newblock \doi{10.2514/2.392}.

\bibitem[{Akervik et~al.(2006)Akervik, Brandt, Henningson, Hoepffner, Marxen,
  and Schlatter}]{ABHHMS2006}
Akervik, E., Brandt, L., Henningson, D.~S., Hoepffner, J., Marxen, O., and
  Schlatter, P., \enquote{Steady solutions of the {Navier-Stokes} equations by
  selective frequency damping,} \emph{Physics of Fluids}, Vol.~18, 2006, p.
  068102.
\newblock \doi{10.1063/1.2211705}.

\bibitem[{Richez et~al.(2016)Richez, Lequille, and Marquet}]{RLM2016}
Richez, F., Lequille, M., and Marquet, O., \enquote{Selective frequency damping
  method for steady RANS solutions of turbulent separated flows around an
  airfoil at stall,} \emph{Computers and Fluids}, Vol. 132, 2016, pp. 51--61.
\newblock \doi{10.1016/j.compfluid.2016.03.027}.

\bibitem[{Visbal and Gordnier(2000)}]{Vetal2000}
Visbal, M.~R., and Gordnier, R., \enquote{A high-order flow solver for
  deforming and moving meshes,} \emph{Fluids 2000 Conference and Exhibit, AIAA
  Paper 2000-2619}, 2000.
\newblock \doi{10.2514/6.2000-2619}.

\bibitem[{CRM([Accessed 19 October 2023])}]{CRM}
\enquote{CRM.65.airfoil sections [Online],}
  \emph{Available:https://commonresearchmodel.larc.nasa.
  gov/crm-65-airfoil-sections}, [Accessed 19 October 2023].

\bibitem[{Sandham et~al.(2014)Sandham, Schülein, Wagner, Willems, and
  Steelant}]{Sandham2014_Transition_SBLI}
Sandham, N., Schülein, E., Wagner, A., Willems, S., and Steelant, J.,
  \enquote{Transitional shock-wave/boundary-layer interactions in hypersonic
  flow,} \emph{Journal of Fluid Mechanics}, Vol. 752, 2014, p. 349–382.
\newblock \doi{10.1017/jfm.2014.333}.

\bibitem[{Sansica et~al.(2014)Sansica, Sandham, and Hu}]{Sansica_PoF_2014}
Sansica, A., Sandham, N.~D., and Hu, Z., \enquote{{Forced response of a laminar
  shock-induced separation bubble},} \emph{Physics of Fluids}, Vol.~26, No.~9,
  2014, p. 093601.
\newblock \doi{10.1063/1.4894427},
  \urlprefix\url{https://doi.org/10.1063/1.4894427}.

\bibitem[{Sansica et~al.(2016)Sansica, Sandham, and Hu}]{Sansica_JFM_2016}
Sansica, A., Sandham, N.~D., and Hu, Z., \enquote{Instability and low-frequency
  unsteadiness in a shock-induced laminar separation bubble,} \emph{Journal of
  Fluid Mechanics}, Vol. 798, 2016, p. 5–26.
\newblock \doi{10.1017/jfm.2016.297}.

\bibitem[{Lusher and Sandham(2020)}]{Lusher2020_FTAC}
Lusher, D.~J., and Sandham, N.~D., \enquote{{Shock-Wave/Boundary-Layer
  Interactions in Transitional Rectangular Duct Flows},} \emph{Flow, Turbulence
  and Combustion}, 2020.
\newblock \doi{10.1007/s10494-020-00134-0}.

\end{thebibliography}

\end{document}